# A SMOOTHED DISSIPATIVE PARTICLE DYNAMICS METHODOLOGY FOR WALL-BOUNDED DOMAINS

by

Jun Yang

A Dissertation
Submitted to the Faculty of
WORCESTER POLYTECHNIC INSTITUTE
in partial fulfillment of the requirement for the degree of
Doctor of Philosophy
in
Mechanical Engineering

____________________________________

April 25, 2013

APPROVED:

Dr. Nikolaos A. Gatsonis, Advisor
Professor, Mechanical Engineering Department, WPI

Dr. David J. Olinger, Committee Member
Associate Professor, Mechanical Engineering Department, WPI

Dr. Simon W. Evans, Committee Member
Assistant Professor, Mechanical Engineering Department, WPI

Dr. Marcus Sarkis, Committee Member
Professor, Department of Mathematics, WPI

Dr. George E. Karniadakis, Committee Member
Professor, Division of Applied Mathematics, Brown University

Dr. Mark W. Richman, Graduate Committee Representative
Associate Professor, Mechanical Engineering Department, WPI

# ABSTRACT

This work presents the mathematical and computational aspects of a smooth dissipative particle dynamics with dynamic virtual particle allocation method (SDPD-DV) for modeling and simulation of mesoscopic fluids in wall-bounded domains. The SDPD-DV method is realized with fluid particles, boundary particles and dynamically allocated virtual particles near solid boundaries. The physical domain in SDPD-DV contains external and internal solid boundaries, periodic inlets and outlets, and the fluid region. The solid boundaries of the domain are represented with boundary particles which have an assigned position, wall velocity, and temperature upon initialization. The fluid domain is discretized with fluid particles placed in a global index. The algorithm for nearest neighbor particle search is based on a combination of the linked-cell and Verlet-list approaches and utilizes large rectangular cells for computational efficiency. The density model of a fluid particle in the proximity of a solid boundary includes the contribution from the virtual particles in its truncated support domain. The thermodynamic properties of a virtual particle are identical to those of the corresponding fluid particle. A periodic boundary particle allocation method is used at periodic inlets and outlets. Models for the conservative and dissipative forces on a fluid particle in the proximity of a solid boundary are presented and include the contributions of the virtual particles in its truncated support domain. The integration of the fluid particle position and momentum equations is accomplished with an implementation of the velocity-Verlet algorithm. The integration is supplemented by a bounce-forward algorithm in cases where the virtual particle force model is not able to prevent particle penetration. The integration of the entropy equation is based on the Runge-Kutta scheme. In isothermal simulations, the pressure of a fluid particle is obtained by an artificial compressibility formulation for liquids and the ideal gas law for compressible fluids. Sampling methods used for particle properties and transport coefficients in SDPD-DV are presented. The self-diffusion coefficient is obtained by an implementation of the generalized Einstein and the Green-Kubo relations. Field properties are obtained by sampling SDPD-DV outputs on a post-processing grid that allows harnessing the particle information on desired spatio-temporal scales.

The isothermal (without the entropy equation) SDPD-DV method is verified and validated with simulations in bounded and periodic domains that cover the hydrodynamic and mesoscopic regimes. Verification is achieved with SDPD-DV simulations of transient, Poiseuille, body-force driven flow of liquid water between plates separated by $10^{-3}$ m. The velocity profiles from the SDPD-DV simulations are in very good agreement with analytical estimates and the field density fluctuation near solid boundaries is shown to be below 5%. Additional verification involves SDPD-DV simulations of transient, planar, Couette liquid water flow. The top plate is moving at $V_{xw} = 1.25 \times 10^{-5} m/s$ and separated by $10^{-3}$ m from the bottom stationary plate. The numerical results are in very good agreement with the analytical solutions. Additional SDPD-DV verification is accomplished with the simulation of a body-force driven, low-Reynolds number flow of water over a cylinder of radius $R = 0.02m$. The SDPD-DV field velocity and pressure are compared with those obtained by FLUENT.

An extensive set of SDPD-DV simulations of liquid water and gaseous nitrogen in mesoscopic periodic domains is presented. For the SDPD-DV simulations of liquid water the mass of the fluid particles is varied between 1.24 and $3.3\times10^{7}$ real molecular masses and their corresponding size is between 1.08 and 323 physical length scales. For SDPD-DV simulations of gaseous nitrogen the mass of the fluid particles is varied between $6.37\times10^{3}$ and $6.37\times10^{6}$ real molecular masses and their corresponding size is between $2.2\times10^{2}$ and $2.2\times10^{3}$ physical length scales. The equilibrium states are obtained and show that the particle speeds scale inversely with particle mass (or size) and that the translational temperature is scale-free. The self-diffusion coefficient for liquid water is obtained through the mean-square displacement and the velocity auto-correlation methods for the range of fluid particle masses (or sizes) considered. Various analytical expressions for the self-diffusivity of the SDPD fluid are developed in analogy to the real fluid. The numerical results are in very good agreement with the SDPD-fluid analytical expressions. The numerical self-diffusivity is shown to be scale dependent. For fluid particles approaching asymptotically the mass of the real particle the self-diffusivity is shown to approach the experimental value. The Schmidt numbers obtained from the SDPD-DV simulations are within the range expected for liquid water.

The SDPD-DV method (with entropy) is verified and validated with simulations with an extensive set of simulations of gaseous nitrogen in mesoscopic, periodic domains in equilibrium. The simulations of $N_2$(g) are performed in rectangular domains with $L_X = L_Y = L_Z$ in the range $0.25\times10^{-6}$~$10\times10^{-6}$ m, with fluid mass $M_F$ in the range $1.85\times10^{-20}$~$1.184\times10^{-15}$kg. The mass of the fluid particles is varied between 118 and $7.5\times10^{6}$ real molecular masses. The self-diffusion coefficient for $N_2$(g) at equilibrium states is obtained through the mean-square displacement for the range of fluid particle masses (or sizes) considered. The numerical self-diffusion is shown to be scale dependent. The simulations show that self-diffusion decreases with increasing mass ratio. For a given mass ratio, increasing the smoothing length, increases the self-diffusion coefficient. The shear viscosity obtained from SDPD-DV is shown to be scale free and in good agreement with the real value. We examine also the effects of timestep in SDPD-DV simulations by examining thermodynamic parameters at equilibrium. These results show that the time step can lead to a significant error depending on the fluid particle mass and smoothing length. Fluctuations in thermodynamic variables obtained from SDPD-DV are compared with analytical estimates.

Additional verification involves SDPD-DV simulations of steady planar thermal Couette flow of $N_2$(g). The top plate at temperature $T_1$ =330K is moving at $V_{xw}$ =30m/s and is separated by $10^{-4}$ m from the bottom stationary plate at $T_2$=300K. The SDPD-DV velocity and temperature fields are in excellent agreement with those obtained by FLUENT.

# ACKNOWLEDGEMENTS

I would like to express the deepest appreciation to my advisor, Prof. Gatsonis, who continually and convincingly conveyed a spirit of adventure in regard to research and scholarship, and an excitement in regard to teaching. You have shaped the professional that I have become and further defined my sense of scientific rigor and engineering creativity. Thank you for the years of support and inspiration.

Great thanks are given to Dr. Raffaele Potami whose initial work on SDPD-DV has been the basis for the current implementation of SDPD-DV. His assistance was essential and invaluable to the advancements of SDPD-DV pursued by me.

I am very appreciative for the interest that Professor Karniadakis expressed in this work. His inputs strengthened this work and deepened my understanding.

I would like to thank also all the members of my committee for their time, patience, and expertise.

Several of the WPI faculty provided invaluable expertise on many aspects of this research. However, I would like to specifically acknowledge the efforts of Sia Najafi not only for his expertise, continued support and endless patience with the Linux uninitiated, but also regarding life overall.

I would like to thank all of the teachers, coaches and professors who have collectively molded the person that I am today. I would like to specifically thank Prof. Dapeng Hu of the Dalian Univ. of Technology in China, it is your discipline and scrutiny of detail that is instilled in all the work that I do, and at times of need, it is your teachings that resound in my mind and enforce the purity of science and the rigor that must define all that it touches.

I would like to extend the sincerest thanks and appreciation to all of the friends that I have made here at WPI throughout the years. You are too numerous to name yet too dear to forget.

Gratitude is extended to the Mechanical Engineering Department, and especially to Gloria Boudreau, Statia Canning, Barbara Edilberti, Barbara Furhman and Patricia Rheaume.

Most of important, I would like to thank my parents. Their continued love, support, and admiration are empowering. From my parents, I learned the value of education both from their instruction and influence. My success is a reflection of their hard work and sacrifice. I thank them for all that I am and all that they have done for me.

Special thanks go to my friends I met in Christian Gospel Church of Worcester for their continuous help and support throughout the years that I had spent at WPI.

This work was partially supported by AFOSR's Computational Mathematics Program under Grant FA9550-10-1-0206. I would also like to acknowledge the support obtained through Teaching Assistantships obtained from the Mechanical Engineering Department at WPI.

***This work is dedicated to my mother and father.***

# TABLE OF CONTENTS

## LIST OF FIGURES

# LIST OF TABLES

## NOMENCLATURE

Boldface denotes a vector. The magnitude of a vector is denoted using the same symbol as the vector, but without boldface. Duplicate use of a symbol, or usage not defined below, will be clarified within the text.

| | |
|---|---|
| $c$ | Sound speed |
| $c_V$ | Specific heat capacity at constant |
| $C_V$ | Heat capacity at constant volume |
| $D_i$ | Individual particle diameter |
| $\rho$ | Density |
| $E$ | Energy |
| $f$ | Body Force |
| $\mathbf{F}^C$ | Total conservative force |
| $\mathbf{F}^D$ | Total dissipative force |
| $\mathbf{F}^R$ | Total random force |
| $h$ | Smoothing length |
| $\kappa$ | Thermal conductivity |
| $k_B$ | Bolztmann constant |
| $Kn$ | Knudsen Number |
| $m$ | Mass |
| $\lambda$ | Mean free path |
| $N$ | Number of particles |
| $\eta$ | Viscosity |
| $P$ | Pressure |
| $\mathbf{r}$ | position |
| $S$ | Entropy |
| $\Delta t$ | Time step |
| $T$ | Temperature |
| $\mathbf{v}$ | velocity |
| $V$ | Volume |
| $\zeta$ | Bulk viscosity |

# 1. INTRODUCTION

The numerical modeling and simulation of gases, liquids and multi-phase systems above the atomic spatiotemporal scales and below macroscopic scales has become increasingly important due to numerous applications in physical, biological and engineering systems. This regime can be characterized as mesoscopic and requires new mathematical and computational methods because the traditional atomistic (microscopic) and hydrodynamic (macroscopic) descriptions are not valid for either computational or theoretical reasons. One of the key characteristics of the mesoscopic flow regime is the presence of thermal fluctuations. The Smooth Dissipative Particle Dynamics (SDPD) developed by Espanol and Revenga (2003) has been proposed as a method appropriate for mescoscopic flows with fluctuations. The SDPD invokes the Smoothed Particle Hydrodynamics (SPH) which is a well-developed method for the Navier-Stokes equations (Liu and Liu, 2003). Using the GENERIC framework developed by Ottinger (2005) to describe hydrodynamic fluctuation, Espanol and Revenga arrived to SDPD discrete equations, which include thermal fluctuations. A three-dimensional implementation of the SDPD method in unbounded domains with dynamic virtual particle allocation (SDPD-DV) has been pursued in the Computational Gas&Plasma Dynamics Lab (CGPL) at WPI (Yang et al., 2011; Yang et al., 2012; Yang et al., 2013; Gatsonis et al., 2013). This work on SDPD-DV is part of the effort at the CGPL to develop simulation methods that apply to the investigation of micro- and nano-scale flows and involves the Direct Simulation Monte Carlo (DSMC) method for gases (Gatsonis et al., 2010; 2013) and particle-in-cell for plasmas (Gatsonis et al., 2009). There are three goals in this work:

- Revise and further implement existing algorithms in SDPD-DV, as well as develop and implement new algorithms.

- Validate and verify the algorithms of the SDPD-DV code by comparisons with experimental, analytical and numerical solutions.
- Investigate thermodynamic properties and transport coefficients with applications of the SDPD-DV code to mesoscopic systems in equilibrium.

We review below characteristic length, time scales and non-dimensional parameters used to characterize liquids and gases in order to establish notions of importance to mesoscale, those of *continuum, local thermodynamic equilibrium,* and *fluctuations*.

### 1.1. Fundamental Spatiotemporal Scales in Fluids

The molecular interaction featured in all states of matter (solids, liquids, and gases) can be represented by a force due to the Lennard-Jones potential (Allen and Tildesley, 1987) given by

$$V_{ij}(r) = 4\varepsilon\left[c_{ij}\left(\frac{r}{\sigma}\right)^{-12} - d_{ij}\left(\frac{r}{\sigma}\right)^{-6}\right], \tag{1.1}$$

where, $r$ is the distance separating the molecules $i$ and $j$, $c_{ij}$ and $d_{ij}$ are parameters particular to the pair of interacting molecules, $\varepsilon$ is a characteristic energy scale, and $\sigma$ is a characteristic length scale. The force between two molecules can be achieved by derivative of Eq.(1.1)

$$F_{ij}(r) = -\frac{\partial V_{ij}(r)}{\partial r} = \frac{48\varepsilon}{\sigma}\left[c_{ij}\left(\frac{r}{\sigma}\right)^{-13} - \frac{d_{ij}}{2}\left(\frac{r}{\sigma}\right)^{-7}\right]. \tag{1.2}$$

The characteristic atomic time scale associated with this Lennard-Jones molecular interaction is

$$\tau = \sigma\sqrt{\frac{m}{\varepsilon}}, \tag{1.3}$$

where $m$ is the mass of the individual molecule. The mean molecular spacing assuming that molecules are in contact with one another is defined as

$$\delta = \rho^{-1/3}, \tag{1.4}$$

where the macroscopic density is $\rho$. For $H_2O$(l) typical scales are shown in Table 1.

For gases additional physical scales can be defined (Vincenti and Kruger, 1975; Gombosi, 1994). The number of molecules in one mole of gas is the Avogadro number $N_A = 6.02214129(27)\times 10^{23}\,\mathrm{mol}^{-1}$. The volume occupied by 1 mole of gas at a given temperature and pressure is constant $V_m = 22.414$ l/mol at STP. Gases obey the perfect gas law

$$P = nk_B T \tag{1.5}$$

where, $n$ is the number density, $k_B = 1.38\times 10^{-23}\,\mathrm{m^2kg/(s^2K)}$ is the Bolztmann constant, $T$ (K) is the temperature. An important length scale in a gas is the mean molecular spacing defined as

$$\delta = n^{-1/3}\,. \tag{1.6}$$

The molecular diameter of a gas molecule $D_i$ is another physical scale which cannot be precisely defined but can be derived from viscosity measurements. Table 1 provides estimates based on the hard-sphere model. For gases undergoing binary collisions (dilute gases), it is required that

$$\delta \gg D_i\,. \tag{1.7}$$

Transport properties (such as viscosity and thermal conductivity) in a gas relate to collisions. The characteristic collision time in a dilute gas derived from the hard-sphere collision frequency is

$$\tau_c = \frac{1}{\nu} = \frac{1}{\sqrt{2}\pi D_i^2 n\bar{v}} \tag{1.8}$$

where $\bar{v}$ is an average relative speed. The average distance between collisions, called the mean free path, is given by

$$\lambda = \frac{1}{\sqrt{2}\pi D_i^2 n} \tag{1.9}$$

The mean-square thermal molecular speed of a molecule for a gas in equilibrium is given

$$\bar{c} = \sqrt{\frac{3k_B T}{m}} \tag{1.10}$$

The properties of a typical gas and liquid at standard conditions are shown in Table 1.

**Table 1. Properties of gaseous nitrogen and liquid water at STP.**

| Property | $N_2$ (g) | $H_2O$ (l) |
|---|---|---|
| Molecular diameter, $D_i$ | $3\times10^{-10}$ m | $3\times10^{-10}$ m |
| Number density, $n$ | $3\times10^{25}$ m$^{-3}$ | $2\times10^{28}$ m$^{-3}$ |
| Intermolecular spacing, $\delta$ | $3\times10^{-9}$ m | $4\times10^{-10}$ m |
| Molecular speed, $\bar{c}$ | 500 m/s | 1,000 m/s |
| Mean free path, λ | $6.54\times10^{-8}$ m | $2.5\times10^{-10}$ m |
| Collision time, $\tau_c$ | $1.667\times10^{-10}$ s | $1.25\times10^{-13}$ s |

While all fluid systems consist of molecules, macroscopic variables are obtained from averages of properties that depend on molecular velocities. These macroscopic averages can be functions of time and space. The *continuum hypothesis* allows us to describe the state of the fluid using a number of thermodynamic variables (or fields) that depend on position $\mathbf{r}$ and time $t$, for example, $\mathbf{v}(\mathbf{r},t)$, $\rho(\mathbf{r},t)$, $T(\mathbf{r},t)$, $P(\mathbf{r},t)$. One must be able to define sampling volumes, infinitesimal subsystems, with enough molecules so that statistical averages can be performed and provide a homogenous thermodynamic field. The *local thermodynamic equilibrium* hypothesis implies that the thermodynamic variables for these infinitesimal subsystems vary with time and space but satisfy the same relations as the equilibrium thermodynamics properties. Once these fields (or point variables) are defined conservation equations can be obtained which contain dissipative fluxes (the stress tensor, the heat flux, and diffusion flux). Phenomenological expressions are used to relate the dissipative fluxes $J_\alpha(\mathbf{r},t)$ with corresponding conjugate thermodynamic forces $X_\beta(\mathbf{r},t)$ in the form given [Ortiz De Zarate and Sengers, 2006]

$$J_\alpha(\mathbf{r},t)=\sum_\beta M_{\alpha\beta}(\mathbf{r},t)X_\beta(\mathbf{r},t), \tag{1.11}$$

The functional derivatives of the dissipative fluxes $M_{\alpha\beta}(\mathbf{r},t)$ are phenomenological coefficients, commonly referred to as the Onsager coefficients, and must be symmetric which implies $M_{\alpha\beta}(\mathbf{r},t)=M_{\beta\alpha}(\mathbf{r},t)$. The dependence of the Onsager coefficients, in principle, on space and time through the local state variables, in most practical applications can be neglected. For example, the heat flux $\mathbf{Q}(\mathbf{r},t)$ is given by Fourier's law with thermal conductivity $\kappa$

$$\mathbf{Q}(\mathbf{r},t)=-\kappa\nabla T(\mathbf{r},t). \tag{1.12}$$

With the phenomenological models defined, conservation laws, such as the Navier-Stokes equations are derived.

It is well known that fluctuations exist in fluids in thermodynamic equilibrium as introduced by Landau and Lifshitz (Statistical Physics, Part 1, Ch XII, 1980) as well as those in non-equilibrium (Ortiz De Zarate and Sengers, 2006). These fluctuations are spontaneous variations of thermodynamic variables about their mean values and are characterized by correlation functions of the fluctuating properties. Fluctuations in systems in thermal equilibrium can be derived from statistical physics (Landau and Lifshitz, Statistcial Physics, Part 1, Ch XII, 1980) and kinetic theory in case of gases (Groot and Mazur, Ch IX, 1962). The fluctuation for a thermodynamic quantity $f$ about its mean $\overline{f}$ measured in a volume $V$ is given as

$$\Delta f = f - \overline{f} , \tag{1.13}$$

where The root-mean-square fluctuation is given by the variance

$$\left\langle (\delta f)^2 \right\rangle^{1/2} = \left( \overline{f^2} - \left( \overline{f} \right)^2 \right)^{1/2} . \tag{1.14}$$

and the relative fluctuation is inversely proportional to $\sqrt{N}$

$$\frac{\left\langle (\delta f)^2 \right\rangle^{1/2}}{\left\langle f \right\rangle} \propto \frac{1}{\sqrt{N}} . \tag{1.15}$$

where the $\langle \ \rangle$ denote the mean values of lengthy expressions. For example, in an infinite fluid in equilibrium the variance in number of particles in a volume $V$ with temperature $T$ is given by (Landau and Lifshitz, 1980, Ch XII; Ortiz de Zarate and Sengers, 2006, Ch.3)

$$\left\langle (\delta N)^2 \right\rangle = -\left( \frac{k_B T N^2}{V^2} \right) \left( \frac{\partial V}{\partial P} \right)_T = \frac{k_B T N^2 \kappa_T}{V} , \tag{1.16}$$

where the isothermal compressibility is

$$\kappa_T = -\frac{1}{V} \left( \frac{\partial V}{\partial P} \right)_T . \tag{1.17}$$

In a given volume element $V$, the variance in macroscopic density is independent of time $t$ and equals to (Ortiz de Zarate and Sengers, 2006, Ch.3)

$$\left\langle \left( \overline{\delta\rho} \right)^2 \right\rangle = \rho m \frac{S_E}{V} = k_B T \rho^2 \frac{\kappa_T}{V} , \tag{1.18}$$

where $m S_E = \rho \kappa_T k_B T$ . Similarly, the variance in pressure is

$$\left\langle (\delta P)^2 \right\rangle = -k_B T \left( \frac{\partial P}{\partial V} \right)_S = \frac{k_B T}{\kappa_T V} . \tag{1.19}$$

The variance in volume is given by

$$\left\langle (\delta V)^2 \right\rangle = -k_B T \left( \frac{\partial V}{\partial P} \right)_T = k_B T \kappa_T V . \tag{1.20}$$

For dilute gas $\kappa_T = 1/P$ and the above expressions become

$$\left\langle (\delta N)^2 \right\rangle = N, \tag{1.21}$$

$$\left\langle \left(\overline{\delta\rho}\right)^2 \right\rangle = \frac{\rho^2}{n}, \tag{1.22}$$

$$\left\langle (\delta P)^2 \right\rangle = k_B T \frac{P}{V}, \tag{1.23}$$

$$\left\langle (\delta V)^2 \right\rangle = \frac{V^2}{n}. \tag{1.24}$$

Meanwhile in general, the variance in temperature is

$$\left\langle (\delta T)^2 \right\rangle = \frac{k_B T^2}{C_V}. \tag{1.25}$$

The variance in entropy is

$$\left\langle (\delta S)^2 \right\rangle = C_P. \tag{1.26}$$

where $C_V$ and $C_P$ are the heat capacity of the body as a whole at constant volume and constant pressure respectively. In general fluctuations increase with decreasing volume $V$, where the property $f$ is measured. It should be noted that the fluctuations of velocity are statistically independent of those of the other thermodynamic quantities. The variance of each Cartesian component of the velocity is equal to

$$\left\langle \delta \mathbf{V}^{\;2} \right\rangle = \frac{3k_B T}{m} \tag{1.27}$$

Fluctuating hydrodynamics for fluids in thermodynamic equilibrium can be described by the usual hydrodynamic equations (e.g. Navier-Stokes) with introduction of random noise terms. One approach introduced by Landau and Lifshitz (Statistical Physics, Part 2, Ch IX, 1980) is referred to as stochastic forcing, and treats the dissipative fluxes as the sum of an average plus a fluctuation term, which is defined as the fluctuating dissipative flux.

$$J_a(\mathbf{r},t)=\sum_{\beta}M_{\alpha\beta}(\mathbf{r},t)X_{\beta}(\mathbf{r},t)+\delta J_{\alpha}(\mathbf{r},t) \tag{1.28}$$

The fluctuation term $\delta J_{\alpha}(\mathbf{r},t)$ should satisfy that

$$\langle\delta J_{\alpha}(\mathbf{r},t)\rangle=0, \tag{1.29}$$

$$\langle\delta J_{\alpha}(\mathbf{r},t)\cdot\delta J_{\beta}(\mathbf{r}',t')\rangle=C_{\alpha\beta}(\mathbf{r},\mathbf{r}')\delta(t-t'), \tag{1.30}$$

where

$$C_{\alpha\beta}(\mathbf{r},\mathbf{r}')=2k_{B}TM_{\alpha\beta}\delta(\mathbf{r}-\mathbf{r}'). \tag{1.31}$$

The expressions in Eq. (1.31) is the so called *fluctuation-dissipation theorem* (FDT). Various extensions of fluctuating hydrodynamics to systems of thermal non-equilibrium under the local equilibrium assumption have been proposed. (de Groot and Mazur, 1962; Ottinger, 2005).

A non-dimensional parameter used to describe the applicability of the continuum approach in a gaseous flow is the local Knudsen Number given by

$$Kn\equiv\frac{\lambda}{L}. \tag{1.32}$$

In this definition $L$ is the local length scale of the gradient of a macroscopic quantity $f(\mathbf{r},t)$

$$L=\left|\frac{f}{\nabla f}\right|, \tag{1.33}$$

(Bird, 2007; Gombosi, 1994). When the Knudsen number is used to define the validity of the Navier-Stokes equations , it is often required that $Kn\leq 10^{-1}$.

In order to simulate mesoscopic flows, new mathematical and computational methods are required because the traditional atomistic (microscopic) and hydrodynamic (macroscopic) descriptions are not valid for either computational or theoretical reasons. The fundamental method appropriate for molecular (atomistic) scales is Molecular Dynamics (MD) (Allen and Tildesley, 1987; Haile, 1997). For example, a simulation of 1 micron volume at STP of $N_2$(g) and $H_2O$(l) contains $3\times10^7$ and $2\times10^{10}$ molecules respectively. A simulation based on MD is certainly not feasible. Various particle and hybrid (particle-fluid) simulation methods covering from atomistic to hydrodynamic scales are reviewed in Koumoutsakos (2005). From the continuum approach, the Navier-Stokes cannot represent the thermal fluctuation in mesoscale. Several models and numerical methods have been proposed for mesoscopic fluid dynamics, derived from either bottom-up (molecular) and from top-down (hydrodynamic) scale. We review below the Dissipative Particle Dynamics (DPD) and Smoothed Particle Hydrodynamics (SPH).

### 1.2. DPD Overview

A fundamental method for mesoscopic domains is the Dissipative Particle Dynamics (DPD) method introduced by Hoogerbrugge and Koelman (1992). The method can be interpreted as a coarsening approach, where the DPD particles represent clusters of molecules interacting by means of repulsive, dissipative and random forces. This clustering permits use of larger integration time steps and allows simulation of spatial scales much larger than those covered by MD (Keaveny et al., 2005). The statistical mechanics context behind DPD is presented by Espanol and Warren (1995).

In DPD the fluid region is discretized by a number of fluid particles each having mass, and gorvened by Newton's equation of motion (Hoogerbrugge and Koelman, 1992)

$$\mathbf{v}_i = \frac{d\mathbf{r}_i}{dt}, \tag{1.34}$$

$$\mathbf{F}_i = m_i \frac{d\mathbf{v}_i}{dt}. \tag{1.35}$$

where $\mathbf{v}_i$ is its velocity, $\mathbf{r}_i$ is its position. $\mathbf{F}_i$ is the total force exerted on particle $i$ by particle $j$, given as the sum of interparticle forces, consisting of a conservative component $\mathbf{F}_{ij}^C$, a dissipative component $\mathbf{F}_{ij}^D$, and a random component $\mathbf{F}_{ij}^R$ (Hoogerbrugge and Koelman, 1992)

$$\mathbf{F}_{ij}^C = F^{(C)}\left(r_{ij}\right)\mathbf{e}_{ij} \tag{1.36}$$

$$\mathbf{F}_{ij}^D = -\gamma\omega^D\left(r_{ij}\right)\left(\left(\mathbf{r}_{ij}/r_{ij}\right)\cdot\mathbf{v}_{ij}\right)\left(\mathbf{r}_{ij}/r_{ij}\right), \tag{1.37}$$

$$\mathbf{F}_{ij}^R = \sigma_R\omega^R\left(r_{ij}\right)\theta_{ij}\left(\mathbf{r}_{ij}/r_{ij}\right). \tag{1.38}$$

All particles $j$ in a sphere of radius $r_c$, which is called cutoff radius, are interacting with particle $i$ as shown in Figure 1. The conservative force is usually a soft repulsion given by $F^{(C)}\left(r_{ij}\right) = a_{ij}\max\left\{1-\left(r_{ij}/r_c\right),0\right\}$. The strength $\omega^D\left(r_{ij}\right)$ and $\omega^R\left(r_{ij}\right)$ in the dissipative force and random force, are coupled by $\omega^D\left(r_{ij}\right) = \left[\omega^R\left(r_{ij}\right)\right]^2 = \max\left\{\left(1-\left(r_{ij}/r_c\right)\right)^2,0\right\}$. The coefficients $\gamma$ and $\sigma_R$ are coupled by $\sigma_R^2 = 2\gamma k_B T$. The $\theta_{ij}$ in Eq. (1.38) is a symmetric Gaussian random variable with zero mean and unit variance.

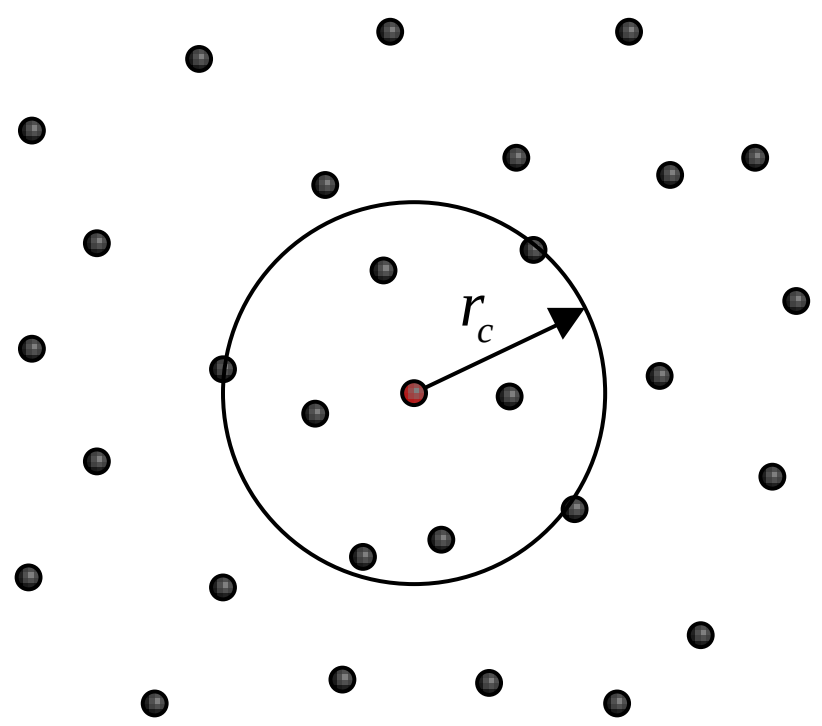

**Figure 1. The DPD cutoff radius used in calculating interparticle forces.**

The total force exerted on particle $i$ is given as (Espanol and Warren, 1995)

$$\mathbf{F}_i = \sum_{j \neq i} \mathbf{F}_{ij}^C + \sum_{j \neq i} \mathbf{F}_{ij}^D + \sum_{j \neq i} \mathbf{F}_{ij}^R . \tag{1.39}$$

The time evolution of the position and momentum equation of a DPD particle using a timestep $dt$ is given by

$$d\mathbf{r} = \mathbf{v}_i dt, \tag{1.40}$$

$$d\mathbf{v}_i = \frac{1}{m}\left(\mathbf{F}_i^C dt + \mathbf{F}_i^D dt + \mathbf{F}_i^R \sqrt{dt}\right). \tag{1.41}$$

There have been many improvements to DPD since its introduction and the method has been applied to a variety of mesoscale systems, including binary immiscible fluids (Novik and Coveney, 1997), colloidal behavior (Dzwinel et al., 2006), DNA in microchannels (Symeonidis et al., 2006), two phase flows (Tiwari and Abraham, 2006a), flow over rotating cylinder (Haber et al., 2006), nanojet breakup (Tiwari and Abraham, 2006b), polymers (Symenodis et al., 2006; Symeonidis and Karniadakis, 2006; Pan et al., 2008), and water in microchannels (Kumar et al., 2009), evaluation of transport properties (Ripoll et al., 2001), and fluids out of equilibrium (Ripoll and Ernst, 2004).

A considerable amount of effort has been devoted to boundary conditions in DPD. These studies include the non-slip boundary condition (Hoogerbrugge and Koelman, 1992; Espanol and Warren, 1995; Willemsen et al., 2000; Xu and Meakin, 2009; Wang et al., 2006; Duong-Hong et al., 2004; Pivkin and Karniadakis, 2005), slip boundary conditions (Smiatek et al., 2008), periodic boundary conditions (Chatterjee, 2007), and wall reflection laws (Revenga et al., 1999).

Attempts to arrive at an energy-conserving DPD have appeared by adding a random heat term (Avalos and Mackie, 1997; Mackie et al., 1999) or mechanical energy (Chaudhri and Lukes, 2009). A generalized form of DPD incorporating an internal energy and a temperature variable for each particle was presented by Espanol, (1997) and Ripoll et al. (1998). The energy-conserving DPD model was used to investigate the heat conduction in nanofluids by He and Qiao (2008).

### 1.3. SPH Overview

The SDPD method has its origins on Smoothed Particle Hydrodynamics (SPH), which was originally developed for modeling of astrophysical phenomena (Lucy, 1977; Benz et al., 1989). SPH was extended to simulate problems of continuum solid and fluid mechanics. Several forms of SPH equations can be derived based on the form of equations and the particle approximation involved. We review the SPH derivation for the compressible, viscous Navier-Stokes equations written in terms of field variables density $\rho(\mathbf{r},t)$, velocity $\mathbf{v}(\mathbf{r},t)$ and internal energy $.e(\mathbf{r},t)$, in the case of free external force (Liu and Liu, 2003)

$$\frac{D\rho(\mathbf{r},t)}{Dt} = -\rho(\mathbf{r},t)\frac{\partial \mathbf{v}^{\beta}(\mathbf{r},t)}{\partial \mathbf{x}^{\beta}}, \tag{1.42}$$

$$\frac{D\mathbf{v}^{\alpha}(\mathbf{r},t)}{Dt} = \frac{1}{\rho(\mathbf{r},t)}\left(-\frac{\partial P}{\partial \mathbf{x}^{\alpha}} + \frac{\partial \tau^{\alpha\beta}}{\partial \mathbf{x}^{\beta}}\right), \tag{1.43}$$

$$\frac{De(\mathbf{r},t)}{Dt} = -\frac{P}{\rho(\mathbf{r},t)}\frac{\partial \mathbf{v}^{\beta}(\mathbf{r},t)}{\partial \mathbf{x}^{\beta}} + \frac{\mu}{2\rho(\mathbf{r},t)}\varepsilon^{\alpha\beta}\varepsilon^{\alpha\beta} \tag{1.44}$$

The superscripts $\alpha$ and $\beta$ denote the coordinate directions. The viscous stress $\tau$ is proportional to shear stress $\varepsilon$ with dynamic viscosity $\mu$ by

$$\tau^{\alpha\beta} = \mu\left(\frac{\partial \mathbf{v}^{\beta}}{\partial \mathbf{x}^{\alpha}} + \frac{\partial \mathbf{v}^{\alpha}}{\partial \mathbf{x}^{\beta}} - \frac{2}{3}(\nabla\cdot\mathbf{v})\delta^{\alpha\beta}\right) \tag{1.45}$$

where the shear stress $\varepsilon$

$$\varepsilon^{\alpha\beta} = \frac{\partial \mathbf{v}^{\beta}}{\partial \mathbf{x}^{\alpha}} + \frac{\partial \mathbf{v}^{\alpha}}{\partial \mathbf{x}^{\beta}} - \frac{2}{3}(\nabla \cdot \mathbf{v})\,\delta^{\alpha\beta} \tag{1.46}$$

The internal energy per unit mass $e(\mathbf{r},t)$ for an ideal gas is given in terms of specific heat capacity $c_V$ by

$$e(\mathbf{r},t) = c_V T \tag{1.47}$$

and the pressure can be expressed as

$$P(\mathbf{r},t) = (\gamma - 1)\,\rho(\mathbf{r},t)e(\mathbf{r},t) \tag{1.48}$$

The energy Eq. (1.44) assumes that heat flux and the body forces are neglected. The derivation of SPH equations encompasses two steps: (a) the integral approximation of fluid fields such as $\rho(\mathbf{r},t)$, $\mathbf{v}(\mathbf{r},t)$, $e(\mathbf{r},t)$ and their derivatives , and (b) the particle approximation of these fields . In SPH the fluid is represented by a finite number of particles each with mass $m_i$ and volume $V_i$, as shown in Figure 2(a). The particle approximation of any field $f(x_i)$ is given by

$$\langle f(x_i) \rangle = \sum_{j=1}^{N} \frac{m_j}{\rho_j} f(x_j) W(x_i - x_j, h) \tag{1.49}$$

where $W(x_i - x_j, h)$ is the smoothing kernel function or smoothing function, $h$ is the smoothing length defining the influence volume of the smoothing function $W(x_i - x_j, h)$ as shown in Figure 2(b).

The smoothing function $W(x - x', h)$ in SPH must be normalized over its support domain as follows

$$\int_{\Omega} W(x - x', h)\, dx' = 1. \tag{1.50}$$

and be symmetric

$$\int (x - x')W(x - x', h)\,dx' = 0. \tag{1.51}$$

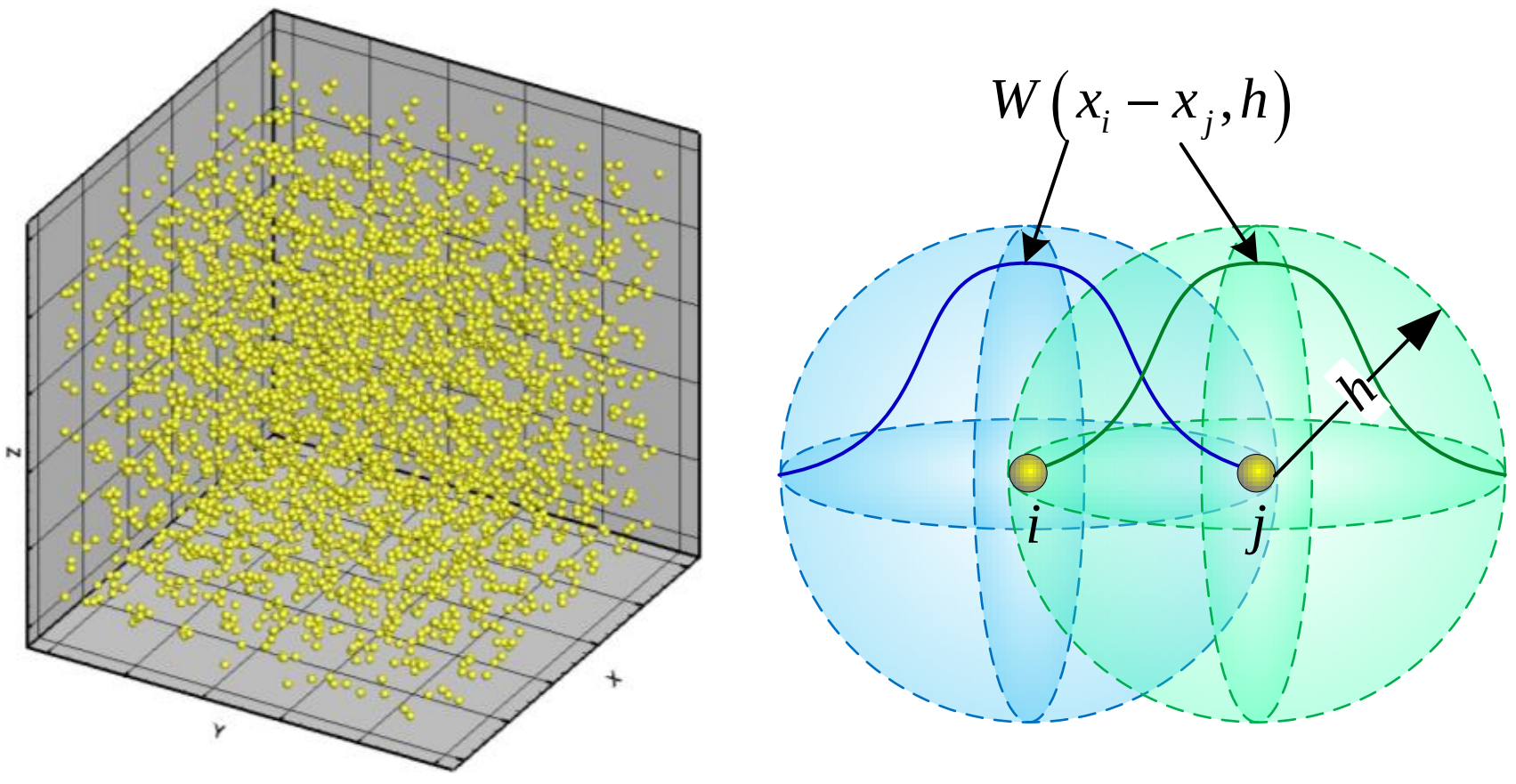

(a) (b)

**Figure 2. Discrete SPH particles representing the fluid and the support domain of the interpolating function around an SPH particle *i*.**

The discrete counterparts of the constant and linear consistency conditions as expressed in Eq. (1.50) and Eq. (1.51) are

$$\sum_{j=1}^{N} W(x_i - x_j, h)\,\Delta x_j = 1, \tag{1.52}$$

$$\sum_{j=1}^{N} (x_i - x_j) W(x_i - x_j, h)\,\Delta x_j = 0. \tag{1.53}$$

As discussed by Liu and Liu (2003), the discretized consistency conditions are not always satisfied due to the unbalanced particle distribution in cases where the support domain intersects with the boundary or support domain is irregularly distributed.

For the Navier-Stokes equations, Eq. (1.42)-(1.44) can be expressed in form of SPH equations (Liu and Liu, 2003) given as

$$\rho_i = \sum_{j=1}^{N} m_j W_{ij}, \tag{1.54}$$

$$\frac{d\mathbf{v}_i^{\alpha}}{dt} = \sum_{j=1}^{N} m_j \left( \frac{p_i}{\rho_i^2} + \frac{p_j}{\rho_j^2} \right) \frac{\partial W_{ij}}{\partial x_i^{\alpha}} + \sum_{j=1}^{N} m_j \left( \frac{\mu_i \varepsilon_i^{\alpha\beta}}{\rho_i^2} + \frac{\mu_j \varepsilon_j^{\alpha\beta}}{\rho_j^2} \right) \frac{\partial W_{ij}}{\partial x_i^{\beta}} \tag{1.55}$$

$$\frac{dE_i}{dt} = \frac{1}{2} \sum_{j=1}^{N} m_j \frac{p_i + p_j}{\rho_i \rho_j} v_{ij}^{\beta} \frac{\partial W_{ij}}{\partial x_i^{\beta}} + \frac{\mu_i}{2\rho_i} \varepsilon_i^{\alpha\beta} \varepsilon_i^{\alpha\beta}. \tag{1.56}$$

SPH is a well-developed method and a plethora of review papers and books have appeared including, Monaghan (1994), Randles and Libersky (1996), Liu and Liu (2003), Monaghan (2005), Li and Liu (2007), Monaghan (2009). There has benn a wide range of SPH applications and a comprehensive review is outside of the scope of this work. Studies include, incompressible fluids (Morris et al., 1997; Ellero et al., 2007), free surface flow (Monaghan, 1994; Fang et al., 2009), viscoelastic flow (Vazquez-Quesada and Ellero, 2012; Ellero et al., 2006), error estimation (Amicarelli et al., 2011; Fatehi and Manzari, 2011), immersed boundaries (Hieber and Koumoutsakos, 2008), multi-phase multiscale flow (Hu and Adams, 2006), thermal fluctuations in viscoelastic fluid (Vazquez-Quesada et al., 2009 c), and phase separating fluid mixture (Thieulot et al., 2005).

### 1.4. SDPD Method: Literature Review and Outstanding Issues

The SDPD was developed by Espanol and Revenga (2003) from a top-bottom approach from SPH as the thermodynamically consistent alternative to DPD. The SDPD invokes the Smoothed Particle Hydrodynamics (SPH), applied to the discretization of the compressible, viscous Navier-Stokes equations written in terms of field variables density $\rho(\mathbf{r},t)$, velocity $\mathbf{v}(\mathbf{r},t)$ and entropy $S(\mathbf{r},t)$ which is given by Batchelor (1967)

$$\frac{D\rho(\mathbf{r},t)}{Dt} = -\rho(\mathbf{r},t)\nabla\cdot\mathbf{v}, \tag{1.57}$$

$$\frac{D\mathbf{v}(\mathbf{r},t)}{Dt} = -\frac{\nabla P}{\rho(\mathbf{r},t)} + \frac{\eta}{\rho(\mathbf{r},t)}\nabla^2\mathbf{v} + \frac{1}{\rho(\mathbf{r},t)}\left(\zeta + \frac{\eta}{3}\right)\nabla\nabla\cdot\mathbf{v} + F\ , \tag{1.58}$$

$$T\frac{DS(\mathbf{r},t)}{Dt} = \frac{\phi}{\rho(\mathbf{r},t)} + \frac{\kappa}{\rho(\mathbf{r},t)}\nabla^2 T\ . \tag{1.59}$$

The material derivative, following a fluid particle, is $D/Dt = \partial/\partial t + \mathbf{v}\cdot\nabla$. In the above system $P$ is pressure, $F$ is the external force exerted per unit mass, $\eta$ is the shear viscosity, $\zeta$ is the bulk viscosity, and $\kappa$ is the thermal conductivity. The viscous heating field $\phi(\mathbf{r},t)$ is defined by

$$\phi = 2\eta\overline{\nabla\mathbf{v}} : \overline{\nabla\mathbf{v}} + \zeta(\nabla\cdot\mathbf{v})^2 \tag{1.60}$$

where the traceless symmetric part of the velocity gradient tensor is

$$\overline{\nabla\mathbf{v}} = \frac{1}{2}[\nabla\cdot\mathbf{v} + \nabla\cdot\mathbf{v}^T] - \frac{1}{3}\nabla\cdot\mathbf{v} \tag{1.61}$$

Through the use of the GENERIC framework developed by Ottinger (2005) to describe hydrodynamic fluctuations, Espanol and Revenga arrived to the SDPD discrete equations that include thermal fluctuations. In SDPD the independent variables are position $\mathbf{r}(t)$, velocity $\mathbf{v}(t)$ and entropy $S(t)$ of each fluid particle, a deviation from most mesoscopic and hydrodynamic models that involve the energy of the particle.

$$d\mathbf{r}_i = \mathbf{v}_i dt\ , \tag{1.62}$$

$$m_i \mathrm{d}\mathbf{v}_i = \mathbf{F}_C \mathrm{d}t + \mathbf{F}_D \mathrm{d}t + \mathbf{F}_R\ , \tag{1.63}$$

$$T_i dS_i = \Delta E_{Vi} dt + \Delta E_C dt + \Delta E_R\ . \tag{1.64}$$

where $\mathbf{F}_C$, $\mathbf{F}_D$ and $\mathbf{F}_R$ are the conservative term, dissipative term and random terms; $\Delta E_{Vi}$, $\Delta E_C$ and $\Delta E_R$ are viscous, conductive and random terms. Details of the formulation are presented in Ch. 2. SDPD requires a state equation to be expressed in the form $E(N,V,S)$. Then, the pressure $P(\mathbf{r},t)$ and temperature $T(\mathbf{r},t)$ are obtained through the Maxwell relations

$$P(\mathbf{r},t) = -\frac{\partial E(N,V,S)}{\partial V}, \tag{1.65}$$

$$T(\mathbf{r},t) = \frac{\partial E(N,V,S)}{\partial S(\mathbf{r},t)} \tag{1.66}$$

The work done by Espanol and Revenga (2003) follows the idea introduced by Espanol and Serrano (1999) about treating an SPH or a DPD particle as moving thermodynamic subsystems. In both of their works, the Sackur-Tetrode equation for internal energy $E(S,N,V)$ is employed to close the system. However, the testing case by Espanol and Revenga (2003) only involved the deterministic part of system equation. Espanol and Serrano (1999) perform test with a full model to show the fluctuation in entropy within equilibrium system are in the order of Boltzmann constant and around a constant value. All of the testing cases are under reduced unit.

The SDPD investigations so far addressed fundamental issues of the method. Serrano (2006) compared the SDPD and the Voronoi fluid particle model in a shear stationary flow. He found the efficiency of the two methods comparable. The accuracy of the Voronoi approach was found superior for regular ordinate systems while SDPD produced more accurate results for arbitrary, disordered configurations.

An incompressible, isothermal SDPD model was used by Litvinov et al. (2008) to model polymer molecules in suspension. The entropy equation was decoupled from the governing equations. A quantic spline kernel function was applied to their model. Periodic boundary conditions are considered to model the bulk fluid. Virtual particles mirrored at the surface of solid boundary are implemented when the support domain of a fluid particle overlaps with the solid wall surface. They concluded that the virtual particles would introduce errors on curved surfaces as Morris et al. (1997) pointed out. In their model, the SDPD interact hydrodynamically as well as by an additionally finitely extendable nonlinear elastic springs. In the test cases, the radius of gyration and the end-to-end radius are compared with analytical solutions at different chain lengths. A static structure factor is employed to extract the static factor exponent. A further test involved the Rouse mode. A form of diffusion coefficient evaluated from the mean square displacement of the center of mass is introduced and compared to an analytical formula for the 2D case. All parameters in these cases are in reduced units. Litvinov et al. (2008) concluded that the confinement of solid boundary affects the polymer configuration statistics and induces anisotropic effects. Their model is a guideline for further handling realistic microfluidic applications.

Vazquez-Quesada et al. (2009) investigated the consistent scaling of thermal fluctuations in SDPD by taking the point of view that the SDPD particles are real portions of fluid material instead of only just simple Lagrangian moving nodes. They used the isothermal SDPD model and normalized units in the simulations. They pointed out that the deterministic part of SDPD governing equation is scale free while the velocity variance from the stochastic part shows dependence on the physical length of SDPD particle. The tests are performed on colloidal particles (radius $R = 0.1$ in reduced unit) suspended in a Newtonian fluid (viscosity $\eta = 0.06$, mass density $\rho = 1$, temperature $T = 1$) with three different particle resolutions (number of particle $N = 484, 1600, 6400$) in a box size $L = 1$). Their results show that the velocity variance vary with the resolution of solvent, however it does not affect the velocity variance of the colloid particle. As a further proof of their result, they examined a polymer molecule in a suspension with different resolution of solvent ($N = 900, 3600, 14400$), and added finitely extensible nonlinear elastic spring forces between the polymer molecules.

The model utilized by Vazquez-Quesada et al. (2009) for viscoelastic fluids is identified as SPH with thermal fluctuations, which is actually the SDPD form added by a dimensionless conformation tensor to characterize the elongation of the polymer molecules within the fluid particles. They provided a full scheme of SDPD with closed equations such as total energy, entropy, conformation tensor. The entropy function is given by the logarithm of the number of microstates and coupled with total energy and conformation tensor. Their simulations are performed in a 2D periodic square box of length $L=1$ and $N=400$ fluid particles. They considered uniform shear flow with density $\rho=1$, kinematic viscosity $\nu=1$, and Kolmogorov flow at Reynolds number $\mathrm{Re}=1$. The dynamics of the conformation tensor are studied in terms of the dynamics of its eigenvalues and eigenvectors and have a non-eligible contribution to thermal fluctuations. They state that the introduction of thermal fluctuations in the viscoelastic fluid particle model is analogous to the stochastic contribution introduced by Landau and Lifshitz (Statistical Physics, part 2, Ch. IX, 1980) in fluctuating hydrodynamics.

Litvinov et al. (2009) provided an analytical expression of the self-diffusion coefficient for isothermal incompressible SDPD model. The coefficient in the self-diffusion coefficient formula depends on the quantic spline smoothing function. Their expression was tested by several simulations performed in a 3D periodic box with $N=3{,}375$ SDPD particles. The simulations assumed $\rho=1$, $k_B T=1$, $m=\rho L^3/N$, $L=1.25$, $h=L/15$ with resulting self-diffusion $D\mu=1$. The simulations varied the dynamic viscosity $\mu$. The actual self-diffusion coefficient and Schmidt number defined as $Sc=\nu/D$ was calculated from the mean-square displacement and was compared with the analytical value under different dynamic viscosity. Schimdt number that depends on fluid particle size does not always agree with the theoretical predictions, and extensive preliminary computations are needed to characterize the diffusion properties of the solvent. They point out that the diffusion properties depend on the choice of the smoothing length. They also state that their simulations do not show the solid-like structures found in DPD simulations at high coarse-graining levels as discussed by Pivkin and Karniadakis (2006).

Bian et al. (2012) utilized the SDPD method to model solid particles in suspension. They start from a reduced set of Navier-Stokes equation for incompressible flow to obtain an SDPD model. Closure is obtained with a state equation in the form of artificial compressibility given by Batchelor (1967). The non-slip velocity boundary condition is embedded on all solid-liquid interface by applying frozen particle as described by Morris et al. (1997). The velocity dynamically assigned to a frozen particle is calculated based on the distance to the tangent plane of the closest point to fluid particle. Therefore, it requires that the cutoff radius of support domain should be smaller than the smallest surface (both of convex and concave) curvature radius. The total force exerted on solid particle would be the summation of all forces exerted on each frozen particles in solid particle. They use the Velocity Verlet algorithm as the time integrator. Their numerical simulations are carried out with reduced units, including cases under non-Brownian and Brownian conditions. Several tests under non-Brownian condition involved flow through a fixed circular or spherical object in a periodic array, a particle moving in a Newtonian fluid under unsteady situations, a particle rotating under shear flow, and hydrodynamic interactions between two approaching spheres. The Brownian disk (in 2D) and Brownian sphere (3D) are investigated with thermal fluctuations producing its ultimate Brownian diffusive dynamics. The mean square displacement in a 2D domain is analyzed, and the diffusional behaviors are studied for both cases. A corrected form of diffusion coefficient related to drag force is provided and is related to the Einstein-Stokes equation for the actual diffusion coefficient from mean square displacement. A more complicated simulation involves a colloidal particle in the vicinity of an external boundary. The split form of the diffusion coefficient is introduced which depends on the distance to the boundary.

### 1.4.1 Boundary Condition for Wall-Bounded Domains

Among the outstanding theoretical and computational issues in SDPD is its implementation in domains with solid boundaries of arbitrary geometry. We address this issue in this work by developing an SDPD method for wall-bounded domains. We review boundary condition approaches in DPD and SPH in order to provide the necessary background for our method.

In mesh-free methods the wall and its effects on the fluid are modeled using various types and layers of ghost particles or frozen particles. Ghost particles can be loaded initially with static properties (Morris et al., 1997) or can be dynamically generated with properties updated during the simulation (Randles and Libersky, 1996). Static ghost or virtual particles are preloaded as uniformly distributed layers (Duong-Hon et al., 2004) or as interacting particles in the flow and solid boundary regions loaded with the same density as the fluid particles. (Hoogerbrugge and Koelman, 1992; Boek et al., 1997; Revenga et al., 1999; Liu and Liu, 2003). Dynamically allocated ghost or virtual particles are generated by reflecting neighboring fluid particles which may lead to an imperfect representation of a curved boundary at low resolution (Morris et al., 1997).

The difficulty arises in assigning the physical properties of such ghost particles and in defining the interaction forces between them and the fluid particles in the proximity of the wall. Such forces must ensure that a fluid particle does not penetrate the wall and that the no-slip condition is enforced. In general, the fluid particle force is composed of a repulsive and a dissipative term. The soft repulsive force acting on a fluid particle near a wall from the neighboring fluid particles may not be sufficiently strong to prevent wall penetration. To overcome this problem a stronger repulsive force between the fluid and the wall ghost particles, in the Lennard-Jones form, has been used in SPH (Monaghan, 1994). Different solutions have been proposed in DPD for imposing wall conditions, such as increasing the repulsive force coefficient for wall/fluid interaction (Pivkin and Karniadakis, 2006), increasing the wall particle density (Fedosov et al., 2008) or imposing bounce-back and bounce-forward boundary conditions (Revenga et al., 1999; Pivkin and Karniadakis, 2005). Some implementations of SDPD used dynamical ghost or virtual particle, including Hu and Adams (2006), Litvinov et al. (2008). A Lees–Edwards boundary condition is applied for solid boundary by Litvinov et al. (2010). Quesada (2009c) used the distance between a fluid particle and the solid wall in order to evaluate the truncated area of the support domain. This area is then used to evaluate the density and force on the fluid particle. This method may lead to an overestimation of the fluid particle density compared to the interior region.

The effectiveness of wall reflections was discussed by Revenga et al. (1999), including specular reflection, bounce-forward reflection, Maxwellian reflection, and bounce-back reflection. Pivkin and Karniadakis (2005) placed an extra thin layer of DPD particles inside the domain and adjacent to the solid boundary with an adjusted wall-fluid conservative force parameter, which is estimated according to the fluid density, to hold the no-slip boundary conditions. A similar model was developed by Pivkin and Karniadakis (2006) to control the density fluctuations near a solid boundary by applying an adaptive force directed perpendicular to the wall. The force is also adjusted according to the density and bins adjacent to the solid boundary. These adapted models are verified and compared by Fedosov et al. (2008). Another algorithm that modifies the boundary force is implemented by Altenhoff et al. (2007), in which the boundary force is estimated by the probability density function of the force contributions in the bin adjacent to solid wall. A phase-field interface representation to DPD for imposing the no-slip boundary condition was proposed by Xu and Meakin (2009), which considers the solid boundary as a phase indicated by a variable.

### 1.4.2 SDPD Self-Density

In this work, we use the summation form in Eq. (1.54) for evaluation of density which is a direct way of the approximation of SPH to the density itself. This form involves the contribution from neighboring particles in support domain by smoothing function. Flebbe et al. (1994) suggested that the contribution of the self-density should be discounted since overestimation arises when self-density in involved. Whitworth et al. (1995) suggested that although an overestimate in density arises from the thermal fluctuation initially with randomly distributed particles, it is eliminated after the system equilibrates.

### 1.4.3 Applications of the Full Non-isothermal SDPD Model

The SDPD investigations so far addressed fundamental issues of the method but considered only the isothermal formulation, which involves continuity and momentum equations alone (Litvinov et al., 2008; Litvinov et al. 2010, Vazquez-Quesada, 2009 b; Bian et al., 2012). One of the difficulties associated with the full SDPD is that it requires a formulation of the state equation indicated in Eq. (1.64)-(1.66). Espanol and Revenga (2003) performed SDPD simulations for an ideal gas and used the Sackur-Tetrode equation for $E(S,N,V)$ with only the deterministic part in Eq.(1.63) and (1.64). The full thermodynamically consistent SDPD model has been validated by Vazquez-Quesada (2009 a) based on a Fourier problem that involves a fluid between two wall at different temperature. For liquids, Vazquez-Quesda (2009) derived a state equation through a second-order Taylor expansion around the equilibrium state and applied it to viscoelastic fluid and colloidal suspension flow.

### 1.5. Objectives and Approach

This work is part of a research effort at the CGPL at WPI to develop a SDPD-DV methodology and apply to the investigation of mesoscopic wall-bounded flows. There are three goals:

- First, to revise and further implement existing algorithms in SDPD-DV, as well as develop and implement new algorithms.
- Second, to validate and verify the algorithms of the SDPD-DV code by comparisons with experimental, analytical and numerical solutions.
- Third, to investigate thermodynamic properties and transport coefficients with applications of the SDPD-DV code to mesoscopic systems in equilibrium.

The objectives and approaches are divided in three main categories.

1. Revise and further implement existing algorithms in SDPD-DV, as well as develop and implement new algorithms in order to achieve a fully functional SDPD-DV methodology:
    a. Revise the periodic boundary conditions algorithm and the implementation of the periodic boundary cells searching list.
    b. Modify evaluation of smoothing function in order to include the contribution of the self-density for each fluid particle.
    c. Modify and further implement the dynamic virtual particle allocation algorithm for the modeling of solid boundary in order to minimize the truncation error of density.
    d. Develop and implement the boundary normal vector algorithm used in the reflection of the dynamically allocated virtual particles.

e. Implement the algorithm for the contribution to the boundary force from the virtual particles.

f. Implement a temperature boundary condition in the dynamic virtual particle allocation model.

g. Implement a bounce-forward algorithm for a solid boundary with arbitrary shape and orientation.

h. Revise and rewrite portions of the Velocity Verlet integration method for the position and momentum equations.

i. Develop and implement a Runge–Kutta integration algorithm for the entropy equation.

j. Implement the artificial incompressibility method for modeling liquid flows.

k. Implement a temperature power law for the shear viscosity, bulk viscosity and heat conductivity that appear in the momentum and energy SDPD equations.

l. Develop and implement algorithms for the evaluation of transport properties such as diffusion coefficient, shear viscosity and heat conductivity based on mean square displacement (MSD) and velocity autocorrelation function (VACF).

m. Develop analytical formulas of the self-diffusion coefficient based on the SDPD-fluid following Litvinov et al. (2009).

2. Validate and verify the SDPD-DV implementation:

   a. Validate the implementation of boundary particle and fluid particle loading, dynamic virtual particle allocation, periodic boundary particle allocation, fluid field properties sampling, and bounce forward reflection. Compare with analytical solutions for transient body-driven Poiseuille flow of water between stationary infinite parallel plates of $10^{-3}$ m height. Calculate the components of the forces attributed to the dynamically allocated virtual particles and compared with previous DPD investigations.

b. Verify moving and no-slip boundary conditions achieved by the dynamic virtual particle method by comparisons of SDPD-DV simulations with transient Couette flow of water between stationary infinite parallel plates of $10^{-3}$ m height.

c. Verify the ability of the SDPD-DV to simulate curved 3D solid boundaries, and the evaluation of pressure by comparisons of SDPD-DV results of steady, low-Re incompressible flow over a cylinder of radius 0.02 m with results from FLUENT.

d. Perform simulations of Poiseuille flow between two infinite parallel plates at different temperature for verification of the non-isothermal SDPD-DV.

e. Perform simulations of Couette flow between two infinite parallel plates at different temperature for verification of the moving wall boundaries with constant temperature.

3. Investigate mesoscopic flows using the isothermal and non-isothermal SDPD-DV implementations.

    a. Perform simulations of $H_2O$(l) and $N_2$(g) at equilibrium states. Evaluate the self-diffusion coefficient, shear viscosity, translational temperature, thermal speed, and thermal fluctuations. Validate and verify by comparisons with analytical and experimental values.

    b. Examine the scale dependence in SDPD-DV and characterize the effects of particle mass, particle volume, smoothing length, and time step.

The presentation of this work is organized in the following manner. In Chapter 2, the overview of the SDPD-DV methodology is presented, as well as the mathematical and numerical aspect, as pertaining to its implementation with dynamic virtual particle and non-isothermal model, is presented in detail for each aforementioned code modification or addition. In Chapter 3, an extensive set of benchmark tests that cover the hydrodynamic and mesoscopic regimes is presented and used for verification, validation and error analysis of the SDPD-DV method. The first verification of SDPD-DV involves comparisons with analytical solutions for a body-force driven, transient, Poiseuille flow of water between parallel plates of $10^{-3}$ m height. The second verification test involves the transient, Couette flow of water between parallel plates of $10^{-3}$ m height. The third test used for verification involves the low-Reynolds number, incompressible flow over a cylinder of radius 0.02 m. An extensive set of SDPD-DV simulations of liquid water and gaseous nitrogen in mesoscopic periodic domains is also presented. In Chapter 4 the SDPD-DV code with entropy equation is applied to microscale non-isothermal case studies. The validation and verification includes simulation of equilibrium gaseous nitrogen in periodic domains and the evaluation of transport coefficients. Fluctuations in thermodynamic variables are evaluated and compared with analytical estimates. Results from SDPD-DV simulation of non-isothermal nitrogen Couette flow are verified with results from FLUENT. Conclusions and recommendations for future work are presented in Chapter 5.

# 2. SDPD-DV METHODOLOGY AND IMPLEMENTATION

We review in this chapter the basic elements of the SDPD model and present the discrete SDPD equations. We then present the major algorithmic features of the SDPD-DV implementation developed for simulation of mesoscopic fluids in wall-bounded domains. Algorithms presented include the particle loading indexing; the neighboring particle search; density and force evaluation on interior domains, solid boundary and periodic boundaries; pressure evaluation; integration of SDPD-DV equations; evaluation of particle transport properties; evaluation of sample-averaged particle properties. Material in this chapter appear in Gatsonis et al. (2013).

## 2.1 Overview of the SDPD Method

The SDPD derivation of Espanol and Revenga (2003) starts with the Navier-Stokes equations written in the material derivative form (i.e. following a fluid particle) with independent field (Eulerian) variables the density $\rho(\mathbf{r},t)$, velocity $\mathbf{v}(\mathbf{r},t)$, and entropy $S(\mathbf{r},t)$ instead of the $E(\mathbf{r},t)$,

$$\frac{D\rho(\mathbf{r},t)}{Dt} = -\rho(\mathbf{r},t)\nabla\cdot\mathbf{v}, \tag{1.57}$$

$$\frac{D\mathbf{v}(\mathbf{r},t)}{Dt} = -\frac{\nabla P}{\rho(\mathbf{r},t)} + \frac{\eta}{\rho(\mathbf{r},t)}\nabla^2\mathbf{v} + \frac{1}{\rho(\mathbf{r},t)}\left(\zeta + \frac{\eta}{3}\right)\nabla\nabla\cdot\mathbf{v}, \tag{1.58}$$

$$T\frac{DS(\mathbf{r},t)}{Dt} = \frac{\phi}{\rho(\mathbf{r},t)} + \frac{\kappa}{\rho(\mathbf{r},t)}\nabla^2 T. \tag{1.59}$$

The derivation arrives first at a discrete SPH-type set of the deterministic Navier-Stokes equations for $\mathbf{r}_i, \mathbf{v}_i$ and $S_i$. The domain containing fluid of total mass $M_F$ and volume $V_T$, following SPH, is discretized with a number of $N_F$ points each one representing a thermodynamic closed system (equivalent to a material volume or a fluid particle). Each particle $i$ has constant mass

$$m_i = \frac{M_F}{N_F}, \tag{2.1}$$

and is described by independent variables position, velocity and entropy at time $t$

$$\mathbf{r}_i(t) = \{x_i(t), y_i(t), z_i(t)\}, \mathbf{v}_i(t) = \{v_{xi}(t), v_{yi}(t), v_{zi}(t)\}, S_i(t). \tag{2.2}$$

The SDPD discretization is consistent with a set of $N_F$ thermodynamics systems (or Lagrangian fluid particles). Alternatively, the SDPD discretization can be considered as a set of $N_F$ grid nodes which are moving with the material (or particle) velocity. These two views are identical and can provide preferable viewpoints during analysis. The number density $d_i$ of the SDPD particle $i$ follows SPH and is defined as a summation of neighboring particles as shown in Figure 3

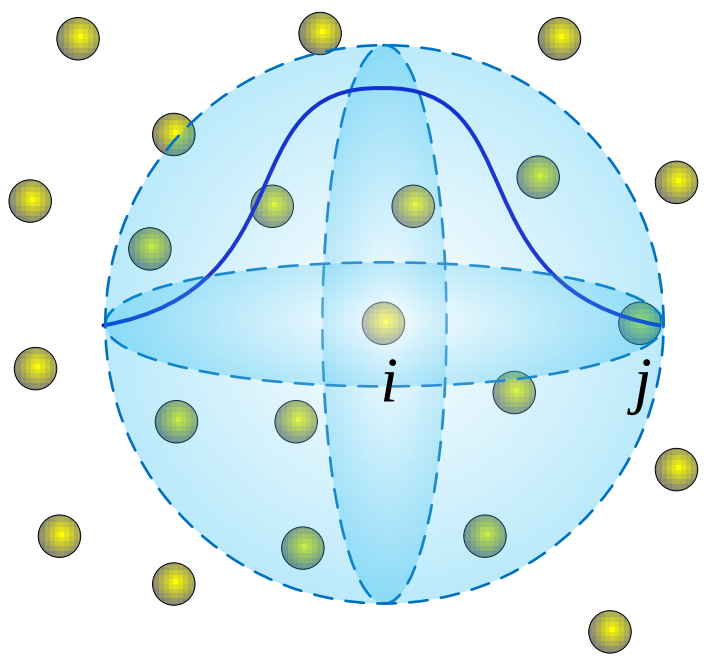

**Figure 3. Contribution of neighboring particles in particle $i$ support domain to the number density evaluation.**

$$d_i = \sum_j W\left(\left|r_i - r_j\right|, h\right), \tag{2.3}$$

where, $W(r,h)$ is the interpolant (smoothing) function with a finite support $h$ satisfying normalization condition

$$\int_{\Omega} W(r,h)dr = 1 . \tag{2.4}$$

Several options are available for the interpolant function and a detailed list of rules to construct smoothing functions can be found on Liu and Liu (2003). The sum in Eq. (2.3) is extended to all the $j$ particles that are within the finite support $h$ of particle $i$, including the contribution from the particles $i$ itself. The distance between particles $i$ and $j$ is

$$\mathbf{r}_i - \mathbf{r}_j = \mathbf{r}_{ij} \equiv \left\{ x_{ij}, y_{ij}, z_{ij} \right\}, \tag{2.5}$$

A dimensionally consistent particle volume $V_i$ is defined as the inverse of the particle number density

$$V_i = \frac{1}{d_i} . \tag{2.6}$$

The field density $\rho(\mathbf{r},t)$ at the particle position $\mathbf{r}_i(t)$ of particle $i$ is obtained as

$$\rho[\mathbf{r}_i(t)] \equiv \rho_i(t) = m_i d_i(t) . \tag{2.7}$$

Similarly to the SPH approach, the discrete value at a point $r_i$ (or position of a particle) for any field variable $\Psi(\mathbf{r},t)$ can be calculated by interpolation

$$\Psi(\mathbf{r}_i) = \frac{\sum_j W(|r_i - r_j|)\Psi_j}{\sum_j W(|r_i - r_j|)} . \tag{2.8}$$

The deterministic SDPD equations consistent with the Naveir-Stokes Eq. (1.57)-(1.59) are:

$$\frac{d\mathbf{r}_i}{dt} = \mathbf{v}_i , \tag{2.9}$$

$$m\frac{d\mathbf{v}_i}{dt}=\sum_j\left[\frac{P_i}{d_i^2}+\frac{P_j}{d_j^2}\right]F_{ij}\mathbf{r}_{ij}-\left(\frac{5\eta}{3}-\zeta\right)\sum_j\frac{F_{ij}}{d_id_j}\mathbf{v}_{ij}-5\left(\zeta+\frac{\eta}{3}\right)\sum_j\frac{F_{ij}}{d_id_j}\mathbf{e}_{ij}\mathbf{e}_{ij}\cdot\mathbf{v}_{ij}\,, \qquad (2.10)$$

$$T_i\frac{dS_i}{dt}=(\phi)_i-2\kappa\sum_j\frac{F_{ij}}{d_id_j}T_{ij}\,. \qquad (2.11)$$

This deterministic dynamics is introduced into the GENERIC framework of Ortega (2005). This is done in order to introduce the stochastic part for the system dynamics. The introduction of thermal fluctuating terms leads to the GENERIC stochastic differential equations that contain (reversible and irreversible) deterministic terms, a term that relates the dissipative (irreversible) dynamics with stochastic terms, and a stochastic term (Eq. (38) in Espanol and Revenga, 2003) as

$$dx=\left[L\frac{\partial E}{\partial x}+M\frac{\partial S}{\partial x}+k_B\frac{\partial}{\partial x}M\right]dt+d\tilde{x}\,, \qquad (2.12)$$

In the above $d\tilde{x}$ is the stochastic term. The term $L\,\partial E/\partial x$ is the reversible part of the dynamics, and the second term $M\,\partial S/\partial x$ is defined as the irreversible part. The matrices $L$ is anti-symmetric, and $M$ is symmetric and positive semi-definite. The following conditions must be satisfied by $L$ and $M$

$$L\frac{\partial S}{\partial x}=0,\quad M\frac{\partial E}{\partial x}=0\,, \qquad (2.13)$$

$$\frac{\partial I}{\partial x}L\frac{\partial E}{\partial x}=0,\quad \frac{\partial I}{\partial x}M\frac{\partial S}{\partial x}=0\,. \qquad (2.14)$$

The stochastic term $d\tilde{x}$ satisfies

$$d\tilde{x}d\tilde{x}^T=2k_BMdt \qquad (2.15)$$

which is an expression of the *fluctuation-dissipation theorem*. It is also required that

$$\frac{\partial E}{\partial x}d\tilde{x}=0,\quad \frac{\partial I}{\partial x}d\tilde{x}=0\,, \qquad (2.16)$$

The term $k_B\left(\partial M/\partial x\right)$ is from the stochastic interpretation of Itô process. Therefore, the matrix $M$ is given by

$$M \rightarrow \mathbf{M}_{ij} = \begin{pmatrix} 0 & 0 & 0 \\ 0 & \dfrac{d\tilde{\mathbf{v}}_i d\tilde{\mathbf{v}}_j^T}{2k_B dt} & \dfrac{d\tilde{\mathbf{v}}_i d\tilde{S}_j}{2k_B dt} \\ 0 & \dfrac{d\tilde{S}_i d\tilde{\mathbf{v}}_j^T}{2k_B dt} & \dfrac{d\tilde{S}_i d\tilde{S}_j}{2k_B dt} \end{pmatrix} \tag{2.17}$$

The matrix $L$ is defined by

$$L \rightarrow \mathbf{L}_{ij} = \frac{1}{m}\begin{pmatrix} 0 & \mathbf{1}\delta_{ij} & 0 \\ -\mathbf{1}\delta_{ij} & 0 & 0 \\ 0 & 0 & 0 \end{pmatrix} \tag{2.18}$$

The derivatives of the energy and entropy with respect to the state variables result in

$$\frac{\partial E}{\partial x} = \begin{pmatrix} \dfrac{\nabla P_i}{d_i} \\ m\mathbf{v}_i \\ T_i \end{pmatrix}, \quad \frac{\partial S}{\partial x} = \begin{pmatrix} 0 \\ 0 \\ 1 \end{pmatrix} \tag{2.19}$$

and the thermal noise is defined as

$$d\tilde{x} = \begin{pmatrix} 0 \\ d\tilde{\mathbf{v}}_i \\ d\tilde{S}_i \end{pmatrix} \tag{2.20}$$

Eq. (2.9) (2.11) can be given in a matrix form as

$$\begin{pmatrix} \dot{\mathbf{r}}_i \\ \dot{\mathbf{v}}_i \\ \dot{S}_i \end{pmatrix} = \sum_j \mathbf{L}_{ij} \begin{pmatrix} \dfrac{\nabla P_i}{d_i} \\ m\mathbf{v}_i \\ T_i \end{pmatrix} + \sum_j \mathbf{M}_{ij} \begin{pmatrix} 0 \\ 0 \\ 1 \end{pmatrix} + k_B \sum_j \begin{pmatrix} 0 \\ \sum_j \dfrac{\partial}{\partial \mathbf{v}_j} \dfrac{d\tilde{\mathbf{v}}_i d\tilde{\mathbf{v}}_j}{2k_B dt} + \sum_j \dfrac{\partial}{\partial S_j} \dfrac{d\tilde{\mathbf{v}}_i d\tilde{S}_j}{2k_B dt} \\ \sum_j \dfrac{\partial}{\partial \mathbf{v}_j} \dfrac{d\tilde{S}_i d\tilde{\mathbf{v}}_j}{2k_B dt} + \sum_j \dfrac{\partial}{\partial S_j} \dfrac{d\tilde{S}_i d\tilde{S}_j}{2k_B dt} \end{pmatrix} + \begin{pmatrix} 0 \\ \dfrac{d\tilde{\mathbf{v}}_i}{dt} \\ \dfrac{d\tilde{S}_i}{dt} \end{pmatrix} \tag{2.21}$$

The final result is a set of discrete particle equations that describe the deterministic and stochastic dynamics, referred to as the SDPD equations (Eq. (63) in Espanol and Revenga, 2003). We rewrite below the SDPD equations for the independent variables $\rho_i, \mathbf{v}_i$ and $S_i$ in a format that is conducive to numerical implementation and allows also direct comparison with SPH and DPD. They are given by

$$d\mathbf{r}_i = \mathbf{v}_i dt , \tag{1.62}$$

$$m_i \mathrm{d}\mathbf{v}_i = \mathbf{F}_C \mathrm{d}t + \mathbf{F}_D \mathrm{d}t + \mathbf{F}_R , \tag{1.63}$$

$$T_i dS_i = \Delta E_{Vi} dt + \Delta E_C dt + \Delta E_R . \tag{1.64}$$

The terms appearing in the momentum equation (1.63) are broken into the conservative, dissipative and velocity random terms given by

$$\mathbf{F}_C = \sum_j \left[ \frac{P_i}{d_i^2} + \frac{P_j}{d_j^2} \right] F_{ij} \mathbf{r}_{ij} , \tag{2.22}$$

$$\begin{aligned} \mathbf{F}_D = & -\sum_j \left( 1 - \frac{T_i T_j}{\left(T_i + T_j\right)^2} \left[ \frac{k_B}{C_i} + \frac{k_B}{C_j} \right] \right) \left( \frac{5\eta}{3} - \zeta \right) \frac{F_{ij}}{d_i d_j} \mathbf{v}_{ij} \\ & -\sum_j \left( 1 - \frac{T_i T_j}{\left(T_i + T_j\right)^2} \left[ \frac{k_B}{C_i} + \frac{k_B}{C_j} \right] \right) 5 \left( \frac{\eta}{3} + \zeta \right) \frac{F_{ij}}{d_i d_j} \mathbf{e}_{ij} \mathbf{e}_{ij} \cdot \mathbf{v}_{ij} \end{aligned} , \tag{2.23}$$

$$\mathbf{F}_R = \sum_j \left( \begin{aligned} & \left[ 8k_B \frac{T_i T_j}{T_i + T_j} \left( \frac{5\eta}{3} - \zeta \right) \frac{F_{ij}}{d_i d_j} \right]^{1/2} d\bar{\mathbf{W}}_{ij} + \\ & \frac{\mathbf{I}}{3} \left[ 8k_B \frac{T_i T_j}{T_i + T_j} \left( \frac{5\eta}{3} + 8\zeta \right) \frac{F_{ij}}{d_i d_j} \right]^{1/2} tr\left[ d\mathbf{W}_{ij} \right] \end{aligned} \right) \cdot \mathbf{e}_{ij} . \tag{2.24}$$

The terms in entropy equation (1.64) are divided into viscous, conductive and random terms.

$$\Delta E_V = \frac{1}{2}\sum_j \left(1 - \frac{T_i T_j}{\left(T_i + T_j\right)^2}\left[\frac{k_B}{C_i} + \frac{k_B}{C_j}\right] - \frac{T_j}{T_i + T_j}\frac{k_B}{C_i}\right)\left[\left(\frac{5\eta}{3} - \zeta\right)\frac{F_{ij}}{d_i d_j}\mathbf{v}_{ij}^2 \right.$$
$$\left. + 5\left(\frac{\eta}{3} + \zeta\right)\frac{F_{ij}}{d_i d_j} \times \left(\mathbf{e}_{ij} \cdot \mathbf{v}_{ij}\right)^2\right] - \frac{2k_B}{m}\sum_j{}' \frac{T_i T_j}{T_i + T_j}\left(\frac{25\eta}{3} + 2\zeta\right)\frac{F_{ij}}{d_i d_j}, \tag{2.25}$$

$$\Delta E_{Cd} = -2\kappa \sum_j \frac{F_{ij}}{d_i d_j} T_{ij} - 2\kappa \frac{k_B}{C_i}\sum_j{}' \frac{F_{ij}}{d_i d_j} T_{ij}, \tag{2.26}$$

$$\Delta E_R = -\frac{1}{2}\sum_j \begin{pmatrix} \left[8k_B \frac{T_i T_j}{T_i + T_j}\left(\frac{5\eta}{3} - \zeta\right)\frac{F_{ij}}{d_i d_j}\right]^{1/2} d\bar{\mathbf{W}}_{ij} + \\ \frac{\mathbf{I}}{3}\left[8k_B \frac{T_i T_j}{T_i + T_j}\left(\frac{5\eta}{3} + 8\zeta\right)\frac{F_{ij}}{d_i d_j}\right]^{1/2} tr\left[d\mathbf{W}_{ij}\right] \end{pmatrix} : \mathbf{e}_{ij}\mathbf{v}_{ij}$$
$$+ \sum_j \left[4\kappa k_B T_i T_j \frac{F_{ij}}{d_i d_j}\right]^{1/2} dV_{ij} . \tag{2.27}$$

A state equation is required to close the system

$$E_i^{eq} = E_i(N, V, S_i) . \tag{2.28}$$

The state equation provides the particle temperature and pressure as

$$T_i = \frac{\partial E^{eq}}{\partial S_i}, \tag{2.29}$$

$$P_i = -\frac{\partial E^{eq}}{\partial V} . \tag{2.30}$$

The term $\mathbf{e}_{ij} = \left(\mathbf{r}_i - \mathbf{r}_j\right) / \left|\mathbf{r}_{ij}\right|$ in the SDPD equations is a unit vector, $\mathbf{v}_{ij} = \mathbf{v}_i - \mathbf{v}_j$, $T_{ij} = T_i - T_j$ and

$$F_{ij} = F\left(\left|\mathbf{r}_i - \mathbf{r}_j\right|\right) = -\nabla W(r)/r . \tag{2.31}$$

The terms $\eta$ and $\xi$ represent the shear and bulk viscosity; $C_i$ is the heat capacity at constant volume of particle $i$ and is an extensive property, $C_i = c_V m_i$ where $c_V$ is the specific heat capacity of fluid; the Boltzmann constant is $k_B$ and $\kappa$ is the thermal conductivity of the fluid.

The velocity and entropy random terms contain $\mathbf{I}$ the identity matrix and $d\overline{\mathbf{W}}_{ij}$ the traceless symmetric part of a matrix of independents increments of the Wiener process $d\mathbf{W}_{ij}$.

$$d\overline{\mathbf{W}}_{ij} = \frac{1}{2}\left[d\mathbf{W}_{ij} + d\mathbf{W}_{ij}^{T}\right] - \frac{\mathbf{I}}{3}tr\left[d\mathbf{W}_{ij}\right]. \tag{2.32}$$

where the trace is defined as

$$tr\left[d\mathbf{W}_{ij}\right] = \sum_{\sigma} d\mathbf{W}_{ij}^{\sigma\sigma}. \tag{2.33}$$

For $k_B/C_i = 0$ the fluctuating terms reduce to zero and the set of Eqs.(1.62)-(1.64) reduce to a discrete SPH-form of the deterministic Navier-Stokes equations for $\mathbf{r}_i$, $\mathbf{v}_i$, and $S_i$ given by Eq. (2.9)-(2.11).

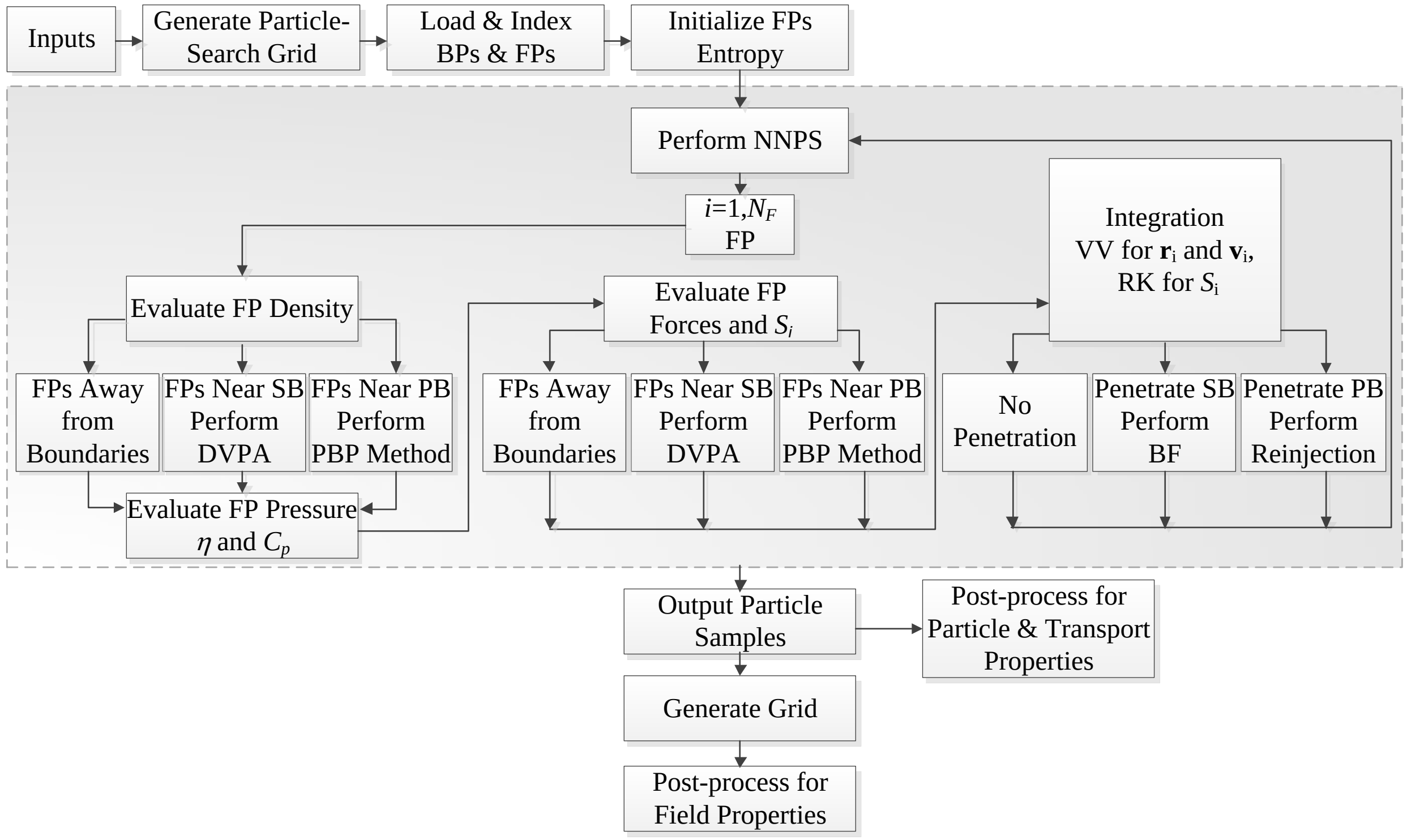

**Figure 4. General flow chart of SDPD-DV and post-processing.**

## 2.2 Mathematical and Computational Aspects of the SDPD-DV method

In this section, we present the mathematical and computational aspects of the SDPD-DV developed in this work for simulation of wall-bounded domains. The general flow chart is presented in Figure 4.

The physical domain is shown in Figure 5 and is characterized by arbitrary external and internal solid-wall boundaries, planar periodic inlets and outlets with equal areas, and the fluid region. A rectangular domain is also constructed to aid in the numerical implementation of the particle search and includes the physical domain as shown in Figure 5.

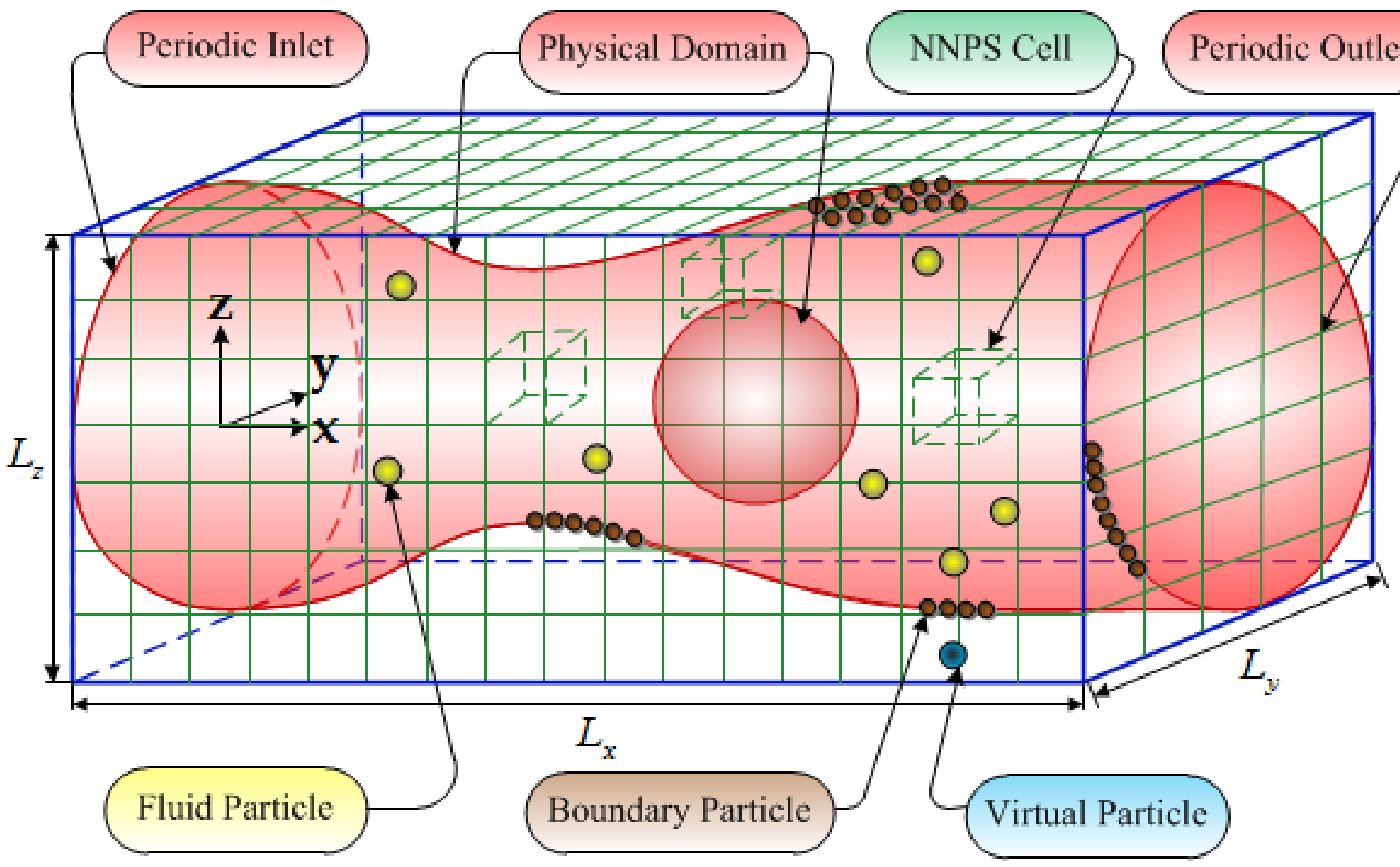

**Figure 5. Physical wall-bounded domain with an exterior wall and an interior solid body showing the fluid particles (FPs), boundary particles (BPs), and virtual particles (VPs) used in the SDPD-DV. Inlets and outlets are considered periodic. The large rectangular cells are used for nearest neighbor particle search (NNPS).**

### 2.2.1 Boundary Particle Loading and Global Indexing

The boundary of the physical domain is discretized with a surface triangulation based on a surface length scale small enough to resolve the physical characteristics of the surface and much smaller than the smoothing length, $h$. A total of $N_\text{B}$ boundary particles (BPs) are placed on the vertices of the surface grid as shown in Figure 5. The BPs position and velocity are stored in the global particle list

$$\begin{aligned}\mathbf{r}_k(t) = \mathbf{r}_k(0) = \{x_k(0), y_k(0), z_k(0)\} \\ \mathbf{v}_k(t) = \mathbf{v}_k(0) = \{v_{kx}(0), v_{ky}(0), v_{kz}(0)\}\end{aligned} \quad k = 1, N_B \tag{2.34}$$

In addition, the normal vector entering the fluid domain at each BP is evaluated and later used in fluid particle-wall interaction. For a surface defined implicitly by $F(x, y, z) = 0$ the normal vector is evaluated analytically by $\mathbf{n}(x_k, y_k, z_k) = \nabla F(x_k, y_k, z_k)$. For an arbitrary solid wall surface the local normal to each surface triangle surrounding a vertex is evaluated as $\mathbf{n}_1 = \mathbf{r}_{12} \times \mathbf{r}_{13}$ and the closest is assigned to the vertex as shown in Figure 6.

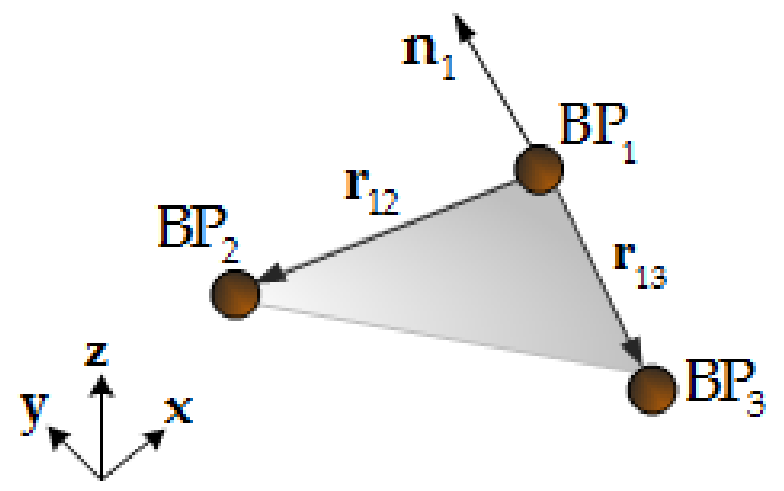

**Figure 6. Normal vector for a BP on a solid boundary.**

### 2.2.2 Fluid Particle Loading and Global Indexing

The computational volume of the physical domain is populated with $N_\text{F}(t = 0)$ fluid particles (FPs) each with mass $m_i$, corresponding to a total mass $M_F$ following Eq.(2.1). The FPs are associated with a global particle list with index $i = N_{\mathrm{B}} + 1, N_{\mathrm{B}} + N_{\mathrm{F}} + 1$. The FPs are assigned upon initialization with position and a velocity as

$$\left.\begin{array}{l}\mathbf{r}_i(0) = \{x_i(0), y_i(0), z_i(0)\} \\ \mathbf{v}_i(0) = \{v_{ix}(0), v_{iy}(0), v_{iz}(0)\}\end{array}\right\} i = N_{\mathrm{B}} + 1, N_{\mathrm{B}} + N_{\mathrm{F}} + 1 \tag{2.35}$$

To simplify FP loading in cases of complex physical geometries, a mesh for the fluid region is generated with the number of tetrahedrons vertices to be equal to the total number of FPs in the domain.

### 2.2.3 Nearest Neighbor Particle Search (NNPS)

The evaluation of the properties of a fluid particle $i$ ($\mathrm{FP}_i$) require the identification of the nearest neighboring particles (NNP) in its support domain shown in Figure 7(c). A fluid particle $j$ is considered as a nearest neighbor of $i$ when it is located within the smoothing domain of particle $i$, and therefore $r_{ij} \leq \kappa h$, with $\alpha$ determined by the smoothing function. An all-pair NNP search among $N_F$ particles in a domain requires $\mathrm{O}(N_F^2)$ operations per computational cycle. Numerous approaches have been developed for various types of particle simulations that require such a search. We implemented in this work, a NNPS approach that uses concepts from the linked-cell (Hockney and Eastwook, 1981), (Putz, 1998) and the Verlet-list algorithm (in't Veld, Plimpton and Grest, 2008). The approach uses a particle-search grid with size $L_X, L_Y, L_Z$ as shown in Figure 7(c). This search-related grid contains $N_C$ rectangular cells generated with lengths $L_{CX}, L_{CY}, L_{CZ} \geq \max(h)$. In case of a periodic (physical) boundary the search-grid face has to align with the physical boundary, as shown in Figure 5(a). The steps for the NNPS algorithm are summarized in Table 1.

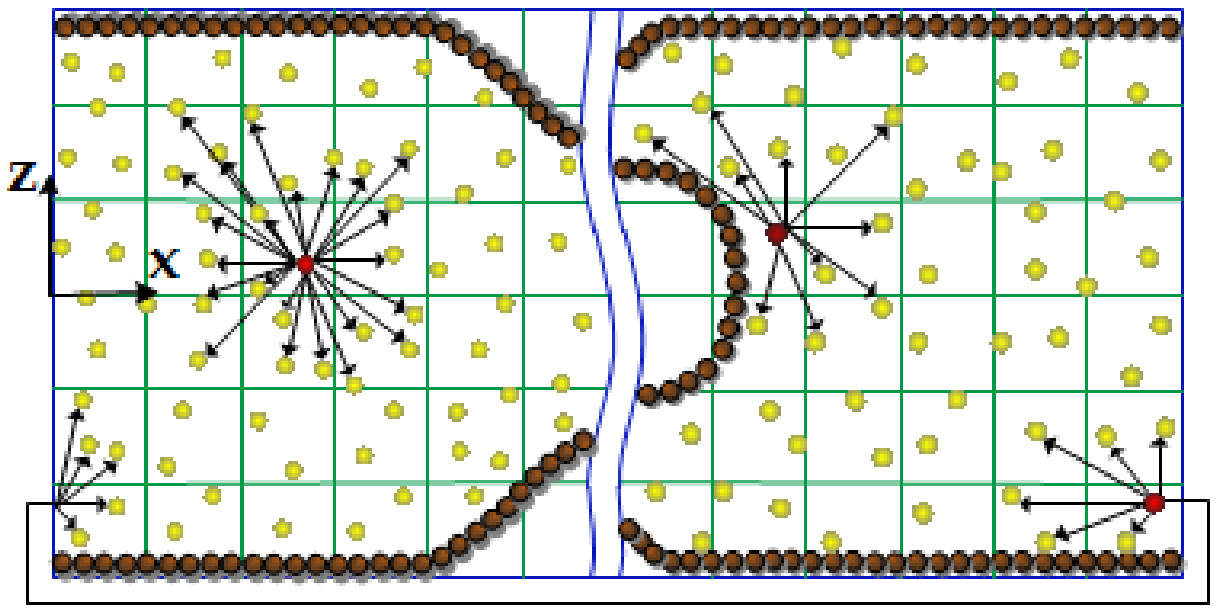

**Figure 7. Rectangular cells used for NNPS in the interior, near a solid and near a periodic boundary.**

**Table 2. Nearest Neighboring Particle Search Algorithm used in SDPD-DV**

S-1. Update in the global particle list the current properties of each FP $i = N_B + 1, N_B + N_F + 1$.

S-2. For each cell $C = 1, N_C$ identify the BPs and FPs contained in it. Generate $N_C$ cell particle lists each with $N_{FC}(t)$ FPs and $N_{BC}(t)$ BPs respectively.

S-3. For each cell particle list $C = 1, N_C$ loop through the $N_{FC}(t)$ FPs and search all FPs and all BPs in the neighboring 27 cells. For each $\mathrm{FP}_i$ create a list with $J_i(t)$ NN FPs and a list with the $K_i(t)$ NN BPs

This particle-search grid is used also in the implementation of periodic boundary conditions. For a FP residing in a cell with a periodic boundary face the NNPS generates the cells on the corresponding periodic boundary, as shown on Figure 7(c). All the FPs in $\mathrm{FP}_i's$ support domain from the corresponding cells are indexed and become available for density and force evaluation.

### 2.2.4 Fluid Particle Density Evaluation

The density evaluation procedure depends on the location of the FP. The three cases considered in this work include a FP away from a boundary (Figure 8(a)), a FP near a solid boundary (Figure 8(b)) and a FP near a periodic boundary (Figure 8(c)).

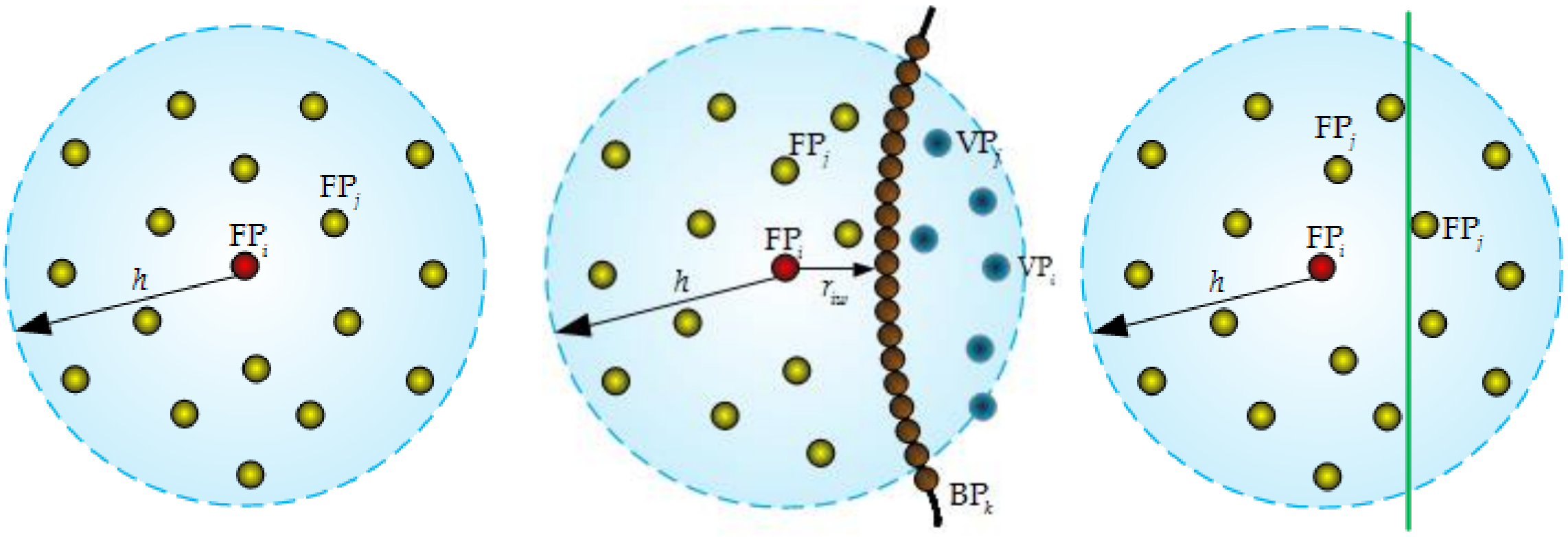

(a) FP away from boundaries. (b) FP near a solid boundary. (c) FP near a periodic boundary.

**Figure 8. Support domain for a fluid particle (FP) for three cases considered in SDPD-DV.**

#### (a) FP Away from Boundaries

The particle number density (number of FPs in unit volume) at the location of the $FP_i$ is defined by the summation over the neighboring FPs in its support domain shown in Figure 8(a) and provides the thermodynamic volume $V_i$,

$$d_i(t) = \sum_{j}^{L_i(t)} W\left(\left|\mathbf{r}_i(t) - \mathbf{r}_j(t)\right|, h\right) = \frac{1}{V_i(t)}, i = 1, N_F \tag{2.36}$$

where $L_i(t)$ is the number of particles in $FP_i$'s support domain. The summation includes the contribution from the $FP_i$ itself. Consequently, the mass density at the position $\mathbf{r}_i$ is given by

$$\rho_i(t) \equiv \rho[\mathbf{r}_i(t)] = m_i d_i(t) = \frac{m_i}{V_i(t)} \tag{2.37}$$

This equation along with the normalization condition Eq. (2.4) satisfies the Lagrangian form of the continuity equation and therefore ensures that mass is conserved in a closed or in a periodic domain ( Espanol and Revenga, 2003). An alternative approach could be based on the continuity equation following SPH by Liu and Liu (2003). A common choice in SPH and SDPD for the interpolant is Lucy's (1977) smoothing function

$$W(r,h) = \frac{105}{16\pi h^3}\left(1+3\frac{r}{h}\right)\left(1-\frac{r}{h}\right)^3, \tag{2.38}$$

which, through Eq. (2.31) gives

$$F_{ij} = \frac{315}{4\pi h^5}\left(1-\frac{r_{ij}}{h}\right)^2. \tag{2.39}$$

**(b) FP Near Solid Boundaries: Dynamic Virtual Particle Allocation Method**

When the support domain of a fluid particle falls outside a boundary (solid or periodic as shown in Figure 8(b) and Figure 8(c) respectively) the smoothing function approximation of Eq. (2.36) will result in error due to the with the unaccounted (truncated) part of the support domain. This boundary truncation error would lead to incorrect density evaluation and would affect the particle momentum and entropy evaluation. This issue has been addressed in SPH (Morris et al., 1997; Randles and Libersky, 1996; Liu and Liu, 2003; Li and Liu, 2007) and SDPD (Litvinov et al., 2008; Bian et al., 2012; Vazquez-Quesada, 2009). With the Lucy smoothing function Eq. (2.38) this occurs when $r_{iw} < h$, where $r_{iw}$ is the distance between the particle and the boundary as shown in Figure 8(b). It should be noted that due to the summation density approach followed in SDPD the density error does not affect the overall mass conservation but affects the total volume.

We describe the density evaluation method developed in this work to compensate for the boundary truncation error. For a FP near a solid wall we developed and implemented the dynamic virtual particle allocation (DVPA) method, by which the truncated portion of the support domain is dynamically filled with virtual particles (VPs). Approaches based on static and dynamic ghost particle allocation appeared in SPH (Morris et al., 1997; Randles and Libersky, 1996; Liu and Liu, 2003; Li and Liu, 2007) and SDPD (Litvinov et al., 2008; Bian et al., 2012). These virtual particles are included in the summation density of Eq. (2.36) following the algorithm described in Table 3. The VPs are generated as mirrored images of the $FP_i$ and its neighbors as shown in Figure 9(a).

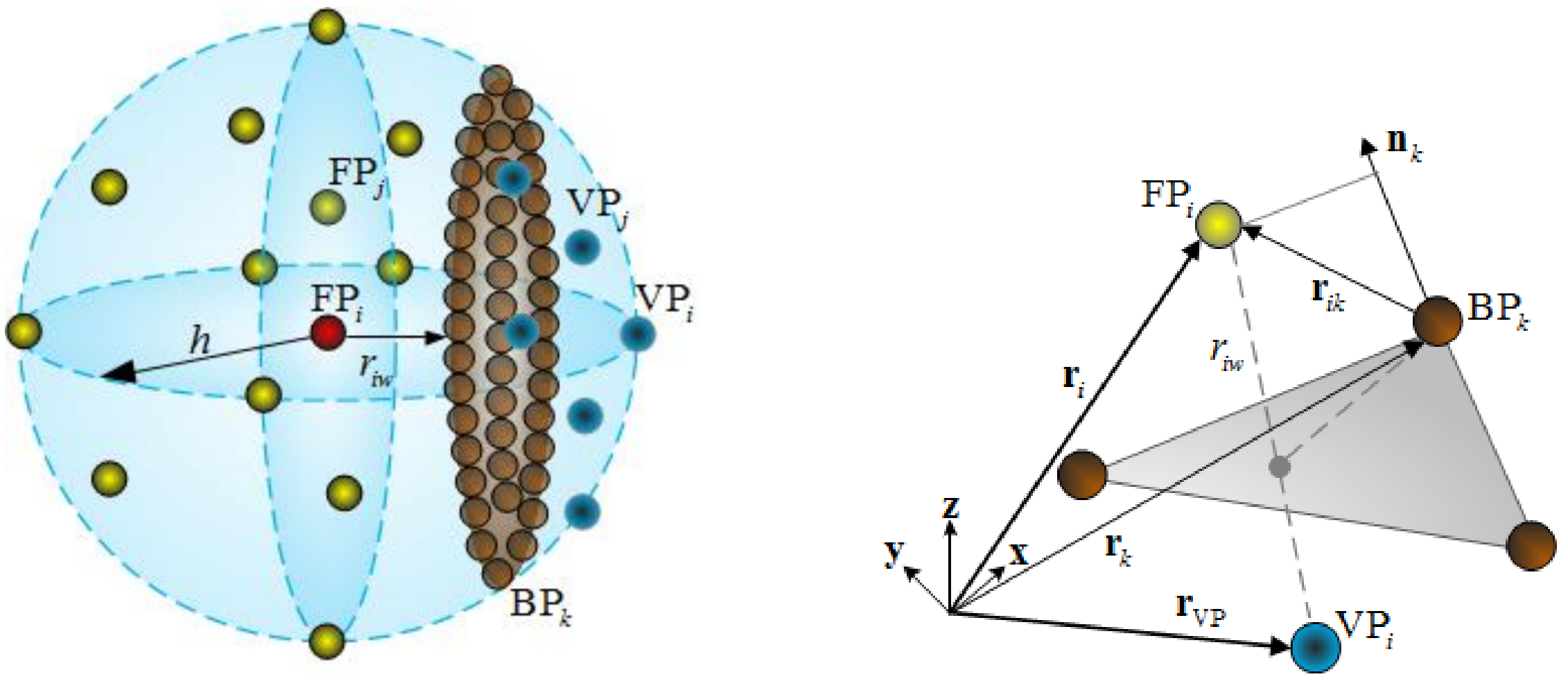

(a) Support domain for a FP near a solid boundary represented by BPs.

(b) Reflection of a FP near a solid boundary to create a virtual particle (VP).

**Figure 9. Dynamic virtual particle allocation (DVPA) method used in SDPD-DV for density and force evaluation. A virtual particle (VP) is generated for a fluid particle (FP) near a solid boundary represented by boundary particles (BP).**

In case of curved walls, we require that

$$h \le aR_{\min} \tag{2.40}$$

where, the coefficient $a \le 0.2$ and $R_{\min}$ is the smallest radius of the curvature for the solid boundary in FP's support domain. With such a constraint it is possible to consider the surface of the wall contained within the smoothing domain of a FP as quasi-flat and its normal vector as locally constant. The algorithm is summarized in Table 3.

**Table 3. Dynamic Virtual Particle Allocation and Density Evaluation Algorithm used in SDPD-DV.**

S-1. Loop through each $\mathrm{FP}_i$.

S-2. Compute the particle number density $d_i$ using the $\mathrm{FP}_j$ neighbors from Eq. (2.36) and $\mathrm{FP}_j$ itself.

S-3. Search the nearest-neighbor BP list to find the closest $\mathrm{BP}_k$. Compute $r_{iw} = (\mathbf{r}_i - \mathbf{r}_k).\mathbf{n}_k$ and compare with the average minimum distance $l_{\mathrm{FPs}}$ between the FPs. A choice is $l_{\min} = 0.1\min(l_{\mathrm{FPs}})$ provided by the surface triangulation. If $r_{iw} \ge l_{\min}$ create a virtual particle $\mathrm{VP}_i$ with its position $\mathbf{r}_{\mathrm{VP}}$ reflected across the normal to the surface plane, $\mathbf{r}_{VP} = \mathbf{r}_i - 2r_{iw}\mathbf{n}_k$. The process is shown in Fig. 4(b).

S-4. If $|\mathbf{r}_i - \mathbf{r}_{\mathrm{VP}}| < h$ add to particle number density $d_i$ the contribution due to the $\mathrm{VP}$.

S-5. Loop through each of the nearest-neighbors $\mathrm{FP}_j$ and create a mirrored $\mathrm{VP}_j$ following S-3.

S-6. If $|\mathbf{r}_j - \mathbf{r}_{\mathrm{VP}}| < h$ then add to particle number density $d_i$ the contribution due to the $\mathrm{VP}_j$.

#### (c) FP Near Periodic Boundaries: Periodic Boundary Particle Allocation Method

The density of a FP containing a periodic boundary within its support domain consists of contributions from the surrounding FPs and the FPs in the truncated part as shown in Figure 8(c). We implemented a periodic boundary particle allocation (PBPA) method, by which the truncated portion of the support domain is filled with copied particles from the periodic cells identified during the NNPS (Sec. 2.2.3). The density of the FP is then obtained using Eq. (2.36) where the $\left|\mathbf{r}_{ij}\right|$ for a copied FP is given by

$$r_{ij} = \sqrt{\left(x_{ij} + L_{Px}\right)^2 + \left(y_{ij} + L_{Py}\right)^2 + \left(z_{ij} + L_{Pz}\right)^2} \tag{2.41}$$

The values for $L_{Px}, L_{Py}, L_{Pz}$ are determined by the location of the periodic boundary at the specified cell face. If there is only one periodic boundary at a cell face, as is the periodic outlet in Figure 5, $L_{Px} = L_X, L_{Py} = 0, L_{Pz} = 0$. In a case of fully periodic boundaries, a corner cell has three periodic boundaries and $L_{Px} = L_X, L_{Py} = L_Y, L_{Pz} = L_Z$ when copying the FP from the corresponding corner.

### 2.2.5 Fluid Particle Pressure and Temperature Evaluation

The full algorithm of SDPD, which include entropy equation, needs a closure by equation of states that provides pressure $P_i = P\left(\rho_i, S_i\right)$ and temperature $T_i = T\left(\rho_i, S_i\right)$ given by Eq. (1.65) and (1.66). Therefore, the internal energy is required as a function of $m_i$, $S_i$, and $V_i$, in the form of

$$E_i = E\left(m_i, V_i, S_i\right). \tag{2.42}$$

For ideal monatomic gases we consider the Sackur-Tetrode entropy equation given by Tetrode (1912) or Laurendeau (2005)

$$S = Nk_B \left[ \ln\left( \frac{V}{N} \left( \frac{4\pi mE}{3Nh^2} \right)^{3/2} \right) + \frac{5}{2} \right], \tag{2.43}$$

where $N$ is the number of molecules in the system, $h$ is the Planck's constant. For an SDPD fluid particle $i$ which is considered as a thermodynamic subsystem, following Eq. (2.43), the entropy is given by

$$S_i = N_i k_B \left[ \ln\left( \frac{V_i}{N_i} \left( \frac{4\pi m_i E_i}{3N_i h^2} \right)^{3/2} \right) + \frac{5}{2} \right], \tag{2.44}$$

where $N_i$ is the number of molecules in FP $i$, with volume $V_i$. Then the internal energy of FP $i$ is given by

$$E_i\left(N_i, V_i, S_i\right) = \frac{3h^2 N_i^{5/3}}{4\pi m_i V_i^{2/3}} \exp\left( \frac{2S_i}{3N_i k_B} - \frac{5}{3} \right), \tag{2.45}$$

Therefore, the temperature and pressure of monatomic ideal gas are given by Espanol and Serrano (1999)

$$T_i = \left( \frac{\partial E_i}{\partial S_i} \right) = \frac{2}{3N_i k_B} E_i\left(N_i, V_i, S_i\right) \tag{2.46}$$

$$P_i = -\left( \frac{\partial E_i}{\partial V_i} \right) = \frac{N_i}{V_i} k_B T_i \tag{2.47}$$

And the heat capacity is given by

$$C_i = \frac{3}{2} N_i k_B \tag{2.48}$$

To initialize the entropy of FP $i$, we follow (2.44) by

$$S_{i0} = \frac{3N_i k_B}{2} \left[ \ln\left( \frac{2\pi m_i V^{2/3} k_B}{h^2 N_i^{2/3}} T_0 \right) + \frac{5}{3} \right] \tag{2.49}$$

In Chapter 3, we consider also the isothermal SDPD implementation, where entropy equation is decoupled from the system, and temperature is therefore kept constant and initially assigned to the FPs. Closure in this case is obtained by an equation of state that provides $P_i = P(\rho_i, T_i)$. For an ideal gas flow we use

$$P_i = \rho_i R T_i \tag{2.50}$$

For a perfect gas with constant specific heats, the pressure is defined as (Batchelor, 1967)

$$P = (\gamma - 1)\rho e \tag{2.51}$$

For an incompressible flow, we follow Batchelor, (1967)

$$P_i = \frac{\rho_0 c^2}{2}\left[\left(\frac{\rho_i}{\rho_0}\right)^2 - 1\right] \tag{2.52}$$

where, $c$ is an artificial sound speed and $\rho_0$ is the initial density. A similar law is given by Morris et al. (1997) and Liu and Liu, (2003),

$$P_i = c^2 \rho_i \tag{2.53}$$

The value for $c$ should be low enough to avoid using very small time steps and high enough to be sufficiently close to the real fluid. As suggested by Morris et al. (1997) and Liu and Liu (2003), $c$ can be estimated by

$$c^2 = \max\left(\frac{V_0^2}{\delta}, \frac{\nu_0 V_0}{L_0 \delta}, \frac{F_0 L_0}{\delta}\right) \tag{2.54}$$

where, $\delta = \Delta\rho/\rho_0$ is the relative density perturbation, $V_0$ is a velocity scale, $L_0$ is a length scale, $\nu_0$ is the kinematic viscosity, and $f_o$ is a body force per unit mass.

Several alternative forms of Eq. (2.52) could be found as

$$P = P_0 \left\{ \left( \frac{\rho}{\rho_0} \right)^{\gamma} - 1 \right\} \tag{2.55}$$

$$P = B \left( \left( \frac{\rho}{\rho_0} \right)^{\gamma} - 1 \right) \tag{2.56}$$

$$P = P_0 \left( \frac{\rho}{\rho_0} \right)^{\gamma} + b \tag{2.57}$$

where $B$ and $b$ are the coefficient chosen to keep the density fluctuation.

In chapter 4, the full algorithm of SDPD-DV will be tested and validated.

### 2.2.6 Fluid Particle Force Evaluation

The force evaluation procedure depends on the location of the $\mathrm{FP}$. As with density we considered three cases depicted in Figure 8.

**(a) FP Away from Boundaries**

For a FP with a position away from a boundary, as shown in Figure 8(a), the forces $\mathbf{F}_C$, $\mathbf{F}_D$ and $\mathbf{F}_R$ are evaluated directly from Eq. (2.22), (2.23) and (2.24). The summation is performed over all FPs in the support domain of the $\mathrm{FP}_i$ provide by the NPPS.

**(b) FP Near Solid Boundaries: Dynamic Virtual Particle Allocation**

When a FP is in the proximity of a solid boundary ($r_{iw} < h$) a correction term is required also in the momentum Eq. (1.63) due to the presence of the wall and truncated domain. This correction term is based only on the local thermodynamic state of the fluid and the local geometry of the wall described by the BPs. For force evaluation, we use the DVPA method outlined for density in Sec. 2.2.4(b). and develop additional force terms to supplement the SDPD equations (2.22)-(2.24).

The thermodynamic properties of each VP are assigned to be identical to the properties of the corresponding FP that is mirrored, specifically

$$\rho_{j_V} = \rho_j \,,\; P_{j_V} = P_j \,. \tag{2.58}$$

For each FP in the proximity of the wall the conservative $\mathbf{F}_C$ in Eq. (2.22) becomes

$$\mathbf{F}_C = \sum_{j=1:\tilde{N}} \left[ \frac{P_i}{d_i^2} + \frac{P_j}{d_j^2} \right] F_{ij} \mathbf{r}_{ij} + \mathbf{F}_C^{VP} \tag{2.59}$$

where the conservative force due to the VPs is

$$\mathbf{F}_C^{VP} = \sum_{j=1:\tilde{N}_{VP}} \left[ \frac{P_i}{d_i^2} + \frac{P_{j_V}}{d_{j_V}^2} \right] F_{ij_V} \mathbf{r}_{ij_V} \tag{2.60}$$

In the above equations, $\tilde{N}$ and $\tilde{N}_V$ are respectively the number of FPs and VPs contained within the support domain of a $\text{FP}_i$. Generally, $\tilde{N}_V < \tilde{N}$ , $F_{ij_V} < F_{ij}$ and the identity sign is achieved only when the $\text{FP}_i$ is located exactly on the wall ( $r_{wall} = 0$ ).

Consistent with the above methodology we also introduce an additional term to the dissipative part of the dynamics to account for the truncated part of the domain. For each FP in the proximity of the wall the $\mathbf{F}_D$ in Eq. (2.22) becomes

$$\begin{aligned} \mathbf{F}_D = -\Bigg( & \sum_{j=1:\tilde{N}} \left( 1 - \frac{T_i T_j}{\left(T_i + T_j\right)^2} \left[ \frac{k_B}{C_i} + \frac{k_B}{C_j} \right] \right) \left( \frac{5\eta}{3} - \zeta \right) \frac{F_{ij}}{d_i d_j} \mathbf{v}_{ij} + \\ & \sum_{j=1:\tilde{N}} \left( 1 - \frac{T_i T_j}{\left(T_i + T_j\right)^2} \left[ \frac{k_B}{C_i} + \frac{k_B}{C_j} \right] \right) 5 \left( \frac{\eta}{3} + \zeta \right) \frac{F_{ij}}{d_i d_j} \mathbf{e}_{ij} \mathbf{e}_{ij} \cdot \mathbf{v}_{ij} \Bigg) + \mathbf{F}_D^{VP} \end{aligned} \tag{2.61}$$

where the dissipative force due to the VPs in the truncated domain is

$$\mathbf{F}_D^{VP} = -\sum_{j=1:\tilde{N}_V}\left(1-\frac{T_iT_j}{\left(T_i+T_j\right)^2}\left[\frac{k_B}{C_i}+\frac{k_B}{C_j}\right]\right)\left(\frac{5\eta}{3}-\zeta\right)\frac{F_{ij}}{d_i d_j}\mathbf{v}_{ij_V} - \sum_{j=1:\tilde{N}_V}\left(1-\frac{T_iT_j}{\left(T_i+T_j\right)^2}\left[\frac{k_B}{C_i}+\frac{k_B}{C_j}\right]\right)5\left(\frac{\eta}{3}+\zeta\right)\frac{F_{ij}}{d_i d_j}\mathbf{e}_{ij_V}\mathbf{e}_{ij_V}\cdot\mathbf{v}_{ij_V} \tag{2.62}$$

For a given wall velocity $\mathbf{v}_W$ the velocity of the VPs required to compute the velocity vector $\mathbf{v}_{ij_V}$ is defined as

$$\mathbf{v}_{j_V} = 2\mathbf{v}_W - \mathbf{v}_j,\ i \in j \tag{2.63}$$

Substituting in the definition of the velocity vector $\mathbf{v}_{ij_V}$ it become

$$\mathbf{v}_{ij_V} = \mathbf{v}_i - \mathbf{v}_{j_V} = \mathbf{v}_i + \mathbf{v}_j - 2\mathbf{v}_W,\ i \in j \tag{2.64}$$

The above formulation is sufficient to impose the non-slip condition at a solid wall boundary. For a planar wall and a FP $i$ located on the wall $\mathbf{e}_{ij_V}^t = \mathbf{e}_{ij}^t, \mathbf{e}_{ij_V}^n = -\mathbf{e}_{ij}^n$, Eq. (2.64) enforces $\mathbf{v}_i = \mathbf{v}_W$.

For a wall without heat flux, we assign the VP temperature equal to the wall temperature given by

$$T_{j_V} = T_w \tag{2.65}$$

If the temperature gradient along wall surface is not equal to zero, the $T_w$ would be the temperature of the nearest BP of FP $j$. Whereas if the heat flux through the surface of the wall is not zero, the VP temperature is assigned by

$$T_{j_V} = 2T_w - T_j \tag{2.66}$$

**(c) FP Near Periodic Boundaries**

The force evaluation of a FP $i$ containing a periodic boundary within its support domain follows the approach discussed in the density evaluation (Sec. 3.4.3). The FPs in the truncated part are copied from the corresponding periodic cells. The force on the $\mathrm{FP}_i$ is then obtained using Eq. (1.63) and Eq. (1.64) where the $\mathbf{r}_{ij}$ for a copied particle is given by

$$\mathbf{r}_{ij} = \left\{ x_i - \left( x_j + L_{Px} \right), y_i - \left( y_j + L_{Py} \right), z_i - \left( z_j + L_{Pz} \right) \right\} \tag{2.67}$$

### 2.2.7 Integration of Fluid Particle Position, Momentum and Entropy Equations

**(a) FP Away from Boundaries**

Once the particle neighbor list is constructed and the density, pressure and force is evaluated the integration of motion and momentum Eqs. (1.62)-(1.64) proceeds as indicated in Figure 4. The integration scheme used in this work is an implementation of the Velocity-Verlet scheme (Nikunen et al., 2003) for momentum equation, coupled with Runge-Kutta scheme for entropy equation, and summarized by the algorithm in Table 4.

The choice for $\delta$ is based on required accuracy in the fractional change of the velocity estimate. The required $\Delta t$ follows standard SPH conditions (Morris et al., 1997; Li and Liu, 2007) and once choice provides a posterior check, after the integration has proceeded,

$$\Delta t \le \frac{h}{v_{\max}} \tag{2.68}$$

**(b) Near Solid Boundaries: Reflective Bounce Forward Condition**

In general the additional repulsive force Eq. (2.59) exerted on the $\mathrm{FP}_i$ from the VPs is not enough to prevent $\mathrm{FP}_i$ penetrating the solid wall. When such event occurs the $\mathrm{FP}_i$ is reinserted in the fluid domain by a bounce forward reflection, this takes place after S-3 in the integration algorithm Table 4. The bounce forward reflection method is depicted in Fig. 5 and summarized by the algorithmic steps in Table 5 which are embedded in S-3 of Table 3.

### (c) FP Near Periodic Boundaries

In case that a FP penetrates a periodic boundary it is re-injected into the related periodic cell. This operation follows the bounce-forward method.

**Table 4. Time integration in SDPD-DV. Velocity-Verlet is used for the particle position and momentum equation, and Runge-Kutta is used for the entropy equation.**

S-1. Calculate $\mathbf{F}_C\left\{\mathbf{r}_i^n, T_i^n\right\}$, $\mathbf{F}_D\left\{\mathbf{r}_i^n, \mathbf{v}_i^n, T_i^n\right\}$, $\mathbf{F}_R\left\{\mathbf{r}_i^n, T_i^n\right\}$ and $\Delta E_{Vi}\left\{\mathbf{r}_i^n, \mathbf{v}_i^n, T_i^n\right\}$, $\Delta E_{Cd}\left\{\mathbf{r}_i^n, T_i^n\right\}$, $\Delta E_R\left\{\mathbf{r}_i^n, \mathbf{v}_i^n, T_i^n\right\}$.

S-2. Update $\mathbf{v}_i^{n+1/2} \leftarrow \mathbf{v}_i^n + \frac{1}{2}\frac{1}{m}\left(\mathbf{F}_C\Delta t + \mathbf{F}_D\Delta t + \mathbf{F}_R\right)$, and calculate

$dS_i^{n+1/2} = \frac{\Delta E_{Vi}}{T_i^n}\frac{dt}{2} + \frac{\Delta E_{Cd}}{T_i^n}\frac{dt}{2} + \frac{\Delta E_R}{2T_i^n}$, then update $T_i^{n+1/2} \leftarrow T_i^n f\left(dS_i^{n+1/2}\right)$.

S-3. Update $\mathbf{r}_i^{n+1} \leftarrow \mathbf{r}_i^n + \mathbf{v}_i^{n+1/2}\Delta t$.

S-4. Calculate $\mathbf{F}_C\left\{\mathbf{r}_i^{n+1}, T_i^{n+1/2}\right\}$, $\mathbf{F}_R\left\{\mathbf{r}_i^{n+1}, T_i^{n+1/2}\right\}$, and $\Delta E_R\left\{\mathbf{r}_i^{n+1}, \mathbf{v}_i^{n+1/2}, T_i^{n+1/2}\right\}$.

S-5. Update $\tilde{\mathbf{v}}_i^{n+1} \leftarrow \mathbf{v}_i^{n+1/2} + \frac{1}{2}\frac{1}{m}\left(\mathbf{F}_C\Delta t + \mathbf{F}_R\right)$, and $\tilde{T}_i^{n+1} = T_i^{n+1/2}$.

S-6. Calculate $\mathbf{F}_D\left\{\mathbf{r}_i^{n+1}, \tilde{\mathbf{v}}_i^{n+1}, \tilde{T}_i^{n+1}\right\}$, $\Delta E_{Vi}\left\{\mathbf{r}_i^{n+1}, \tilde{\mathbf{v}}_i^{n+1}, \tilde{T}_i^{n+1}\right\}$ and $\Delta E_{Cd}\left\{\mathbf{r}_i^{n+1}, \tilde{T}_i^{n+1}\right\}$.

S-7. Update $\mathbf{v}_i^{n+1} \leftarrow \tilde{\mathbf{v}}_i^{n+1} + \frac{1}{2}\frac{1}{m}\left(\mathbf{F}_D\Delta t\right)$, and calculate $dS_i^{n+1} = \frac{\Delta E_{Vi}}{\tilde{T}_i^{n+1}}\frac{dt}{2} + \frac{\Delta E_{Cd}}{\tilde{T}_i^{n+1}}\frac{dt}{2} + \frac{\Delta E_R}{2T_i^{n+1/2}}$, then update $T_i^{n+1} \leftarrow T_i^{n+1/2} f\left(dS_i^{n+1}\right)$.

S-8. If $\left|\frac{\mathbf{v}_i^{n+1} - \tilde{\mathbf{v}}_i^{n+1}}{\mathbf{v}_i^{n+1}}\right| \geq \delta$ , update $\tilde{T}_i^{n+1} = T_i^{n+1}$, loop over step 6.

Calculate $\mathbf{F}_D\left\{\mathbf{r}_i^{n+1}, \mathbf{v}_i^{n+1}, T_i^{n+1}\right\}$, $\Delta E_{Vi}\left\{\mathbf{r}_i^{n+1}, \mathbf{v}_i^{n+1}, T_i^{n+1}\right\}$, $\Delta E_{Cd}\left\{\mathbf{r}_i^{n+1}, T_i^{n+1}\right\}$ and go to step S-2.

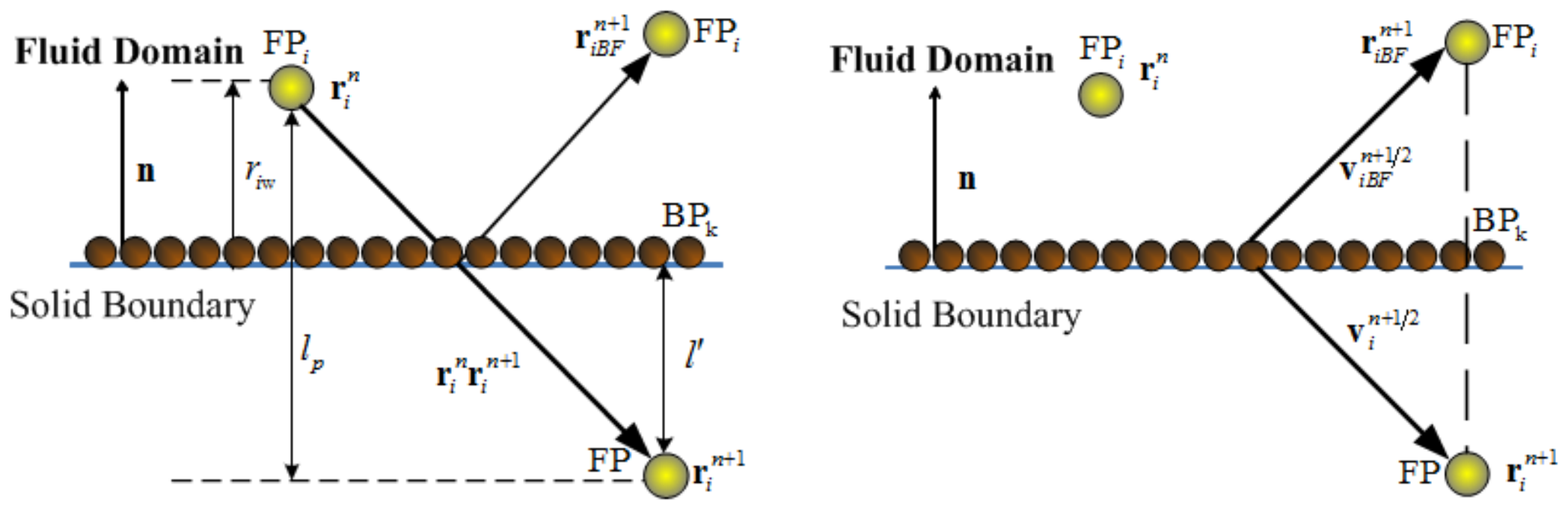

(a) Position reflection of a penetrating FP. (b) Velocity reflection of a penetrating FP.

**Figure 10. Reflective bounce-forward method in SDPD-DV applied in a case where a fluid particle (FP) penetrates a solid boundary represented by boundary particles (BP).**

**Table 5. Bounce-forward Method used in SDPD-DV in case of particle penetration.**

S-1. After the FPs position has been updated to the new time step n+1, check the sign of the scalar product ($\left[\mathbf{P}_{i\mathrm{BP}}^{n}\mathbf{P}_{i}^{n+1}\right]\bullet\mathbf{n}$), where $\left[\mathbf{P}_{i\mathrm{BP}}^{n}\mathbf{P}_{i}^{n+1}\right]$ is the vector between the FP n+1 position and the closest BP of n position, $\mathbf{n}$ is the wall normal unit. If $\left[\mathbf{P}_{i\mathrm{BP}}^{n}\mathbf{P}_{i}^{n+1}\right]\bullet\mathbf{n} < 0$, particle penetrates, and proceed to S-2.

S-2. If $\mathrm{FP}_i$ penetrates the wall, compute the scalar product between the vector $\mathbf{P}_{i}^{n}\mathbf{P}_{i}^{n+1}$ and the unit normal wall vector $\mathbf{n}$ obtaining the distance $l_p = \left|\left[\mathbf{P}_{i}^{n}\mathbf{P}_{i}^{n+1}\right]\bullet\mathbf{n}\right|$.

S-3. Compute the distance between the wall surface and $\mathrm{FP}_i$ new position (inside the wall) $\mathbf{P}_{i}^{n+1}$ as $l' = l_p - r_W$ .

S-4. Reflect the $\mathrm{FP}_i$ inside the domain and compute its new position as $\mathbf{P}_{iBF}^{n+1} = \mathbf{P}_{i}^{n+1} + 2l'\mathbf{n}$.

S-5. Impose the velocity component of $\mathrm{FP}_i$ normal to the wall to be opposite in sign. This is achieved by imposing the velocity of $\mathrm{FP}_i$ after the reflection to be $\mathbf{v}_{iBF}^{n+1/2} = \mathbf{v}_{i}^{n+1/2} + 2\left|\mathbf{v}_{i}^{n+1/2}\cdot\mathbf{n}\right|$ .

### 2.2.8 Particle Properties and Transport Coefficients

#### (a) Self-Diffusion Coefficient

A property of a FP from the SDPD-DV simulation at a discrete time $t = k\Delta t$ is designated as $X_i^k \equiv \mathbf{r}_i(t), \mathbf{v}_i(t), S_i(t), T_i(t)$ etc. An SDPD-DV output (or sample) consists therefore of all particle properties $X_i^k, i = 1, N_F$. For steady SDPD-DV simulations we gather after reaching steady-state, $m = 1, M$ independent SDPD-DV samples. For unsteady simulations, a number of $M$ individual runs are performed in order to generate a sufficient number of independent samples.

Transport coefficients such as diffusivity, shear viscosity are dynamical properties desired from SDPD simulations. The self-diffusion coefficient $D$ can be evaluated for a system with dimensionality $d_D$ from the mean-square displacement (MSD) through the Generalized Einstein formula as shown by Allen and Tildesley (1997)

$$D = \lim_{\tau \to \infty} \frac{\left\langle \left[ \mathbf{r}_i(t_0 + \tau) - \mathbf{r}_i(t_0) \right]^2 \right\rangle}{2 d_D \tau} \tag{2.69}$$

where, $\tau$ is the delay time $\tau = nK\Delta t$, $n$ is the time interval of each particle sample, and $K$ the number of particle samples in delay time $\tau$ which is $n\Delta t = t_{k+1} - t_k$ and $\tau = t_{k+K} - t_k$. The schema is shown in Figure 11. It can also be evaluated by the Green-Kubo formula through the velocity autocorrelation function (VACF) as introduced by Allen and Tildesley (1997)

$$D = \frac{1}{d_D} \int_0^\infty d\tau \left\langle \mathbf{v}_i(t_0) \mathbf{v}_i(t_0 + \tau) \right\rangle \tag{2.70}$$

The term in the brackets in Eq. (2.69) and Eq. (2.70) denotes the time correlation function for time-dependent signals $A(t), B(t)$ (Haile, 1997)

$$\Psi(t) = \langle A(t_0)B(t_0+\tau)\rangle = \lim_{t\to\infty}\frac{1}{t}\int_0^t A(t_0)B(t_0+\tau)dt_o \tag{2.71}$$

The function $A(t_0)$ is sampled at $t_0$ and $B(t_0+\tau)$ after a delay time $\tau$. The integral is evaluated over many time origins $t_0$ shown in Figure 11. For a given number of time samples $M$ and a number of particle samples $K$ in delay times $\tau$ the time correlation can be approximated as a summation over $M-K+1$ number of available time origins (Haile, 1997)

$$\Psi(t) \simeq \frac{1}{M-K+1}\sum_{m=1}^{M-K+1} A(t_0)B(t_0+\tau) \tag{2.72}$$

The calculation can be improved by averaging over all particles in each sample,

$$\langle A(t_0)B(t_0)\rangle = \frac{1}{(M-K+1)N_P}\sum_{m=1}^{M-K+1}\sum_{i=1}^{N_P} A_i(t_0)B_i(t_0+\tau) \tag{2.73}$$

Then the self-diffusion coefficient based on MSD is provided by

$$D = \frac{1}{2d_D t(M-K+1)N_P}\lim_{\tau\to\infty}\sum_{m=1}^{M-K+1}\sum_{i=1}^{N_P}\left[\mathbf{r}_i(t_0+\tau)-\mathbf{r}_i(t_0)\right]^2 \tag{2.74}$$

And the self-diffusion coefficient based on VACF is given by

$$D = \frac{1}{d_D M N_P}\int_0^\infty dt\left(\sum_{m=1}^{M-K+1}\sum_{i=1}^{N_P}\mathbf{v}_i(t_0)\mathbf{v}_i(t_0+\tau)\right) \tag{2.75}$$

The integral is approximated with a summation over all the discrete time in delay time $\tau$ shown by

$$D = \frac{1}{d_D M N_P}\left(\sum_{t=t_0}^{nK\Delta t}\sum_{m=1}^{M-K+1}\sum_{i=1}^{N_P}\mathbf{v}_i(t_0)\mathbf{v}_i(t_0+t)t\right) \tag{2.76}$$

In this work, we implemented the VAC algorithms following Haile (1997) and the MSD algorithm following Rapaport (1995).

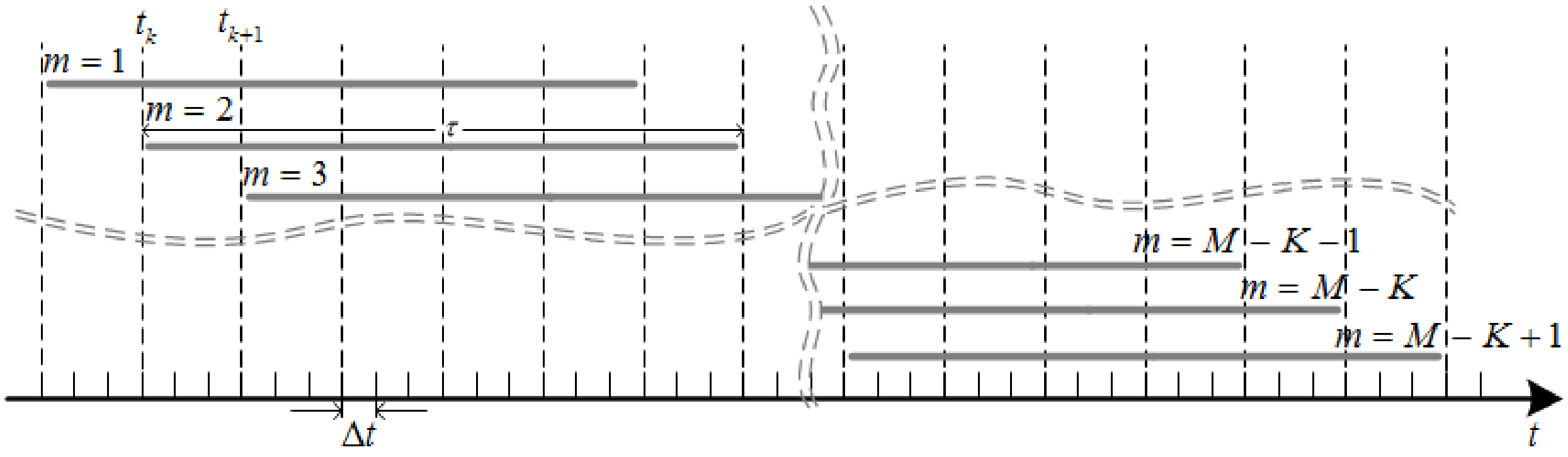

**Figure 11. Sampling used inn SDPD-DV for evaluation of the self-diffusion coefficient. The delay time is *τ*. The total number of samples is *M*, and *K* is the number of samples in delay time $\tau$ . The time origins are indexed from *m*=1 ~ (*M*-*K*+1).**

### (b) Shear Viscosity

Another transport coefficient we studied in this work is the bulk viscosity $\eta$. An expression analogous to the Einstein diffusion Eq. (2.69) lead to the viscosity expression given by Allen and Tildesley (1987)

$$\eta = \lim_{\tau\to\infty} \frac{1}{6k_B T V \tau} \left\langle \sum_{x<y} \left[ \sum_i m_i v_{yi}(t_0+\tau) r_{xi}(t_0+\tau) - \sum_i m_i v_{yi}(t_0) r_{xi}(t_0) \right]^2 \right\rangle, \qquad (2.77)$$

where $\sum_{x<y}$ denotes a sum over the three pairs of distinct vector components ( $xy$ , $yz$ , and $zx$ ) which improve the statistics. The alternative Green-Kubo form is given by Allen and Tildesley (1987)

$$\eta = \frac{V}{3k_B T} \int_0^\infty \left\langle \sum_{x<y} P_{xy}(t) P_{xy}(0) \right\rangle dt \, , \qquad (2.78)$$

where

$$P_{xy} = \frac{1}{V} \left[ \sum_j m_j v_{xj} v_{yj} + \frac{1}{2} \sum_{i\neq j} r_{xij} f_{yij} \right] \qquad (2.79)$$

The second term in $P_{xy}$ can be fulfilled with the force computation, and for Lennard-John potential the weighting function is $\mathbf{f}_{ij} = f\left(r_{ij}\right)\mathbf{r}_{ij}/r_{ij}$, which leads to $P_{xy} = P_{yx}$.

The numerical implementation follows section 2.2.8.(a) as shown in Figure 11.

### 2.2.9 Instantaneous and Sample-Averaged Fluid Field Properties

For analysis and visualization we construct also Eulerian (field) properties $\mathbf{V}(\mathbf{r},t), S(\mathbf{r},t), T(\mathbf{r},t)$. To obtain such properties we first generate a tetrahedral mesh over the fluid domain of interest. The vertices of the tetrahedron are designated as the nodes of the domain, each with an assigned index $d = 1,..,G_d$ and coordinates $\mathbf{r}_d = (x_d, y_d, z_d)$. An instantaneous fluid property associated with a node $d$ at $t = k\Delta t$, is obtained using the smoothing function approximation Eq. (2.8) shown in Figure 12, as

$$ {}^{m}X(\mathbf{r}_d, t) \equiv {}^{m}X^{k}(d) \equiv {}^{m}X_d^k = \frac{\sum_i W(|r_d - r_i|)\, {}^{m}X_i^k}{\sum_i W(|r_d - r_i|)} \tag{2.80} $$

The summation is extended over all the FPs within a set distance $l_d$ from the vertex $\mathbf{r}_d$. For unsteady simulations, we obtain after reaching steady-state the sample-averaged field property at a node $d$ at $t = k\Delta t$, is given by

$$ X(\mathbf{r}_d, t) \equiv X_d^k = \sum_{m=1}^{\mathrm{M}} {}^{m}X_d^k \Big/ \mathrm{M} \tag{2.81} $$

For steady simulations, we obtain after reaching steady-state the sample-averaged field property at a node $d$ is given by

$$ X(\mathbf{r}_d) \equiv X_d = \sum_{m=1}^{\mathrm{M}} {}^{m}X_d^k \Big/ \mathrm{M} \tag{2.82} $$

The choice of the Eulerian grid size and thus the interpolating distance $l_d$ must be compatible with the length scale of the phenomena under consideration. An excessively coarse (Eulerian) grid or interpolating distance could smooth out small scales of interest.

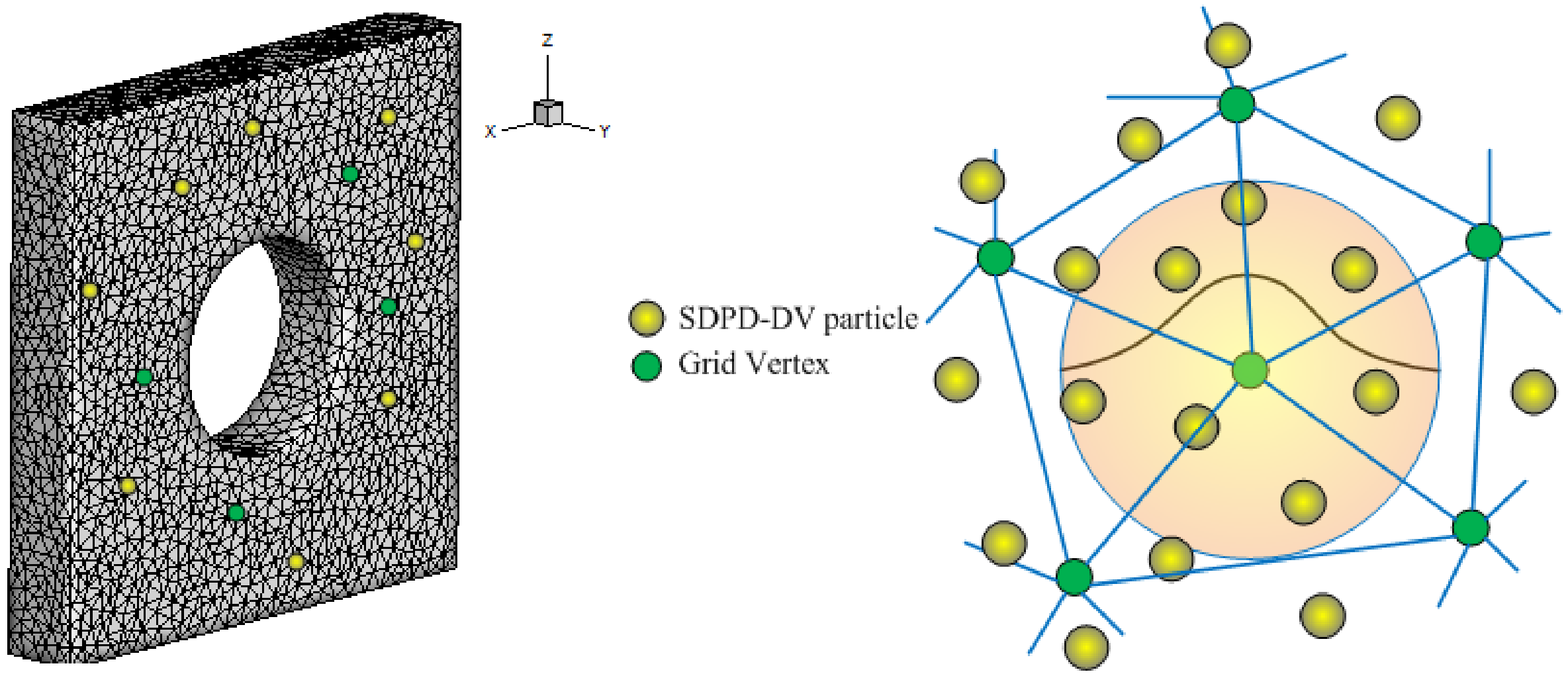

(a) A typical post-processing grid. (b) Sampling region near vertex of the grid.

**Figure 12. Post-processing grid used for evaluation of field (Eulerian) instantaneous and time-averaged properties from SDPD-DV particle samples.**

# 3. VERIFICATION, VALIDATION, AND ERROR OF THE ISOTHERMAL SDPD-DV

In this chapter we utilize the isothermal SDPD-DV code and perform verification and validation tests that cover the hydrodynamic and mesoscopic regimes. The first verification of SDPD-DV involves comparisons with analytical solutions for a body-force driven, transient, Poiseuille flow of water between parallel plates of $10^{-3}$ m height. Physical insights on the force method in SDPD-DV are obtained by calculating the components of the forces attributed to the dynamically allocated virtual particles and compared with previous DPD investigations by Altenhoff et al. (2007) and Fedosov et al. (2008). The second verification test involves the transient, Couette flow of water between parallel plates of $10^{-3}$ m height. The third test used for verification involves the low-Reynolds number, incompressible flow over a cylinder of radius 0.02 m. This benchmark test has been used in SPH by Morris et al. (1997), Vazquez-Quesada and Ellero (2012) and in DPD simulations by Kim and Phillips (2004). Our SDPD-DV results are compared with those obtained from ANSYS FLUENT (FLUENT 6.3.26, help system, ANSYS Inc.). The final set of benchmark tests involves calculation of equilibrium states for liquids and gases in mesoscopic domains. This extended set of SDPD-DV simulations evaluates the translational temperature for liquid water and gaseous nitrogen, and the self-diffusion coefficient of liquid water. The SDPD-DV results are compared with analytical expressions introduced by Litvinov et al. (2009), Bird et al. (2007) and experiments Holz and Sacco (2000). The SDPD-DV liquid water simulations examine also the scale effects on the self-diffusion coefficient and the Schmidt number by varying the mass and size of the fluid particles (Vazquez-Quesada et al., 2009), (Litvinov et al., 2009). The material from this chapter can be found in Gatsonis et al. (2013).

### 3.1 Transient Body-Force Driven Planar Poiseuille Flow

The first test case involves an incompressible Poiseuille flow between two stationary infinite plane parallel plates as shown in Figure 13. The velocity profiles as predicted by SDPD-DV are compared to theoretical formulations. The SDPD-DV density is plotted along with the standard deviation. In addition, the boundary forces due to virtual particles are evaluated, and the distributions are plotted. This test case verifies the ability of the SDPD-DV to minimize density fluctuations and enforce the no-slip condition on solid boundaries.

#### 3.1.1 Input Conditions and Computational Parameters

The test involves an incompressible Poiseuille flow with density $\rho$ across two infinite parallel, stationary walls with separation height $L_Z$ as depicted in Figure 13. The fluid considered in the SDPD-DV simulations is $H_2O$(l) with $\rho_a = 1{,}000 \text{ kg} \cdot \text{m}^{-3}$ and $\eta = 10^{-3} \text{ kg} \cdot \text{m}^{-1}\text{s}^{-1}$.

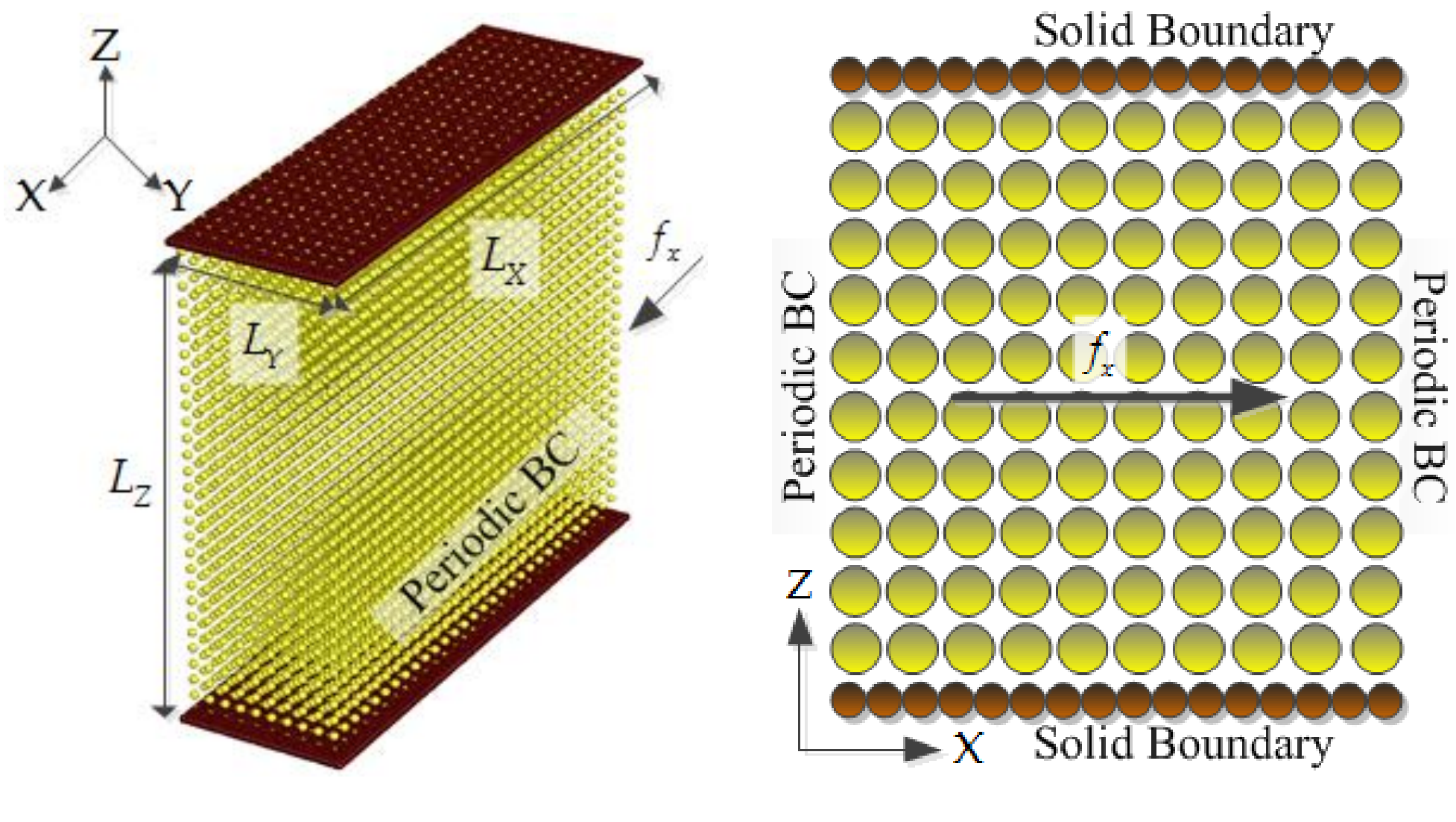

(a) 3D view (b) X-Z plane 2D view.

**Figure 13. SDPD-DV simulations of transient, body-force driven, planar Poiseuille flow. Physical domain showing the BPs and FPs.**

**Table 6: Input parameters used in SDPD-DV simulations of transient planar Poiseuille flow, transient planar Couette flow, and flow over a cylinder.**

| Input Parameters | Case | | |
|---|---|---|---|
| | Transient Poiseuille | Transient Couette | Flow Over Cylinder |
| $L_X$ (m) | $10^{-3}$ | $10^{-3}$ | 0.1 |
| $L_Y$ (m) | $3\times10^{-4}$ | $3\times10^{-4}$ | 0.015 |
| $L_Z$ (m) | $10^{-3}$ | $10^{-3}$ | 0.1 |
| $R$ (m) | N/A | N/A | 0.02 |
| $f_x$ (ms$^{-2}$) | $10^{-4}$ | N/A | $1.5\times10^{-7}$ |
| $M_F$ (kg) | $3\times10^{-7}$ | $3\times10^{-7}$ | 0.1312 |
| $T$ (K) | 300 | 300 | 300 |
| $\rho_a\left(\mathrm{kg\cdot m^{-3}}\right)$ | 1,000 | 1,000 | 1,000 |
| $\eta\ \left(\mathrm{kg\cdot m^{-1}s^{-1}}\right)$ | $10^{-3}$ | $10^{-3}$ | $10^{-3}$ |
| $\zeta\left(\mathrm{kg\cdot m^{-1}s^{-1}}\right)$ | 0 | 0 | 0 |
| $c_V\left(\mathrm{J\cdot kg^{-1}K^{-1}}\right)$ | 4,140 | 4140 | 4140 |
| $\kappa\ \left(\mathrm{W\cdot m^{-1}K^{-1}}\right)$ | 0.58 | 0.58 | 0.58 |
| $V_{xw}$ (ms$^{-1}$) | N/A | $1.25\times10^{-5}$ | N/A |
| $N_\mathrm{F}$ | 9,000 | 9,000 | 17,632 |
| $N_\mathrm{B}$ | 8,282 | 8,282 | 7,440 |
| $N_C$ | 300 | 300 | 1323 |
| $h$ (m) | $8.5\times10^{-5}$ | $8.5\times10^{-5}$ | $4\times10^{-3}$ |
| $c$(ms$^{-2}$) | $1.25\times10^{-4}$ | $1.25\times10^{-4}$ | $1.25\times10^{-4}$ |
| $\Delta t$ (s) | $10^{-4}$ | $10^{-4}$ | $10^{-2}$ |
| $l_{\min}$ (m) | $1\times10^{-6}$ | $1\times10^{-6}$ | $1.9\times10^{-4}$ |
| $l_d$ (m) | $10^{-4}$ | $10^{-4}$ | $5\times10^{-3}$ |
| $M$ | 50 | 50 | 100 |

The physical domain has $L_X = 10^{-3}$ m, $L_Y = 3\times10^{-4}$ m, $L_Z = 10^{-3}$ m. The total mass of $M_F = 3\times10^{-7}$ kg in the domain is represented by 9,000 fluid particles with a constant temperature of $T_i = 300$ K. The solid walls are represented by 8,282 boundary particles as shown in Figure 13. Periodic boundary conditions are imposed along the $x$-axis and $y$-axis.

Closure for pressure and density is obtained by using the artificial compressibility relation Eq. (2.52). Integration is carried out following Sec. 2.2.7 and a constant body force of $f = 10^{-4}$ m/s$^2$ is imposed on each fluid particle. The DVPA method (Sec 2.2.4(b) and 2.2.6(b)) is applied for density and force evaluation. Over $M = 50$ simulations were performed to generate instantaneous samples used to derive the instantaneous particle properties. Fluid properties were then sampled on a structured grid with edge length $l_d = 10^{-4}$ m. Input parameters in the SDPD-DV are listed in Table 6 (Gatsonis et al., 2013).

### 3.1.2 Results and Discussion

The classic Poiseuille flow is due to an applied pressure gradient which can be replaced by a body force per unit mass $f$ (ms$^{-2}$) parallel to the x-axis. The flow starts from rest and develops a velocity profile given by Morris et al. (1997) and Papanastasiou et al. (1999)

$$V_x(z,t) = \frac{f\rho}{2\eta} z(z - L_Z) + \sum_{n=0}^{\infty} \frac{4fL_Z{}^2\rho}{\eta\pi^3(2n+1)^3} \sin\left(\frac{\pi z}{L_Z}(2n+1)\right) \exp\left(-\frac{(2n+1)^2\pi^2\eta}{L_Z{}^2\rho} t\right) \quad (3.1)$$

The sample-averaged field density $\rho(\mathbf{r}_d, t)$ and velocity $v(\mathbf{r}_d, t)$ profiles are plotted in Figure 14 along with the standard deviation. The SDPD-DV density $\rho(Z_d)$ shown in Figure 14(a) has an error $|\rho_a - \rho|/\rho_a$ of less than 4% in the interior and less than 5% near the wall boundary. The field density averaged over the entire channel, $\bar{\rho} = 966$ kg/m$^3$ is used Eq. (3.1) to evaluate the analytical velocity profile. Comparisons between the analytical and SDPD-DV velocity profiles across the channel is plotted in Figure 14(b) and show them to be in excellent agreement for all times considered. The density and velocity results demonstrate the ability of our DVPA-based density and force evaluation method as well as integration algorithm implemented in SDPD-DV.

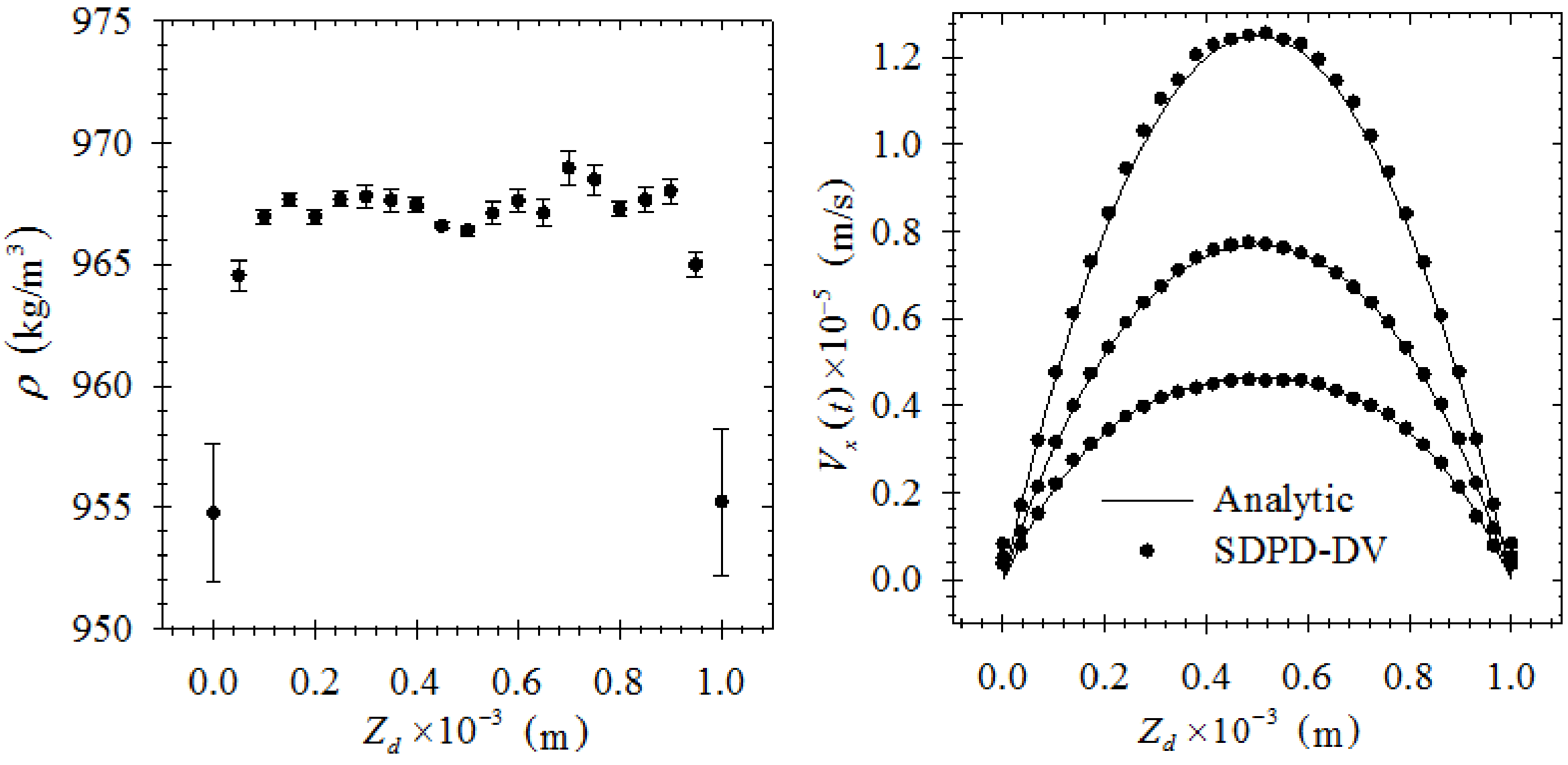

(a) Density distribution with standard deviation. (b) Velocity distribution.

**Figure 14. Sample-averaged fluid density $\rho(Z_d)$ and velocity $Vx(Z_d,t)$ from SDPD-DV simulations of transient, body-force driven, planar Poiseuille flow. The domain has $L_X$ =$10^{-3}$m, $L_Y$ =$3\times10^{-4}$m, $L_Z$ =$10^{-3}$m, the fluid is $H_2O$(l) with $\rho_a$ =1,000 kg/m$^3$, $T_i$=300 K, $\eta$ =$10^{-3}$ kg·m$^{-1}$s$^{-1}$, and $f_x$=$10^{-4}$ m/s$^2$. The analytical profiles are plotted for verification (Morris et al. 1997).**

To further investigate the DVPA method and its ability to enforce the no-slip boundary conditions we evaluate the boundary forces due to the virtual particles using Eq. (2.60) and Eq. (2.62). The distribution of the virtual particle force components are shown in Figure 15. As shown in Figure 15(a) and Figure 15(b), the normal components of conservative and dissipative VP forces are decreasing as the distance from solid wall increases. The negative and positive sign of the conservative normal virtual particle forces shows that they are perpendicular to the plates and pointing to the inner domain. Figure 15(c) and Figure 15(d) shows the tangential component of the conservative and dissipative virtual particle forces. Both contribute to the imposition of the non-slip condition. Our results show that the forces due to virtual particles in SDPD-DV have the same qualitative behavior as the DPD boundary forces [Fig. 4, Altenhoff et al., 2007] and to the average wall forces [Fig. 4, Fedosov et al., 2008].

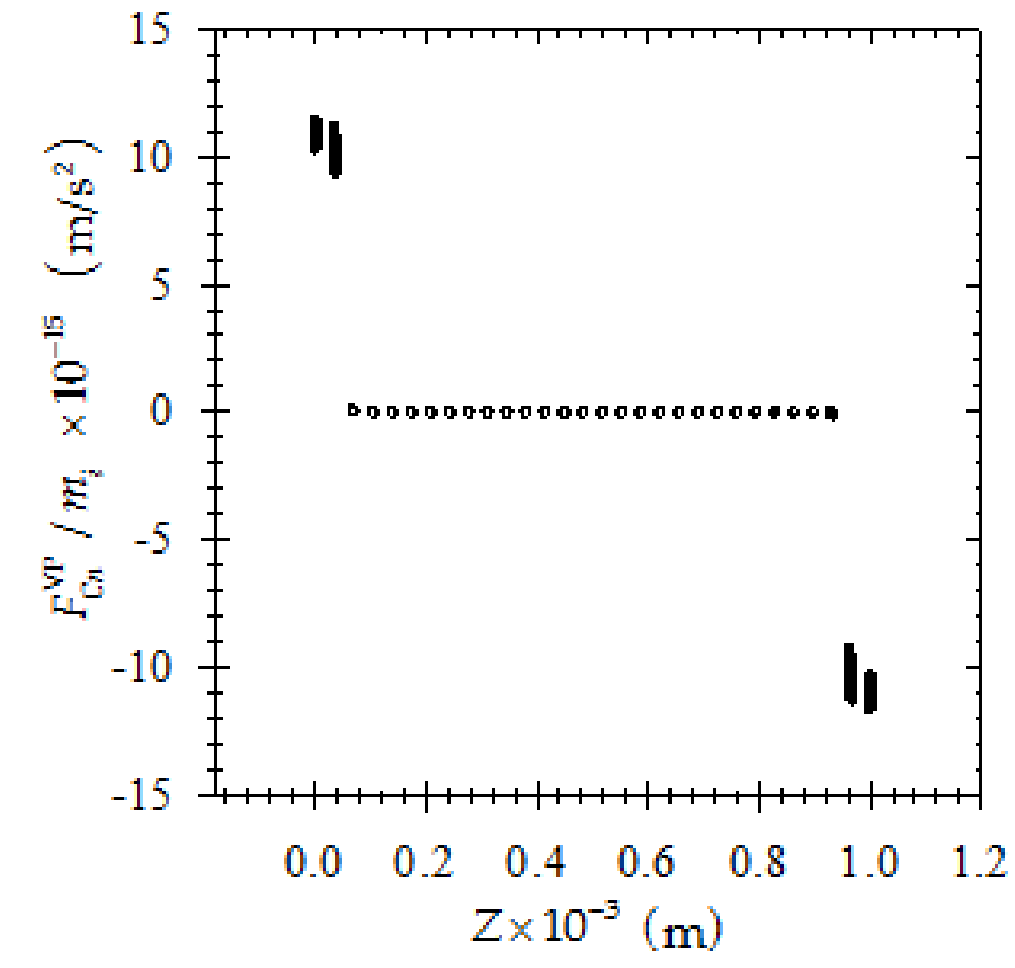

(a) Conservative normal component.

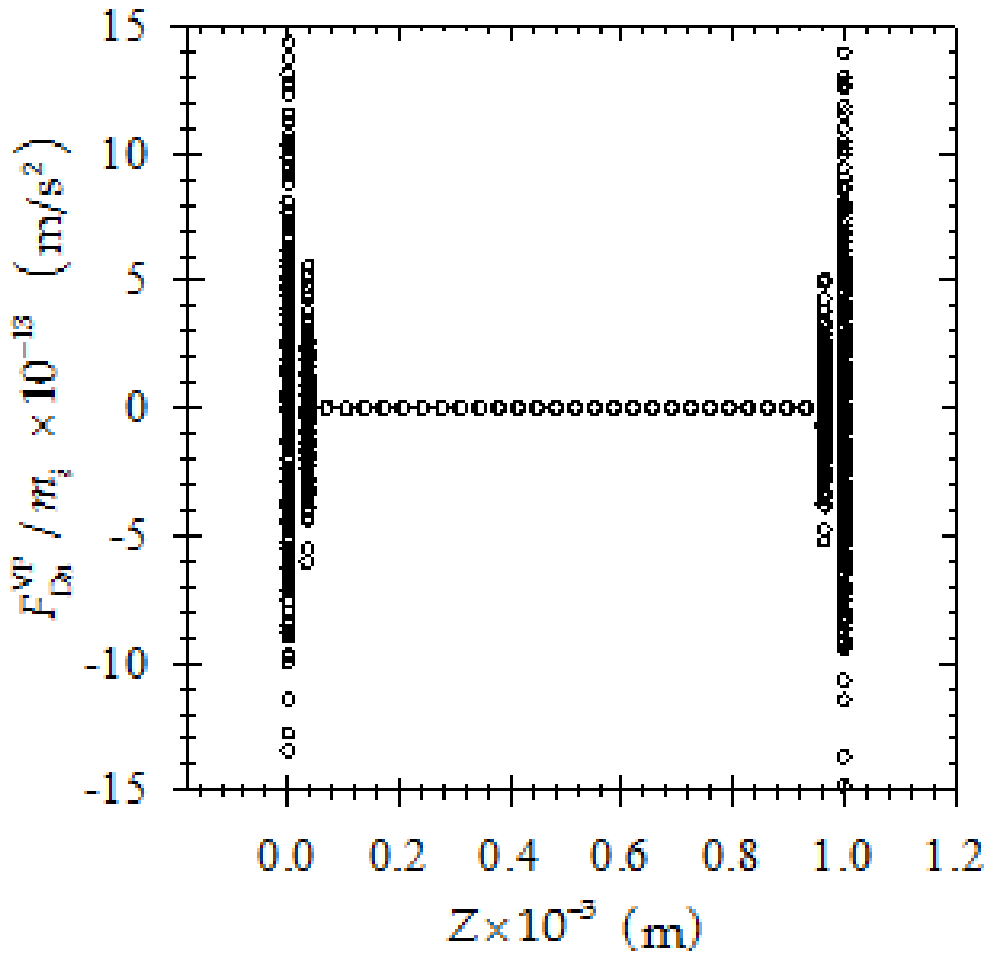

(b) Dissipative normal component.

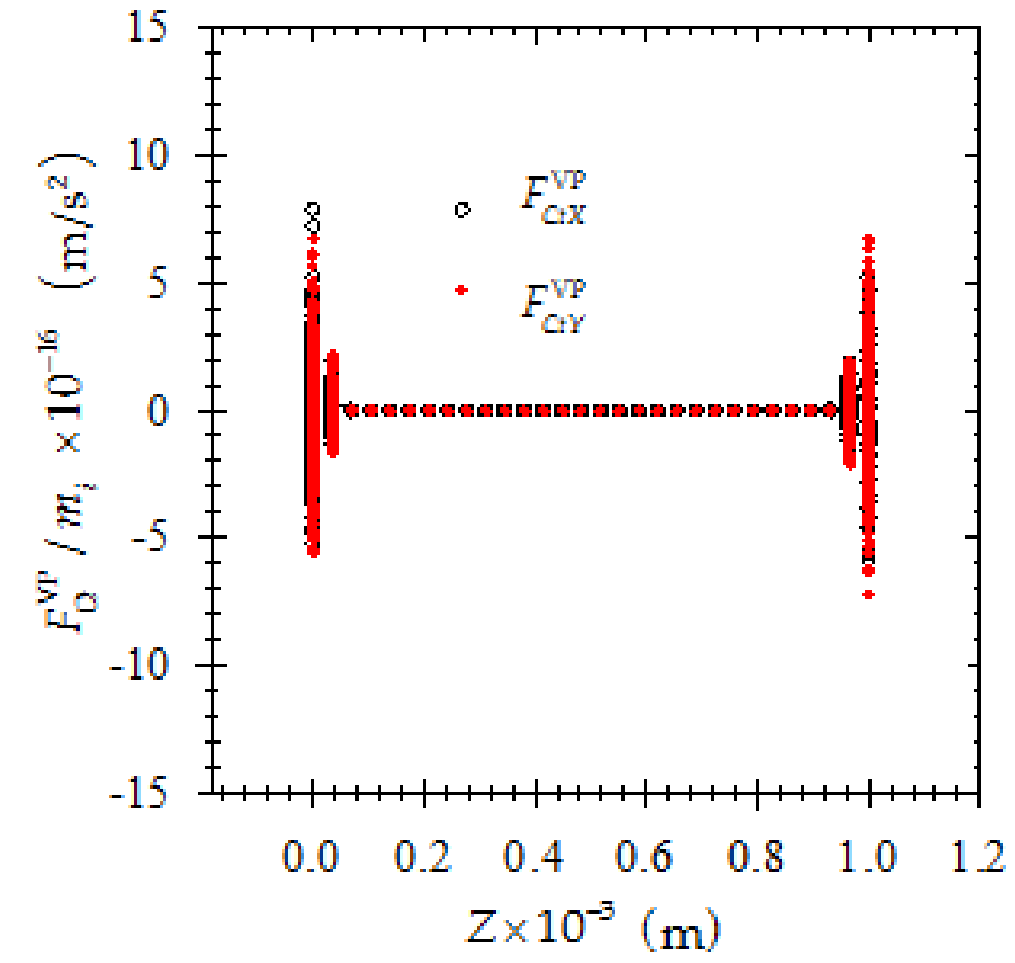

(c) Conservative tangential component.

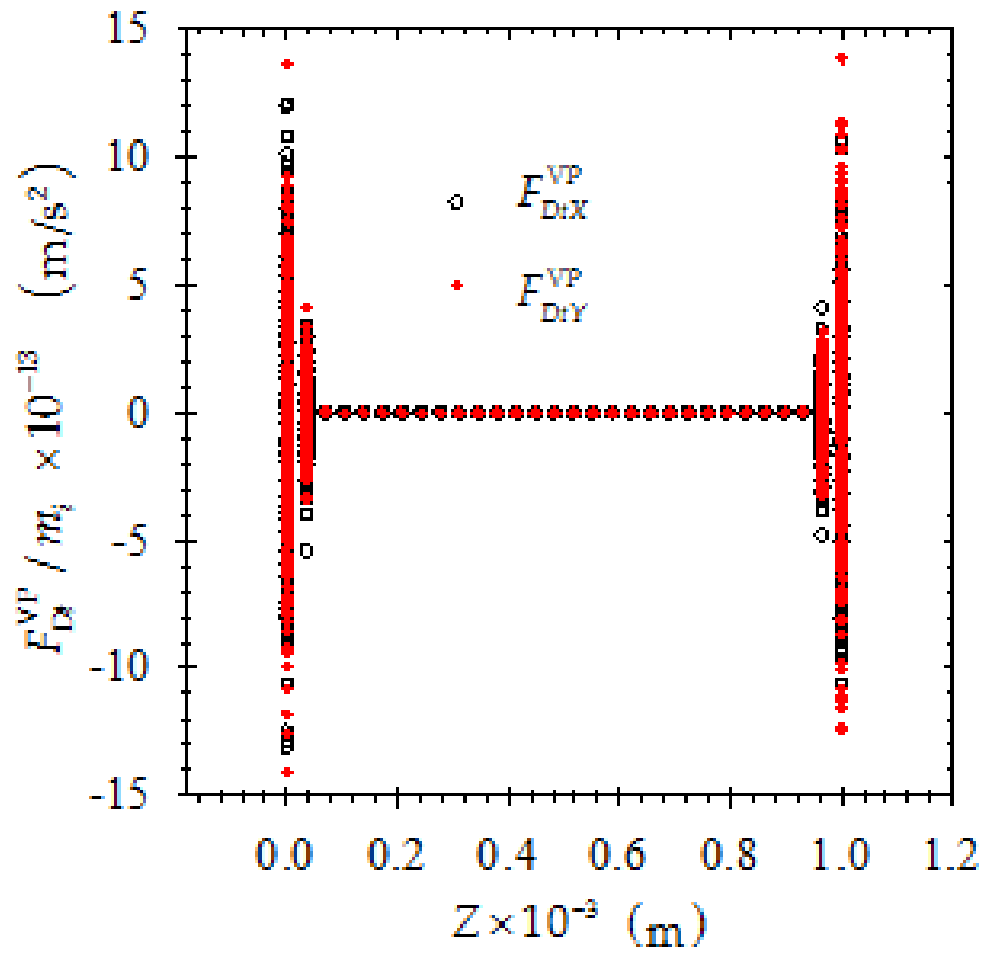

(d) Dissipative tangential component.

**Figure 15. Forces due to virtual particles from SDPD-DV simulations of transient, body-force driven, planar Poiseuille flow. The domain has $L_X$ =$10^{-3}$m, $L_Y$ =$3\times10^{-4}$m, $L_Z$ =$10^{-3}$m, the fluid is $H_2O$(l) with $\rho_a$ =1,000 kg/m$^3$, $\eta$ =$10^{-3}$ kg·m$^{-1}$s$^{-1}$, and $f_x$=$10^{-4}$ m/s$^2$.**

## 3.2 Transient Planar Couette Flow

The test involves an incompressible flow across two infinite parallel walls with the top wall moving with a constant velocity as shown in Figure 16. The velocity profiles as predicted by SDPD-DV are compared to theoretical formulations. The SDPD-DV density is plotted along with the standard deviation. This benchmark test is used as further verification of the SDPD-DV and its ability to enforce no-slip and no penetration on a moving solid boundary.

### 3.2.1 Input Conditions and Computational Parameters

The test involves an incompressible flow with density $\rho$ across two infinite parallel walls as depicted in Figure 13 with the top wall moving with a constant velocity $V_{xw}$.The SDPD-DV simulation considers $H_2O$(l) with $\rho_a = 1{,}000\ \text{kgm}^{-3}$ and $\eta = 10^{-3}\ \text{kg}\cdot\text{m}^{-1}\text{s}^{-1}$. The physical domain has $L_X = 10^{-3}\ \text{m}$, $L_Y = 3\times10^{-4}\ \text{m}$, and $L_Z = 10^{-3}\ \text{m}$. The total mass of $M_F = 3\times10^{-7}\ \text{kg}$ in the domain is represented by 9,000 fluid particles with a constant temperature of $T_i = 300\ \text{K}$. The solid walls are represented by 8,282 boundary particles as shown in Figure 16 each with $v_{xw} = V_{xw} = 1.25\times10^{-5}\ \text{ms}^{-1}$. Periodic boundary conditions are imposed along the $x$-axis and $y$-axis. Closure for pressure and density is obtained by using the artificial compressibility relation Eq. (2.52). Input parameters are listed in Table 6.

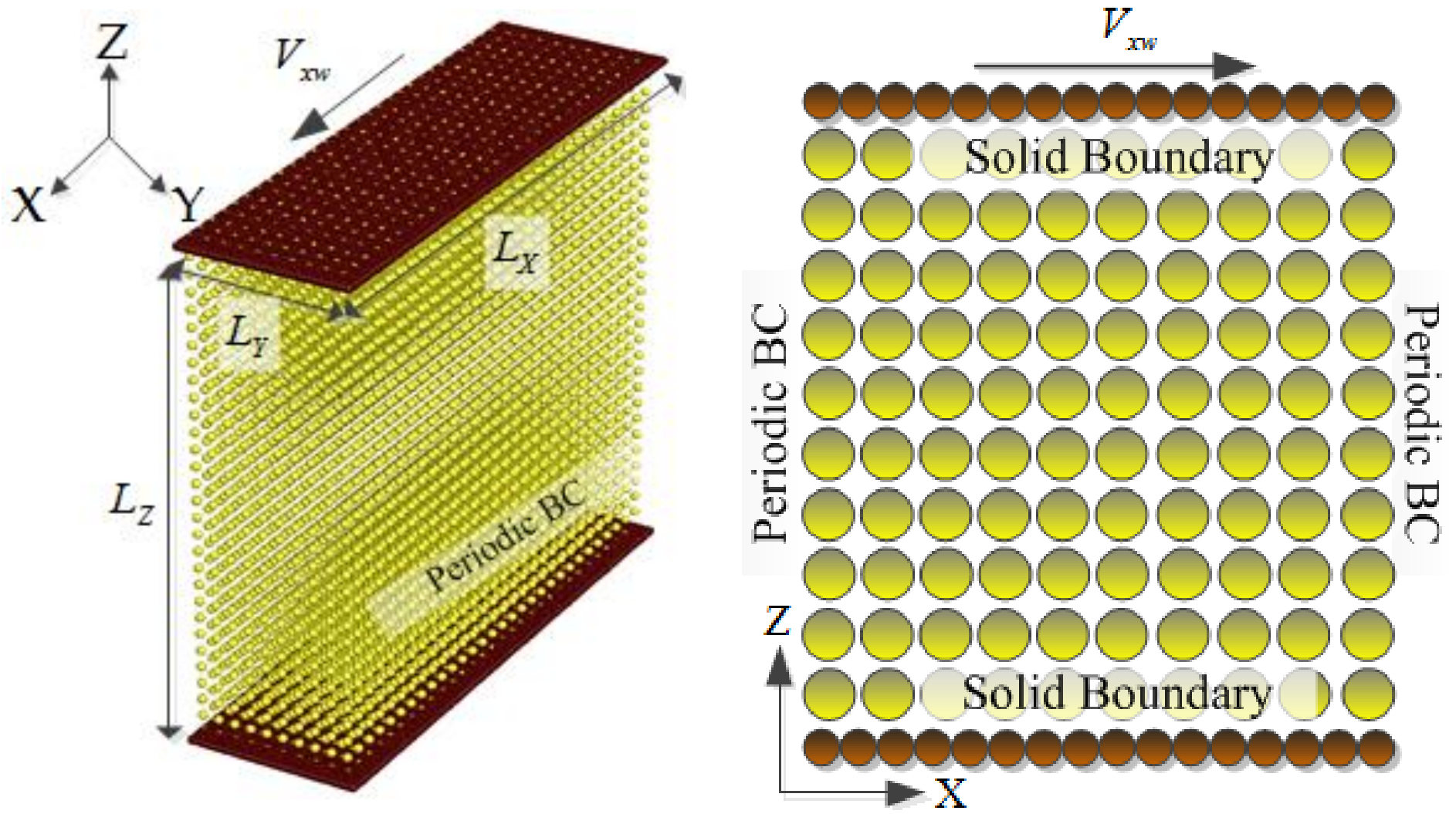

(a) 3D view (b) X-Z plane 2D view.

**Figure 16. SDPD-DV simulations of transient planar Couette flow. Physical domain showing the BPs and FPs.**

### 3.2.2 Results and Discussion

The flow starts from rest and develops a velocity profile given by Morris et al. (1997) and Papanastasiou et al. (1999)

$$V_x(z,t) = \frac{V_{xw}}{L_Z} z + \sum_{n=1}^{\infty} \frac{2V_{xw}}{n\pi} (-1)^n \sin\left(\frac{n\pi}{L_Z} z\right) \exp\left(-\frac{n^2 \pi^2 \eta}{{L_Z}^2 \rho} t\right) \tag{3.2}$$

The sample-averaged field density $\rho(\mathbf{r}_d, t)$ and velocity $\mathbf{V}(\mathbf{r}_d, t)$ profiles are plotted in Figure 17(a),(b) along with the standard deviation. The field density shown in Figure 17(a) has an error $|\rho_a - \rho|/\rho_a$ of less than 4% in the interior and less than 5% near the wall boundary. The analytical velocity profile from Eq. (3.2) using the average density $\bar{\rho} = 966\ \text{kg/m}^3$ is plotted in Figure 17(b). A comparison between the analytical and SDPD-DV velocity profiles across the channel is plotted in Figure 17(b). The numerical results from SDPD-DV are in very good agreement with the analytical solution for all times considered.

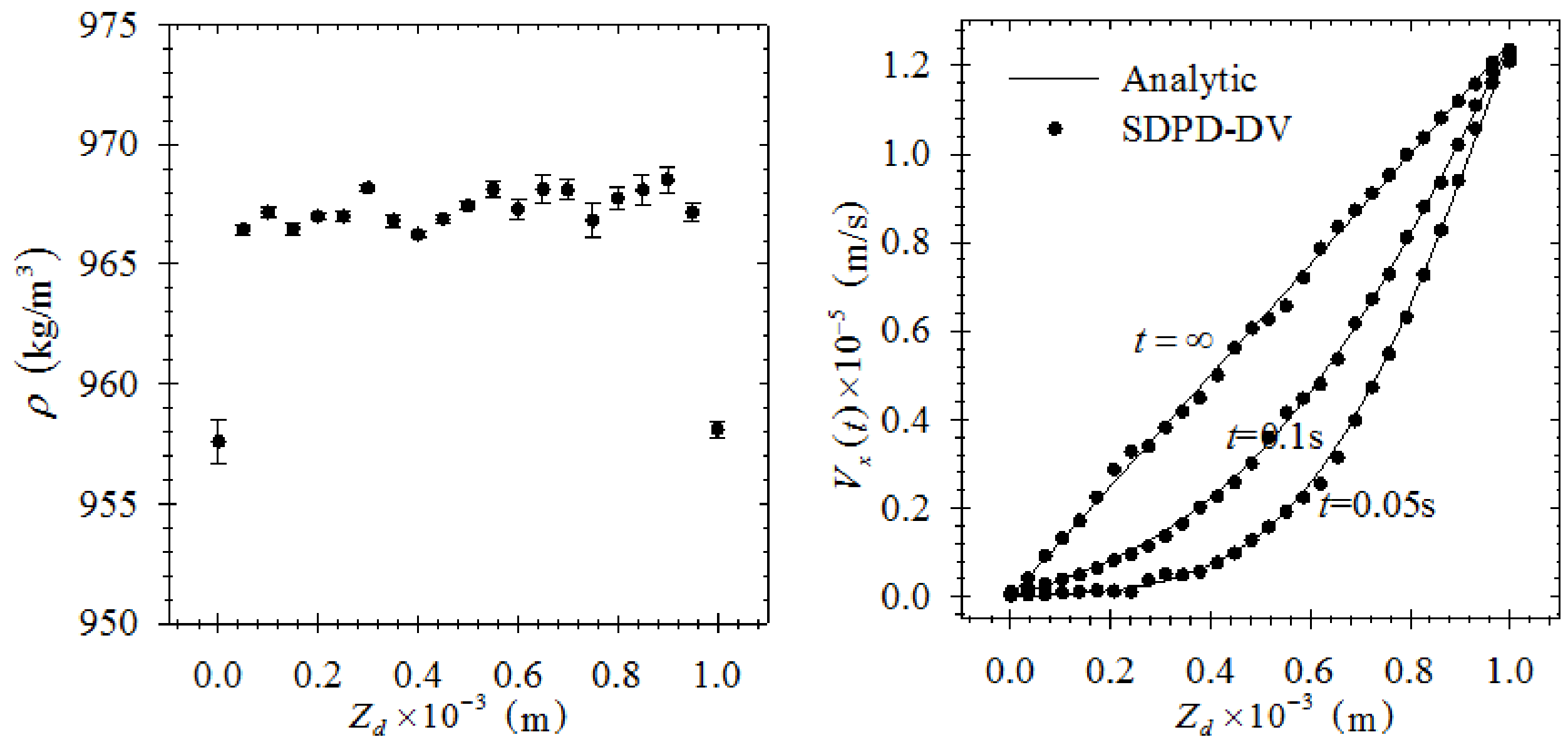

(a) Density distribution with standard devian.　　　(b) Velocity distribution.

**Figure 17. Sample-averaged fluid density $\rho(Z_d)$and velocity $V_x(Z_d,t)$ from SDPD-DV simulations of transient, planar, Couette flow. The domain has $L_X$ =$10^{-3}$m, $L_Y$ =$3\times10^{-4}$m, $L_Z$ =$10^{-3}$m, the fluid is $H_2O$(l) with $\rho_a$ =1000 kg/m$^3$, $T_i$=300 K, $\eta$ =$10^{-3}$ kg·m$^{-1}$s$^{-1}$, and $V_{xw}$=1.25x$10^{-5}$ m/s. The analytical profiles are plotted for verification (Morris, Fox and Zhu, 1997).**

### 3.3 Steady Low-Re Incompressible Flow over a Cylinder

The third benchmark test involves the SDPD-DV simulation of flow over a cylinder. This test appeared in several SPH (Morris et al., 1997; Vazquez-Quesada and Ellero, 2012) and DPD (Keaveny et al., 2005; Kim and Phillips, 2004) simulations. The sample-averaged, steady velocity and pressure fields from SDPD-DV and comparison with FLUENT are shown. The SDPD-DV flow field exhibits the anticipated behavior for the low-Reynolds number flow over a cylinder (Keaveny et al., 2005; Morris et al., 1997; Vazquez-Quesada and Ellero, 2012; Kim and Phillips, 2004). For further direct quantitative comparison the SDPD-DV and FLUENT, velocity and pressure fields are superimposed and plotted $(x, y = 7.5\times10^{-3}\text{m}, z)$. Additional verification is obtained by comparing velocity and pressure field along contours. This test further verifies the ability of SDPD-DV to simulate curved solid boundaries and the implementation of the artificial compressibility method in SDPD-DV.

### 3.3.1 Input Conditions and Computational Parameters

The incompressible fluid of density $\rho$ is driven by a body-force $f_x$ over a cylinder of radius $R$. Periodic boundary conditions are assumed in the x, y and z direction as shown in Figure 18(a). The periodic boundary conditions can be realized as a flow in an infinite array of cylinders as depicted in Figure 18(b). For verification of the SDPD-DV results we also performed simulations using FLUENT over a domain shown in Figure 18(b).

The SDPD-DV simulation considers $H_2O$(l) with $\rho_a = 1{,}000\ \mathrm{kgm^{-3}}$ and $\eta = 10^{-3}\ \mathrm{kg \cdot m^{-1} s^{-1}}$ driven by a body force $f_x = 1.5 \times 10^{-7}\ \mathrm{ms^{-2}}$. The physical domain has $L_X = 0.1\ \mathrm{m}$, $L_Y = 0.015\ \mathrm{m}$, $L_Z = 0.1\ \mathrm{m}$ and the cylinder radius $R = 0.02\ \mathrm{m}$. The total fluid mass of $M_F = 0.1312\ \mathrm{kg}$ in the domain is represented by 17,632 fluid particles with a constant temperature of $T_i = 300\ \mathrm{K}$. The solid walls of the cylinder are represented by 7,440 boundary particles as shown in Figure 18(a). The initial fluid particle distribution is shown in Figure 18(a). The Reynolds number $\mathrm{Re} = R\rho V_0/\eta$ based on the cylinder radius and $V_0$ in the average velocity field is $\mathrm{Re} \sim 1$. Closure is obtained by using the artificial compressibility relation Eq. (2.52). Using Eq. (2.68) integration is carried out with $\Delta t = 1 \times 10^{-2}\ \mathrm{s}$ . Steady state was reached and $M = 100$ independent samples were used to evaluate fluid properties on a grid with $l_d = 5 \times 10^{-3}\ \mathrm{m}$. The input parameters for the SDPD-DV simulation are listed in Table 6.

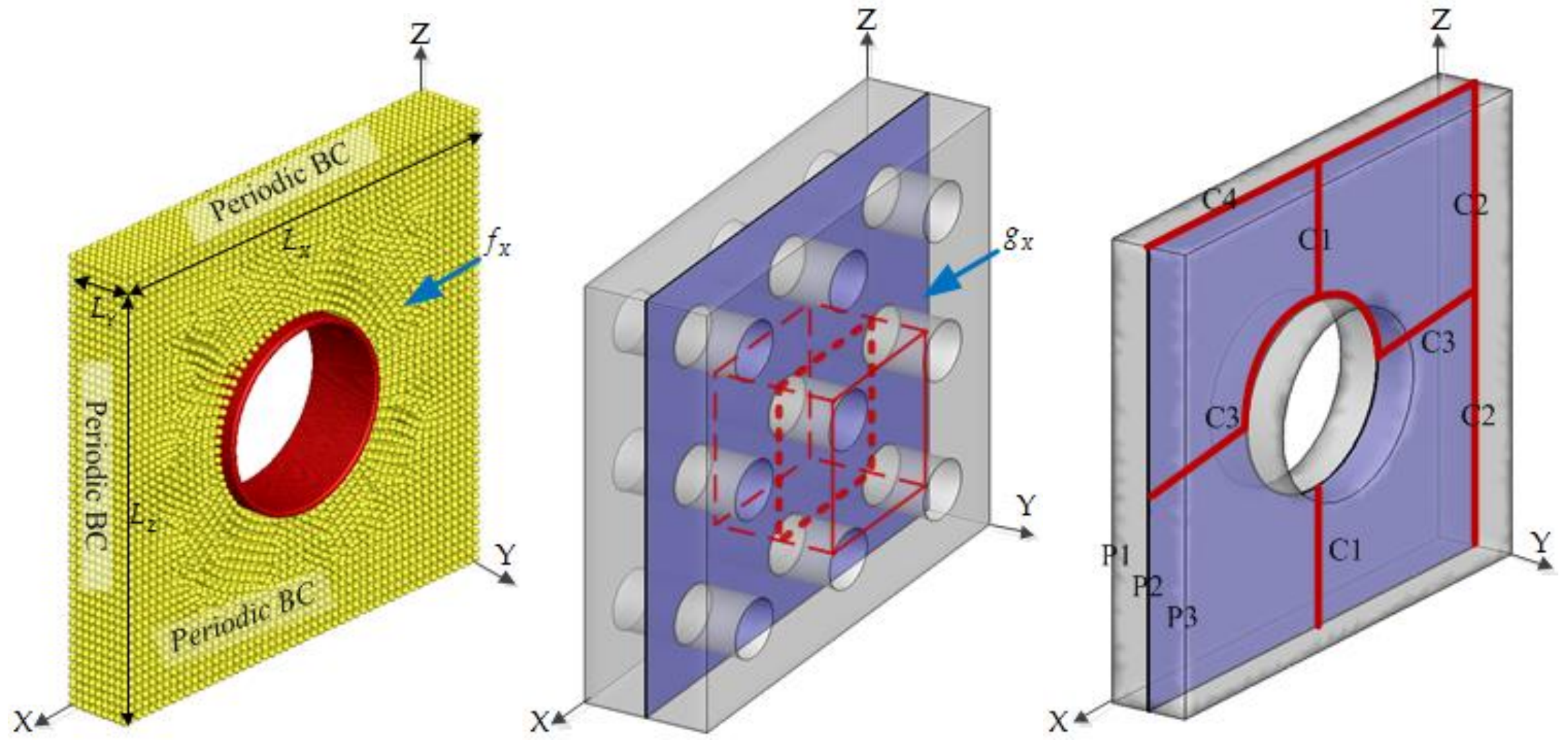

(a) (b) (c)

**Figure 18. Computational domains used in SDPD-DV and FLUENT simulations of steady, low-Reynolds number water flow with $\rho_a$=1,000 kg/m$^3$, $T_i$ = 300 K, $\eta$ =10$^{-3}$ kg·m$^{-1}$s$^{-1}$ . The cylinder has $R$=0.02m and the body force per unit mass is $f_x$=$g_x$=1.5x10$^{-7}$ m/s$^2$. (a) SDPD-DV domain with $L_X$=0.1 m, $L_Y$=0.015 m, $L_Z$=0.1 m showing the BPs on the cylinder and FPs in the domain. (b) FLUENT domain $L_X$=0.3 m, $L_Y$=0.015 m, $L_Z$=0.3 m includes an array of cylinders to simulate the periodic flow. The insert shows the extend of the SDPD-DV physical domain. (c) Planes P1,P2,P3 and contours C1,C2,C3,C4 used for comparison of SDPD-DV and FLUENT results.**

In order to achieve periodic boundary conditions in ANSYS FLUENT, we placed 9 cylinders in a lattice as shown in Figure 18(b). The cylinders were placed apart so that there is no influence and simulation was performed with 246,132 cells. The physical domain for the FLUENT simulation has $L_X = 0.1$ m, $L_Y = 0.015$ m, $L_Z = 0.1$ m. The gravitational acceleration per unit mass $g_x$ in ANSYS FLUENT was equal to the SDPD-DV body force $f_x$. For direct comparison we sampled field properties on planes P1 ($y$=0 m), P2 ($y$=7.5×10$^{-3}$ m), P3 ($y$=0.015 m) and along contours C1, C2, C3 and C4 shown in Figure 18(c).

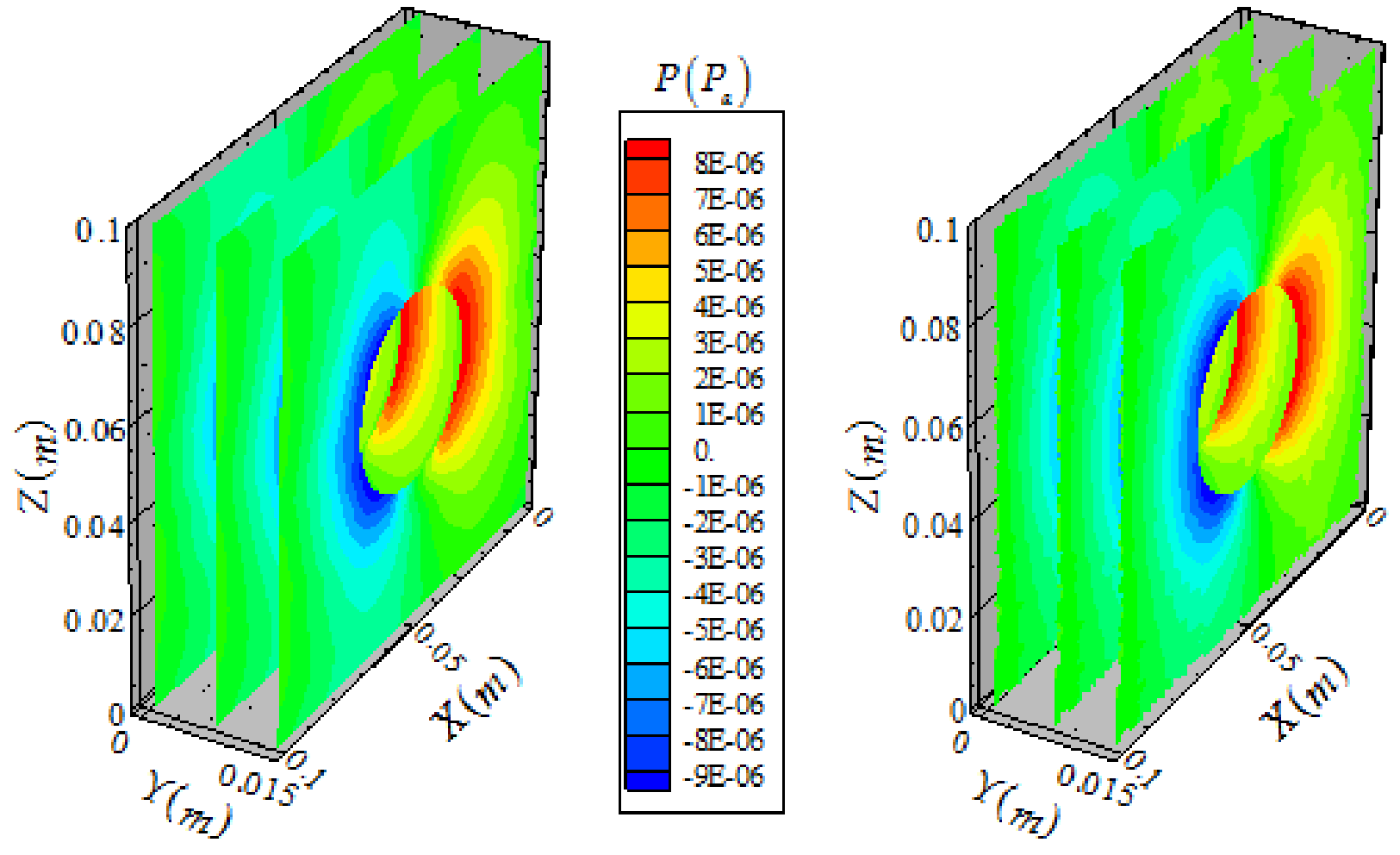

(a) SDPD-DV (b) FLUENT

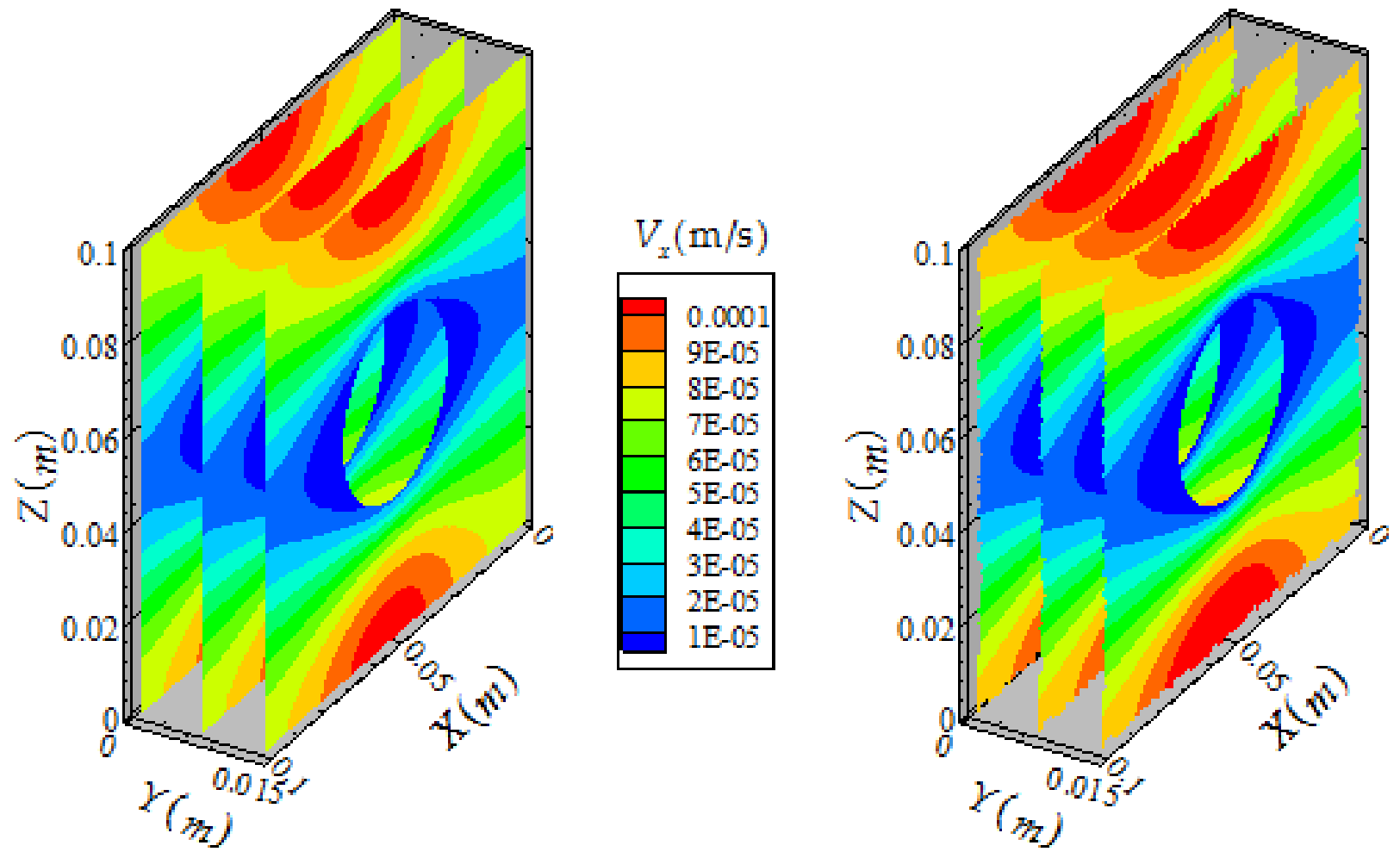

(a) SDPD-DV (b) FLUENT

**Figure 19. Sample-averaged $V_x$ (r) and $P$(r) on $y$ = 0 m, $y$ = 7.5×10$^{-3}$ m, $y$ = 1.5×10$^{-2}$ m planes from SDPD-DV simulations. $V_x(x,z)$ and $P$(r) from FLUENT simulation. Steady, low-Reynolds number water flow with $\rho_a$=1,000 kg/m$^3$, $T_i$ = 300 K, $\eta$ =10$^{-3}$ kg·m$^{-1}$s$^{-1}$ . The cylinder has $R$=0.02m and the body force per unit mass is $f_x$=$g_x$=1.5x10$^{-7}$ m/s$^2$.**

### 3.3.2 Results and Discussion

The overall flow field characteristics are shown in Figure 19. The sample-averaged, steady $V_x(\mathbf{r}_d)$ and $P(\mathbf{r}_d)$ pressure fields from SPDP-BV and $V_x(\mathbf{r})$ and $P(\mathbf{r})$ from FLUENT are shown on planes P1, P2, P3 parallel to $x$ axis. The SDPD-DV flow field exhibits the anticipated behavior for the low-Reynolds flow over a cylinder (Keaveny et al., 2005; Morris et al., 1997; Vazquez-Quesada and Ellero, 2012; Kim and Phillips, 2004). The SDPD-DV properties are qualitatively and quantitatively similar to those obtained from FLUENT. For further direct quantitative comparison the SDPD-DV and FLUENT velocity and pressure fields are superimposed and plotted in Figure 20(a)-(b) on the plane $(x, y = 7.5\times10^{-3}\text{m}, z)$. The non-smooth character of the SDPD-DV pressure contours are due to fluctuations introduced by the artificial compressibility model applied to this incompressible flow.

It is important to note also that velocities and pressures considered are very small and susceptible to numerical perturbations. Additional verification is obtained by comparing velocity and pressure field along contours C1, C2, C3, and C4 shown in Figure 20(c)-(d). The pressure comparison along the centerline C3 in Figure 20(d) shows the ability of our solid boundary density and force evaluation to accurately predict pressure in the ram and wake side of the cylinder where fluid particles are impinging the curved boundary. The overall very good agreement demonstrates the ability of our SDPD-DV implementation to simulate curved solid boundaries.

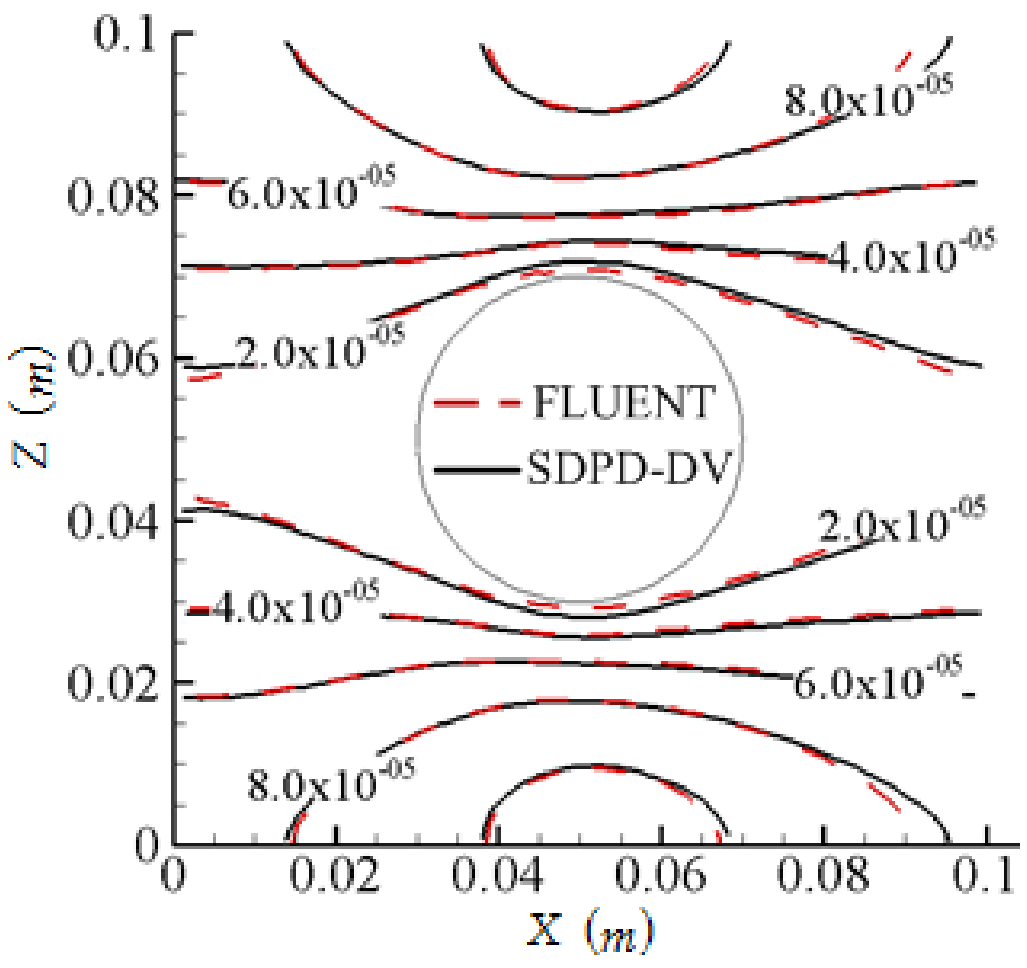

(a) $V_X(\mathbf{r})$ on $y$=7.5×10$^{-3}$ m plane.

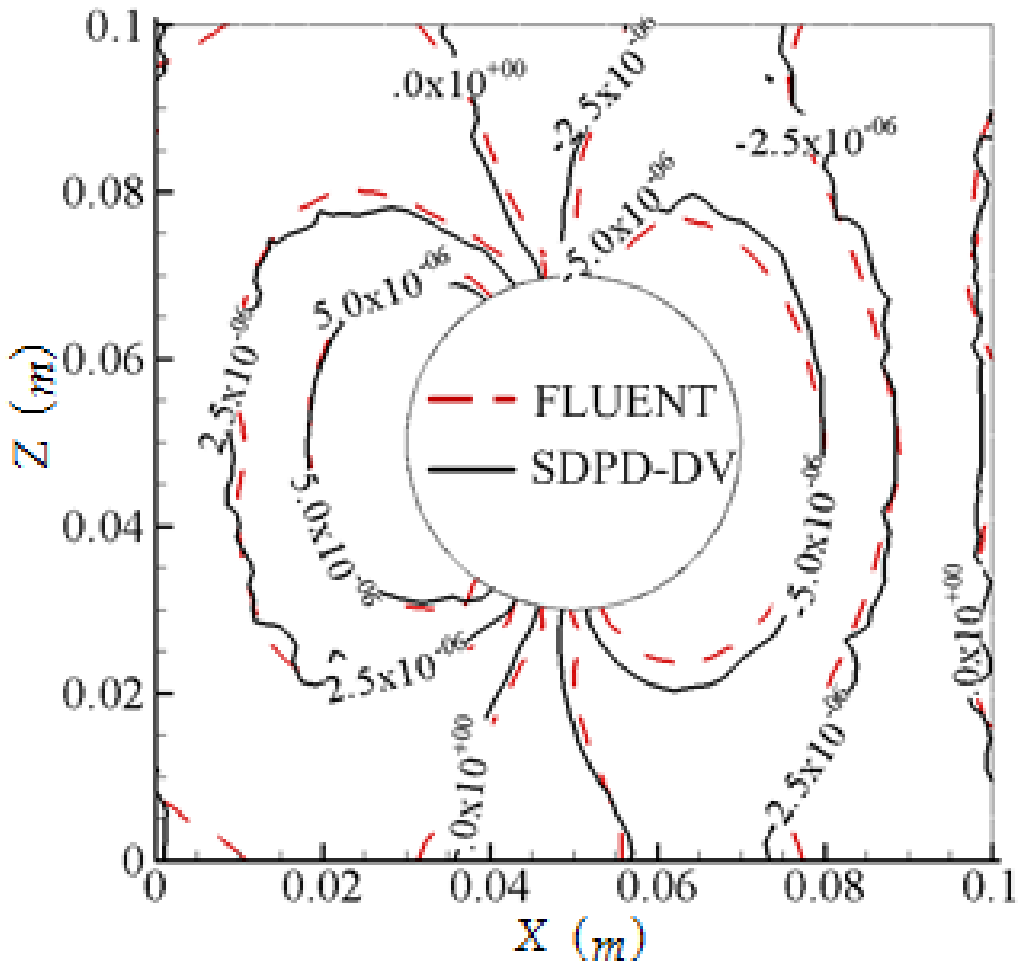

(b) $P(\mathbf{r})$ on $y$=7.5×10$^{-3}$ m plane.

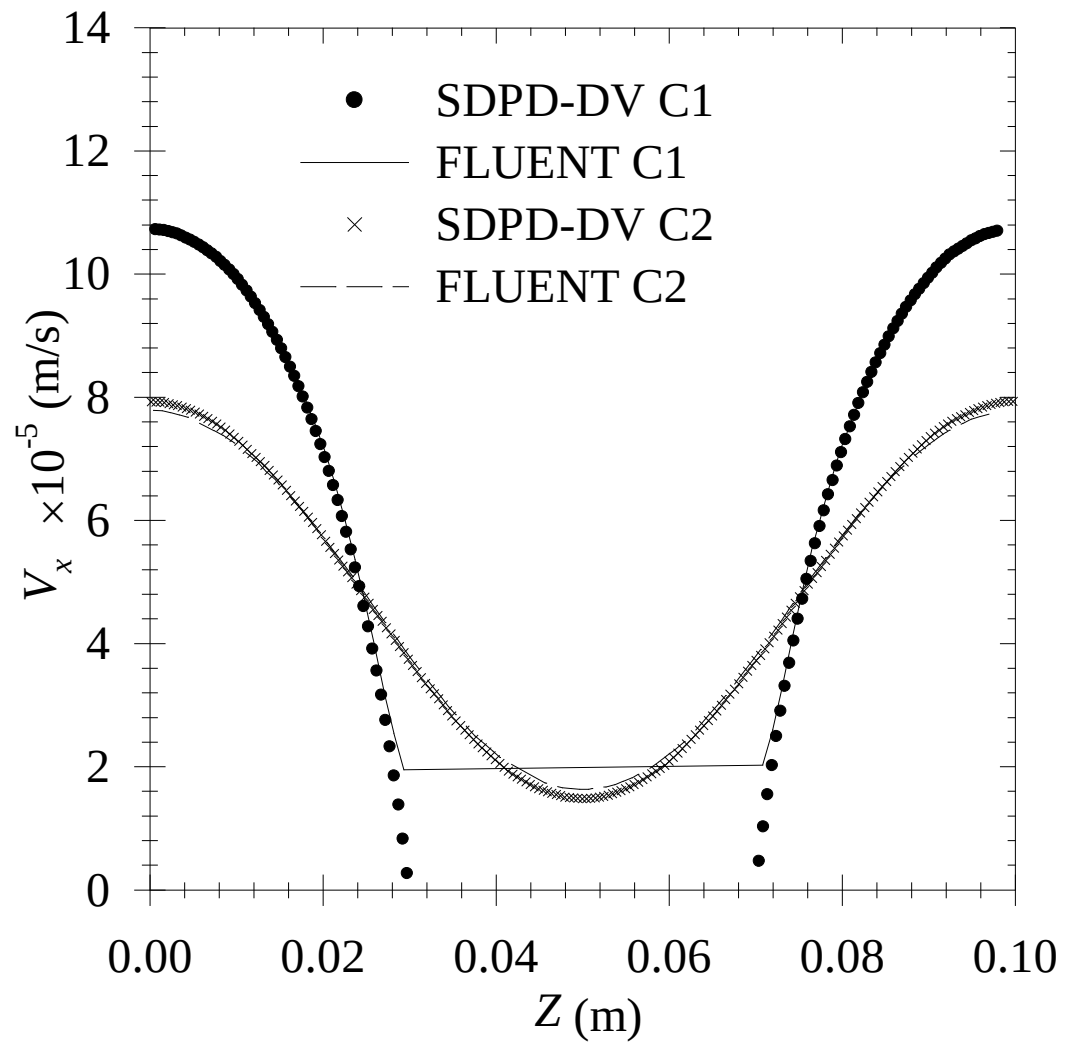

(c) $V_X(\mathbf{r})$ along C1 and C2 .

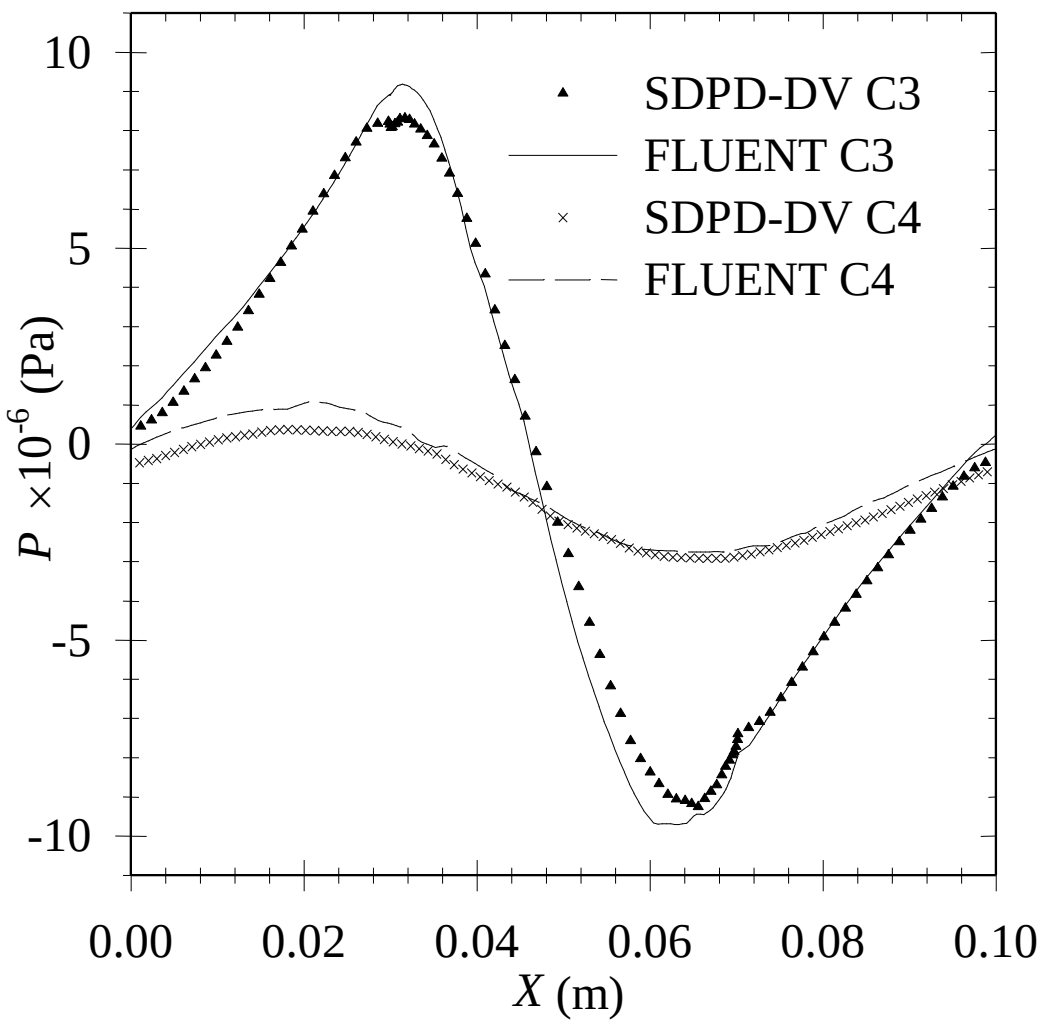

(d) $P(\mathbf{r})$ along C3 and C4.

**Figure 20. Sample-averaged $V_x$ (r) and $P$(r) on $y$ = 7.5×10$^{-3}$ m plane and along C1,C2,C3,C4 from SDPD-DV simulations. $V_x(x,z)$ and $P$(r) from FLUENT simulation. Steady, low-Reynolds number water flow with $\rho_a$=1,000 kg/m$^3$, $T_i$ = 300 K, $\eta$ =10$^{-3}$ kg·m$^{-1}$s$^{-1}$ . The cylinder has $R$=0.02m and the body force per unit mass is $f_x$=$g_x$=1.5x10$^{-7}$ m/s$^2$.**

### 3.4 Equilibrium State and Self-Diffusion Coefficient

The last series of tests also serve for validation and verification since the SDPD-DV results for the self-diffusion coefficient are compared with data and analytical formulas. The tests also provide the opportunity to examine the scale-dependence of our SDPD-DV implementation and compare our results with those of Litvinov et al. (2009) and Vazquez-Quesada et al. (2009). These tests are intended as a demonstration of our SDPD-DV to simulate mesoscale flows at equilibrium states.

#### 3.4.1 Input Conditions and Computational Parameters

We performed SDPD-DV simulations of $H_2O$(l) with $M_F$ in the range $1.25\times10^{-22} \sim 1\times10^{-15}$ kg, $\rho_a = 1{,}000$ kg/m$^3$, $\eta = 10^{-3}$ kg$\cdot$m$^{-1}$s$^{-1}$ and $T_i = 300$ K. We performed also SDPD-DV simulations of $N_2$(g) with $M_F$ in the range $1\times10^{-18} \sim 1\times10^{-15}$ kg with $\rho_a = 1.184$ kg/m$^3$, $\eta = 10^{-5}$ kg$\cdot$m$^{-1}$s$^{-1}$ and $T_i = 300$ K. In order to examine fluid particle scale effects we follow Vazquez-Quesada et al. (2009) and assume that the "size" of the fluid particle is given in terms of the SDPD variables as,

$$D_i \sim (V_i)^{1/3} = (m_i / \rho_i)^{1/3} \tag{3.3}$$

For comparison, we need also length scales for the real liquid and gas molecules. For a liquid with viscosity $\eta_A$, temperature $T_A$ and assuming that the fluid occupies a cubic lattice where the molecules simply touch each other, the appropriate scale is given by Bird et al. (2007)

$$2R_A = (\tilde{V}_A / \tilde{N}_A)^{1/3} = (m_A / \rho_A)^{1/3} \tag{3.4}$$

where, $\tilde{V}_A$ is the molar volume and $\tilde{N}_A$ is the Avogadro number, $m_A$ is the mass of the molecule A. For a gas a relevant length scale can be considered the molecular diameter $D_g$. The input parameters and some derived variables from the SDPD-DV simulations are listed in Table 7 and Table 8.

**Table 7. Input and derived parameters in SDPD-DV simulations of mesoscale flows of liquid water at equilibrium states in rectangular domains with periodic boundaries.**

| | Inputs, $H_2O$(l), $T = 300$ K | | | | | Derived SDPD-DV | | | | |
|---|---|---|---|---|---|---|---|---|---|---|
| Case | $L_{X,Y,Z}$ ( m) | $\frac{m_i}{m_{H_2O}}$ | $N_{FP}$ | $\Delta t$ (×$10^{-16}$ s) | $h$ (m) | $\bar{\rho}_i$ | $\sqrt{\overline{c_i^2}}$ | $\frac{(\bar{V}_i)^{1/3}}{(2R_{H_20})}$ | $D_{ii}$ (×$m^2s^{-1}$) | $D_{ii}^h$ (×$m^2s^{-1}$) |
| 1 | 5×$10^{-9}$ | 1.24 | 3,375 | 1 | $10^{-9}$ | 1050 | 576 | 1.08 | 1.98×$10^{-9}$ | 1.86×$10^{-9}$ |
| 2 | 8×$10^{-9}$ | 5.07 | 3,375 | 1 | 1.4×$10^{-9}$ | 1040 | 288 | 1.72 | 1.24×$10^{-9}$ | 8.83×$10^{-10}$ |
| 3 | 1×$10^{-8}$ | 9.91 | 3,375 | $10^2$ | 1.9×$10^{-9}$ | 1050 | 202 | 2.15 | 9.88×$10^{-10}$ | 7.55×$10^{-10}$ |
| 4 | 5×$10^{-8}$ | 1240 | 3,375 | $10^3$ | $10^{-8}$ | 1,020 | 17.9 | 10.8 | 1.98×$10^{-10}$ | 1.81×$10^{-10}$ |
| 5 | 1×$10^{-7}$ | 9910 | 3,375 | $10^3$ | 1.9×$10^{-8}$ | 1,050 | 6.5 | 21.5 | 9.88×$10^{-11}$ | 8.41×$10^{-11}$ |
| 6 | 1×$10^{-6}$ | 1.24×$10^6$ | 27,000 | $10^5$ | $10^{-7}$ | 1,015 | 0.578 | 108 | 1.98×$10^{-11}$ | 1.80×$10^{-11}$ |
| 7 | 1×$10^{-6}$ | 2.14×$10^6$ | 15,625 | $10^5$ | 1.1×$10^{-7}$ | 1,015 | 0.44 | 129 | 1.65×$10^{-11}$ | 1.26×$10^{-11}$ |
| 8 | 1×$10^{-6}$ | 4.18×$10^6$ | 8,000 | $10^5$ | 1.4×$10^{-6}$ | 1,039 | 0.314 | 161 | 1.32×$10^{-11}$ | 1.07×$10^{-11}$ |
| 9 | 1×$10^{-6}$ | 9.91×$10^6$ | 3,375 | $10^5$ | 1.9×$10^{-6}$ | 1,034 | 0.204 | 215 | 9.88×$10^{-12}$ | 8.28×$10^{-12}$ |
| 10 | 1×$10^{-6}$ | 3.34×$10^7$ | 1,000 | $10^5$ | 3×$10^{-7}$ | 1,055 | 0.11 | 323 | 6.59×$10^{-12}$ | 6.24×$10^{-12}$ |

**Table 8. Input and derived parameters in SDPD-DV simulations of mesoscale flows of $N_2$(g) at equilibrium states in rectangular domains with periodic boundaries.**

| Inputs $N_2$(g), $\rho_a = 1.184\,\text{kg/m}^3, T = 300$ K | | | | | | Derived SDPD-DV | | |
|---|---|---|---|---|---|---|---|---|
| Case | $L_{X,Y,Z}$ ( m) | $\frac{m_i}{m_{N_2}}$ | $N_{FP}$ | $\Delta t$ ($\times 10^{-15}$ s) | $h$ (m) | $\bar{\rho}_i$ | $\sqrt{\overline{c_i^2}}$ | $\frac{(\bar{V}_i)^{1/3}}{D_{N_2}}$ |
| 1 | $1\times10^{-6}$ | 6,371 | 3,375 | $10^2$ | $2\times10^{-7}$ | 1.21 | 6.2 | 222.22 |
| 2 | $1\times10^{-6}$ | 21,505 | 1,000 | $10^3$ | $3\times10^{-7}$ | 1.20 | 3.36 | 333.33 |
| 3 | $1\times10^{-5}$ | $6.37\times10^6$ | 3,375 | $10^4$ | $1.9\times10^{-6}$ | 1.22 | 0.201 | 2222.22 |

### 3.4.2 Results and Discussion

Figure 21 shows the $(v_{xi} - v_{yi})$ phase plots for the smallest and largest scales considered. The plots show that the equilibrium states sampled have particle velocities that are not scale free. To gain further insight, we use the fluid particle velocities and field mean velocities obtained from the SDPD-DV simulations to evaluate the average translational temperature for the entire system. Following the standard description for the thermal (or random) particle velocity $\mathbf{C}_i(t) = \mathbf{v}_i(t) - \mathbf{V}(\mathbf{r},t)$ (Gombosi, 1994)

$$\bar{T}_t = \frac{m_i}{3k_B} \frac{\sum_{i=1}^{N_F} (v_{xi} - V_x)^2 + (v_{yi} - V_y)^2 + (v_{zi} - V_z)^2}{N_F} = \frac{m_i}{3k_B} \overline{c_i^2} \tag{3.5}$$

Table 7 and Table 8 show the rms thermal speeds, $\sqrt{\overline{c_i^2}}$ evaluated from the SDPD-DV simulations. These thermal speeds are inversely related to the fluid particle masses considered and they are equal to those obtained with the equilibrium Maxwellian formula $\sqrt{3k_B T_i / m_i}$ for the SDPD fluid. Figure 22(a) shows the evolution of the translational temperature for Case 10. This behavior is typical for all cases considered and shows that the system quickly reaches equilibrium and the imposed FP temperature of $T_i = 300\,\mathrm{K}$. Figure 22(b) and Figure 22(c) depict the average translational temperature for all cases simulated in Table 7 and Table 8. It is clear that for the range of fluid particle masses and sizes considered the average translational temperature is scale-free and matches the imposed fluid particle temperature for both the $H_2O$(l) and $N_2$(g).

We also evaluate the self-diffusion coefficient for the $H_2O$(l) simulations following the MSD Eq. (2.74) and VAC Eq. (2.76) methods. Various analytical formulas for the self-diffusivity are used for validation. The self-diffusion coefficient is given by Bird et al. (2007) by

$$D_{AA} = \frac{k_B T_A}{2\pi\eta_A(2R_A)} \tag{3.6}$$

Assuming by analogy, that the SDPD liquid has a lattice scale $D_i$ from Eq. (3.3), then Eq. (3.6) can be expressed as

$$D_{ii} = \frac{k_B T_i}{2\pi\eta D_i} \tag{3.7}$$

An additional expression for the self-diffusivity in a SDPD liquid is obtained following Litvinov et al. (2009) and using the smoothing function (2.38) used in this work. It provides

$$D_{ii}^h = \frac{\rho_i h^2 k_B T_i}{63 m_i \eta} \tag{3.8}$$

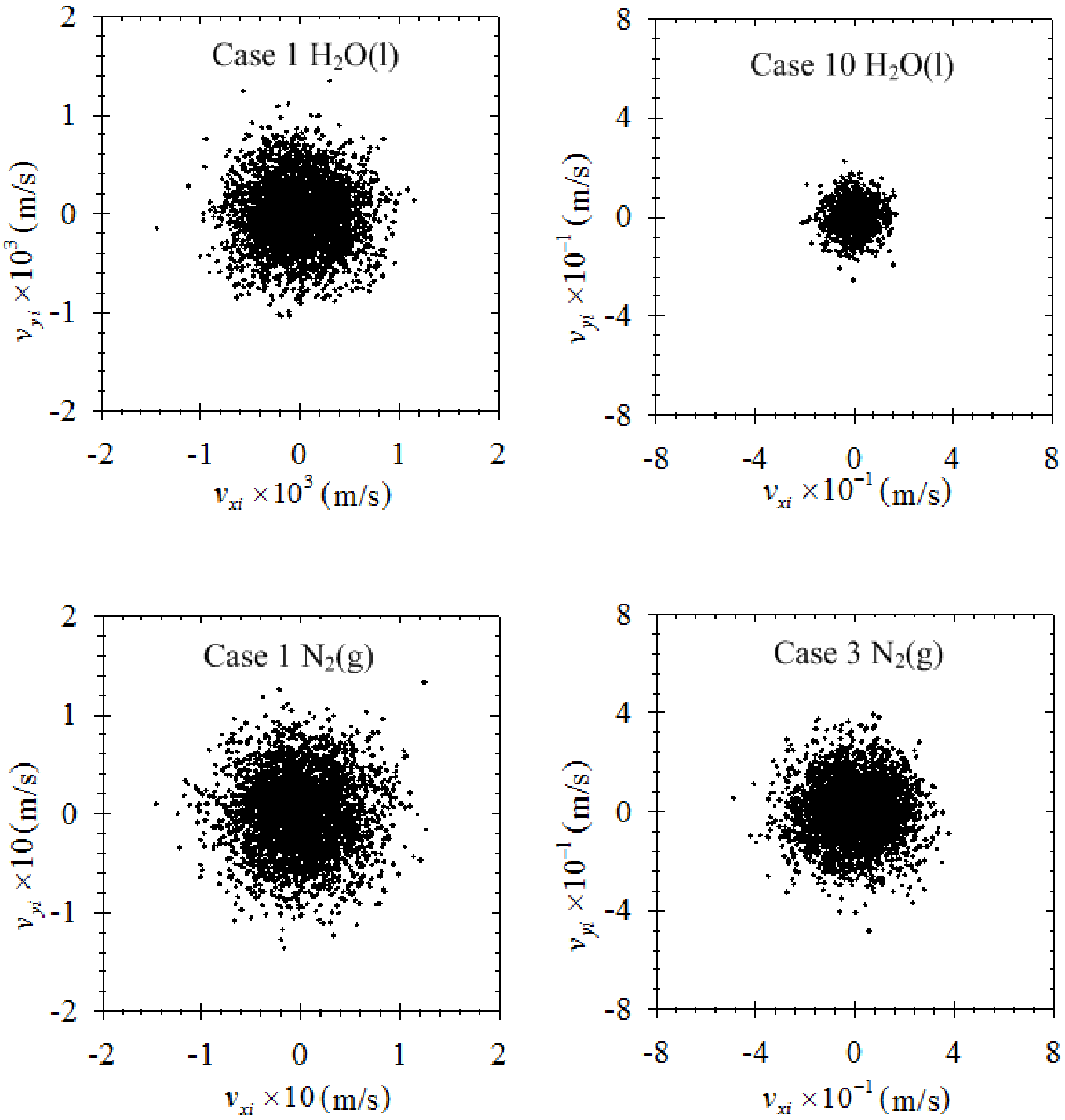

**Figure 21. Phase plot $v_{xi}$-$v_{yi}$ from SDPD-DV simulations of $H_2O$(l) with $\rho_a$=1,000 kg/m$^3$, $T_i$ = 300 K, $\eta$ =10$^{-3}$ kg·m$^{-1}$s$^{-1}$ and $N_2$(g) with $\rho_a$=1,184 kg/m$^3$, $T_i$ = 300 K, $\eta$ =10$^{-5}$ kg·m$^{-1}$s$^{-1}$. Results show the scale effects of fluid particle size on velocity.**

The Schmidt number is then evaluated by

$$\mathrm{Sc} = \frac{D}{\rho\eta} \tag{3.9}$$

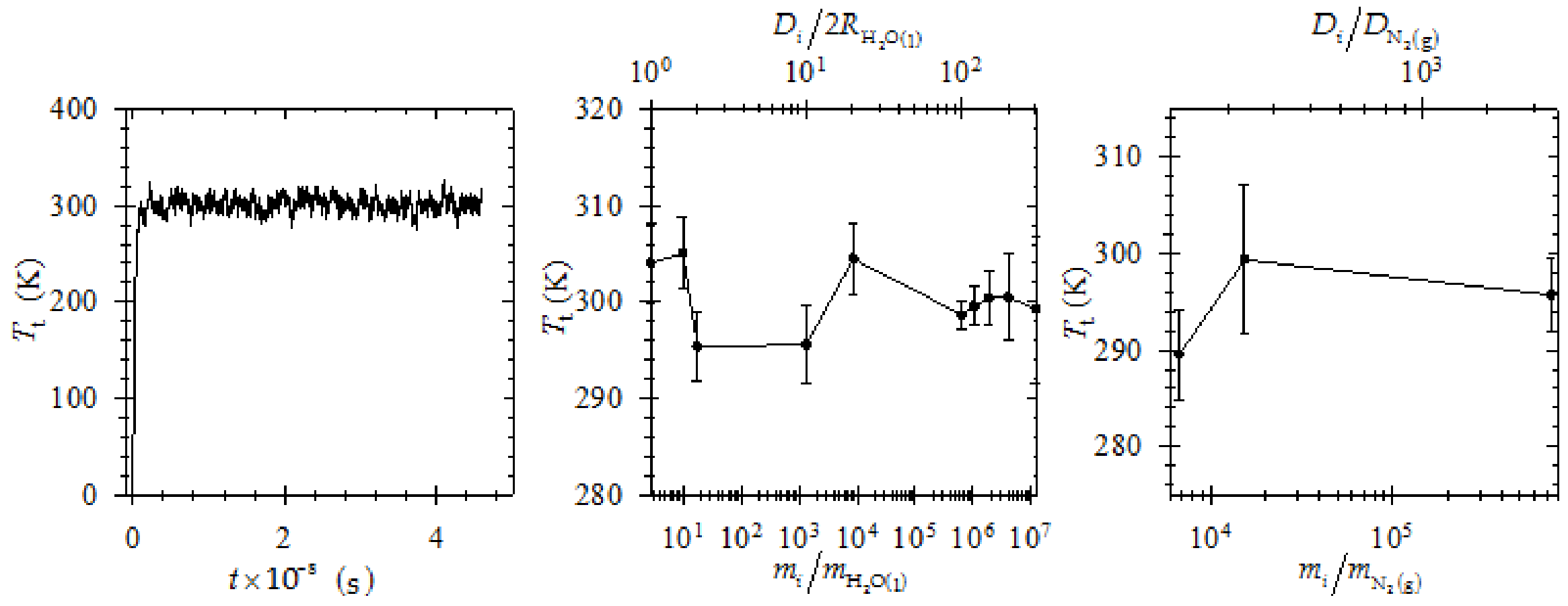

(a) (b) (c)

**Figure 22. Average translational temperature from SDPD-DV simulations of $H_2O$(l) with $\rho_a$=1,000 kg/m$^3$, $T_i$ = 300 K, $\eta$ =$10^{-3}$ kg·m$^{-1}$s$^{-1}$ and $N_2$(g) with $\rho_a$=1,184 kg/m$^3$, $T_i$ = 300 K, $\eta$ =$10^{-5}$ kg·m$^{-1}$s$^{-1}$ . (a) Average translational temperature as a function of time (Case 1, $H_2O$(l)). (b) Average translational temperature and standard deviation as a function of $m_i$ and $D_i$ for $H_2O$(l). (c) Average translational temperature and standard deviation as a function of $m_i$ and $D_i$ for $N_2$(g).**

In Figure 23(a) we plot the self-diffusion coefficient for $H_2O$(l). The experimental value $D_{\mathrm{H_2O,H_2O}} = 2.5\times10^{-9}$ m$^2$s$^{-1}$ is from Holz and Sacco (2009). With $m_{\mathrm{H_2O}} = 2.99\times10^{-26}$ kg and $\rho_{\mathrm{H_2O}} = 1{,}000\,\mathrm{kg/m^3}$ then Eq. (3.4) provides $2R_{\mathrm{H_2O}} = 3.1\times10^{-10}\mathrm{m}$ and the analytical value $D_{AA}$ is evaluated from Eq. (3.6). The values of $D_{ii}$ from the MSD Eq. (2.74) and VAC Eq. (2.76) are also plotted in Figure 23(a). For comparison we also evaluate $D_{ii}$ from Eq. (3.7) and $D_{ii}^h$ from Eq. (3.8) using SDPD-DV results for all the variables entering these expressions. Figure 23(a) shows that the self-diffusion coefficients obtained from MSD, VAC and the SDPD-liquid formulas are in good agreement been within a factor of about 3 of each other. The experimental and analytical value of the self-diffusion coefficient is approached by SDPD-DV, as the mass of the SDPD particles approaches asymptotically the relevant physical scale.

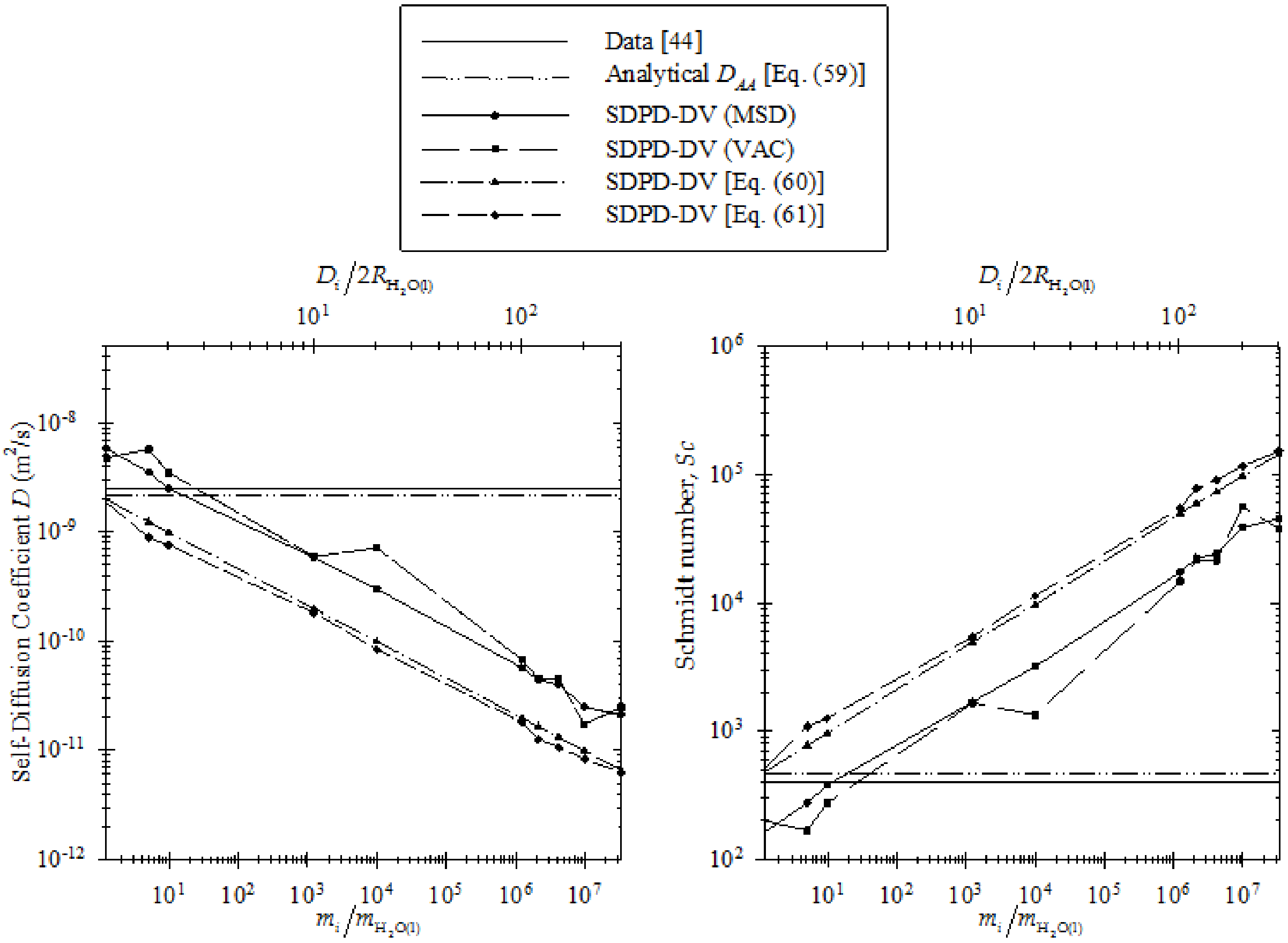

(a) (b)

**Figure 23. Self-diffusion coefficient and Schmidt number from SDPD-DV simulations of $H_2O(l)$ , $\rho_a$=1,000 kg/m$^3$, $\eta$=10$^{-3}$kg·m$^{-1}$s$^{-1}$ , *T*=300K for various sizes of the SDPD-DV fluid particles. Experimental and analytical estimates, as well as analytical estimates using SDPD-DV parameters are shown for validation and verification.**

It should be noted, that at the real molecule level the applicability of the SDPD model becomes tenuous. These results also show that for the fluid particle sizes considered in this work, the self-diffusion coefficient from SDPD-DV is not scale-free, a result that corroborates earlier results of Litvinov et al. (2009) and Vazquez-Quesada et al. (2009). The Sc numbers from Eq. (3.9) using the various diffusion-coefficients are plotted in Figure 23(b). Similar to previous results, we show that the Sc number scales with the mass (or size) of the SDPD particles, identified in our work with either the SDPD-fluid size $D_i$ through Eq. (3.7) or the smoothing length $h$ through Eq. (3.8). The results show also that with SDPD-DV we achieve Sc values close to the realistic ones.

# 4. VERIFICATION, VALIDATION AND ERROR ANALYSIS OF SDPD-DV

In this chapter, we perform validation and verification of SDPD-DV using the full-set of SDPD Eq. (1.62)-(1.64) as implemented in the SDPD-DV method discussed in Chapter 2. The validation and verification includes an extensive set of SDPD-DV simulations of gaseous nitrogen in mesoscopic periodic domains in equilibrium. The self-diffusion coefficient for $N_2$(g) and shear viscosity at equilibrium states are obtained through the mean-square displacement for the range of fluid particle masses (or sizes) considered.

Additional verification involves SDPD-DV simulations of steady Couette $N_2$(g) flow between parallel plates. The top plate is moving at $V_{xw}$=30m/s and separated by $10^{-4}$ m from the bottom stationary plate. The top plate is assigned a constant temperature $T_1$=330K and bottom plate $T_2$=300K. The SDPD-DV field velocity and temperature profiles are compared with those obtained by FLUENT.

## 4.1 Equilibrium State and Transport Coefficients

We consider first $N_2$(g) systems in thermal equilibrium. The full algorithm of SDPD equations is considered with entropy provided by Eq. (1.64) and closed by Eq. (2.44)-(2.49).

### 4.1.1 Input Conditions and Computational Parameters

The simulations of $N_2$(g) were performed in rectangular domains with $L_X = L_Y = L_Z$ in the range $0.25\times10^{-6}$~$10\times10^{-6}$ m, with mass $M_F$ in the range $1.85\times10^{-20} \sim 1.184\times10^{-15}$ kg. The simulations examine the effects of fluid particle mass $m_i / m_{\mathrm{N}_2}$ and the smoothing length $h$.

**Table 9: Input parameters in full-set SDPD-DV simulations of mesoscale flows of $N_2$(g) at equilibrium states in rectangular domains with periodic boundaries.**

| Case | Inputs $N_2$(g), $\rho_a = 1.184$ kg/m$^3$, $T = 300$ K, $P_0$=101942.4 | | | | | | | |
|---|---|---|---|---|---|---|---|---|
| | $L_{XYZ}$ ($\times 10^{-6}$m) | $\frac{m_i}{m_{N_2}}$ | $N_{FP}$ | $(V_i)^{1/3}$ ($\times 10^{-8}$m) | $h$ ($\times 10^{-8}$m) | $\Delta t$ ($\times 10^{-15}$s) | $\eta_0$ ($\times 10^{-5}$kg m$^{-1}$s$^{-1}$) | $k_0$ (W m$^{-1}$K$^{-1}$) |
| 1 | 0.25 | 118 | 3375 | 1.67 | 3.7 | 1 | 1.79 | 0.026 |
| 2 | 0.25 | 118 | 3375 | 1.67 | 4.7 | 1 | 1.79 | 0.026 |
| 3 | 0.25 | 118 | 3375 | 1.67 | 5.9 | 1 | 1.79 | 0.026 |
| 4 | 0.25 | 118 | 3375 | 1.67 | 7.5 | 1 | 1.79 | 0.026 |
| 5 | 0.55 | 1255 | 3375 | 3.67 | 9 | 10 | 1.79 | 0.026 |
| 6 | 0.55 | 1255 | 3375 | 3.67 | 11 | 10 | 1.79 | 0.026 |
| 7 | 0.55 | 1255 | 3375 | 3.67 | 12.9 | 10 | 1.79 | 0.026 |
| 8 | 0.55 | 1255 | 3375 | 3.67 | 16.5 | 10 | 1.79 | 0.026 |
| 9 | 1 | 7544 | 3375 | 6.67 | 15 | 50 | 1.79 | 0.026 |
| 10 | 1 | 7544 | 3375 | 6.67 | 19 | 50 | 1.79 | 0.026 |
| 11 | 1 | 7544 | 3375 | 6.67 | 24 | 50 | 1.79 | 0.026 |
| 12 | 1 | 7544 | 3375 | 6.67 | 29.5 | 50 | 1.79 | 0.026 |
| 13 | 2.5 | $1.2\times10^5$ | 3375 | 16.7 | 38 | 50 | 1.79 | 0.026 |
| 14 | 2.5 | $1.2\times10^5$ | 3375 | 16.7 | 50 | 50 | 1.79 | 0.026 |
| 15 | 2.5 | $1.2\times10^5$ | 3375 | 16.7 | 59 | 50 | 1.79 | 0.026 |
| 16 | 2.5 | $1.2\times10^5$ | 3375 | 16.7 | 75 | 50 | 1.79 | 0.026 |
| 17 | 10 | $7.5\times10^6$ | 3375 | 66.7 | 150 | 100 | 1.79 | 0.026 |
| 18 | 10 | $7.5\times10^6$ | 3375 | 66.7 | 190 | 100 | 1.79 | 0.026 |
| 19 | 10 | $7.5\times10^6$ | 3375 | 66.7 | 240 | 100 | 1.79 | 0.026 |
| 20 | 10 | $7.5\times10^6$ | 3375 | 66.7 | 290 | 100 | 1.79 | 0.026 |

**Table 10: Derived parameters in SDPD-DV simulations of mesoscale flows of $N_2$(g) at equilibrium states in rectangular domains with periodic boundaries.**

| Case | Derived SDPD-DV | | | | | | |
|---|---|---|---|---|---|---|---|
| | $(\bar{V}_i)^{1/3}$ ($\times 10^{-8}$ m) | $\bar{\rho}_i$ (kg/m$^3$) | $\bar{T}$ (K) | $\bar{P}$ (Pa) | $\sqrt{\overline{C_i^2}}$ | $D$ ($\times 10^{-9}$ m$^2$/s) | $\eta$ ($\times 10^{-5}$ kg m$^{-1}$s$^{-1}$) |
| 1 | 1.66 | 1.193 | 299.2 | 102410 | 47.64 | 3.1 | 4 |
| 2 | 1.67 | 1.188 | 299.25 | 102030 | 47.64 | 4.1 | 2 |
| 3 | 1.67 | 1.185 | 299.25 | 101800 | 47.75 | 5.8 | 3 |
| 4 | 1.67 | 1.184 | 299.3 | 101750 | 47.64 | 8.5 | 2 |
| 5 | 3.667 | 1.184 | 299.41 | 101752 | 14.59 | 1.4 | 4.5 |
| 6 | 3.662 | 1.188 | 299.5 | 102155 | 14.63 | 2.4 | 1.5 |
| 7 | 3.665 | 1.185 | 299.43 | 101867 | 14.59 | 2.7 | 2 |
| 8 | 3.666 | 1.184 | 299.40 | 101775 | 14.56 | 4.2 | 1.5 |
| 9 | 6.678 | 1.178 | 299.41 | 101227 | 5.96 | 0.3 | 1 |
| 10 | 6.678 | 1.178 | 299.43 | 101228 | 5.96 | 0.7 | 1.5 |
| 11 | 6.668 | 1.183 | 299.42 | 101684 | 5.96 | 1.4 | 1 |
| 12 | 6.667 | 1.184 | 299.47 | 101729 | 5.96 | 2.9 | 1.5 |
| 13 | 16.7 | 1.177 | 299.41 | 101160 | 1.53 | 0.21 | 3 |
| 14 | 16.7 | 1.176 | 299.41 | 101052 | 1.52 | 0.3 | 3 |
| 15 | 16.7 | 1.176 | 299.42 | 101065 | 1.61 | 0.4 | 2 |
| 16 | 16.7 | 1.172 | 299.40 | 100752 | 1.52 | 0.65 | 2.5 |
| 17 | 66.9 | 1.173 | 299.4 | 100803 | 0.195 | 0.0365 | 1 |
| 18 | 66.7 | 1.171 | 299.4 | 100664 | 0.190 | 0.0259 | 1 |
| 19 | 66.7 | 1.181 | 299.4 | 101506 | 0.187 | 0.044 | 1 |
| 20 | 66.9 | 1.172 | 299.4 | 100670 | 0.187 | 0.1 | 3 |

For each $m_i / m_{\mathrm{N}_2}$ we considered four $h$'s so that the resulting number of fluid particles within the support domain, $L_i$, is 50, 100, 200, 360 upon initialization. Periodic boundary conditions are imposed on each side of the rectangular domain. Upon initialization we set $\rho_a = 1.184\ \mathrm{kg/m^3}$, $T_i(t=0) = 300\mathrm{K}$, $\eta_i(t=0) = 1.79\times 10^{-5}\ \mathrm{kg/m\cdot s}$ and $\kappa_i(t=0) = 0.026\ \mathrm{W/m\cdot K}$. During the computation the transport coefficients appearing in the SDPD equations (1.62)-(1.64) are evaluated as functions of particle temperature $T_i(t)$ according to power law (White, 1974)

$$\eta_i(t) \approx \eta_0 \left(\frac{T_i(t)}{T_0}\right)^n \tag{4.1}$$

$$k_i(t) \approx k_0 \left(\frac{T_i(t)}{T_0}\right)^n \tag{4.2}$$

where $n$ is the power law coefficient of the order of 0.7, $T_0 = 300\mathrm{K}$ the reference temperature, $\kappa_0 = 0.026\mathrm{W\,m^{-1}K^{-1}}$ the heat conductivity of $N_2$(g) at $T_0$, and $\eta_0 = 1.79\times 10^{-5}\mathrm{kgm^{-1}s^{-1}}$ is the viscosity of $N_2$(g) at $T_0$. The heat capacity (J/K) is evaluated as Eq. (2.48).

$$C_i = \frac{3}{2} N_i k_B \tag{2.48}$$

In order to examine fluid particle scale effects we follow Vazquez-Quesada et al (2009) discussion and assume that the "size" of the fluid particle is given in terms of the SDPD variables as,

$$D_i \sim \left(V_i\right)^{1/3} = \left(m_i / \rho_i\right)^{1/3} \tag{4.3}$$

Input and some derived parameters from the SDPD-DV simulations are shown in Table 9 and Table 10.

### 4.1.2 Results and Discussion

Table 10 and

Table 11 show that the average density $\bar{\rho}$, temperature $\bar{T}$ and pressure $\bar{P}$ exhibit minimal perturbations from the input value for the entire range of parameters considered. These rms thermal speeds, $\sqrt{\overline{c_i^2}}$ evaluated from the SDPD-DV simulations, are inversely related to the fluid particle masses considered and they are equal to those obtained with the equilibrium Maxwellian formula $\sqrt{3k_B T_i / m_i}$ for the SDPD fluid.

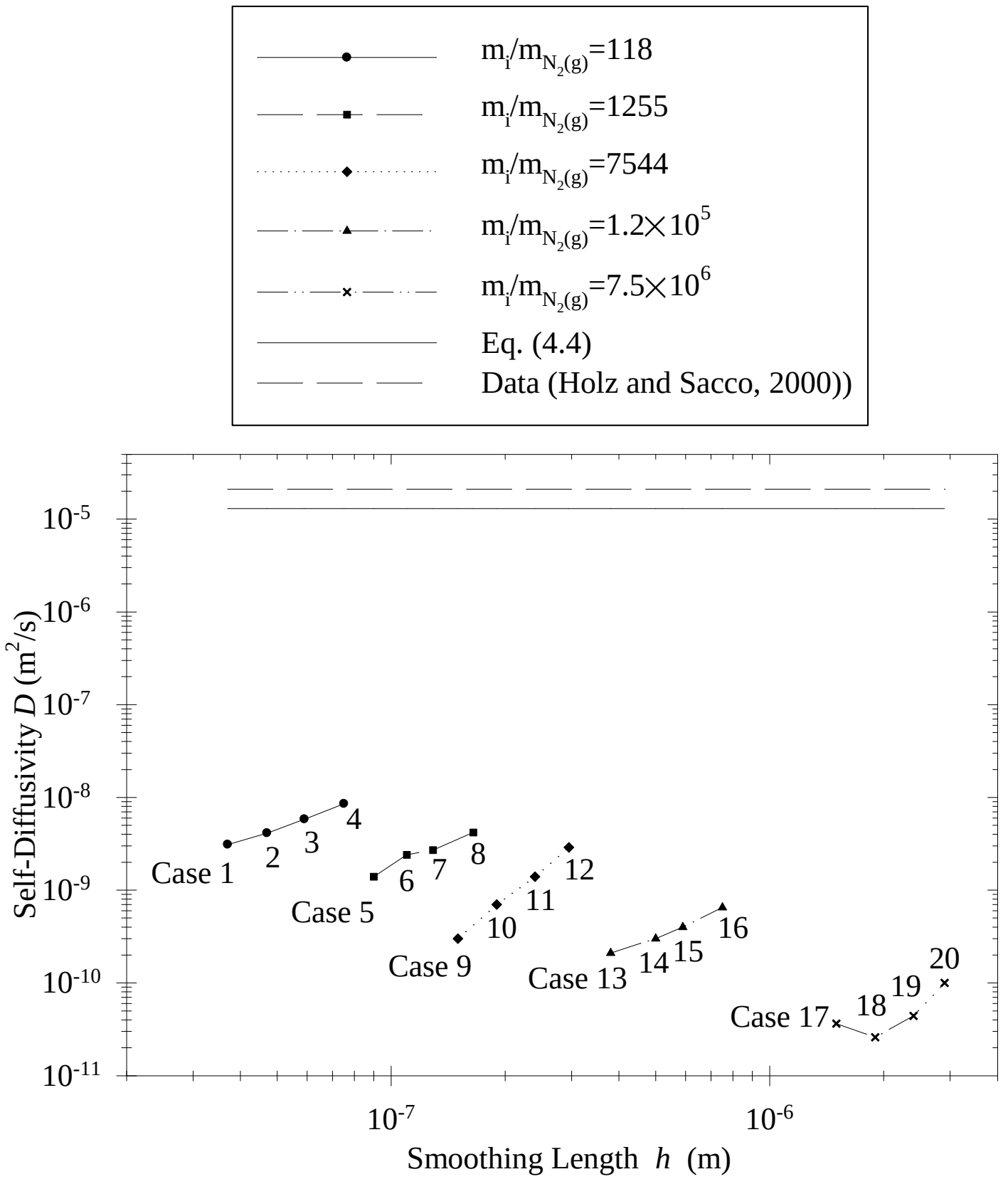

**Figure 24: Self-diffusion coefficient from SDPD-DV simulations of $N_2$(g) for values of $h$ and $m_i/m_{N_2(g)}$ used in Cases 1-20. Analytical estimates are from Eq.(4.4) and data by Holz and Sacco (2000).**

We also evaluate the transport coefficient for the $N_2$(g) simulations following the MSD Eq. (2.74) for self-diffusion coefficient and Eq.(2.77) for viscosity. Validation and verification is obtained by comparison with experimental (Holz and Sacco, 2000) and analytical values. For a gas with molecular diameter $D_A$, molecular mass $m_A$ and mass density $\rho_A$ the self-diffusivity is given by Bird et al. (2007) as

$$D_{AA} = \frac{2}{3\pi} \frac{\sqrt{m_A \pi k_B T_A}}{\pi D_A^2} \frac{1}{\rho_A}. \tag{4.4}$$

The self-diffusion coefficient of SDPD-DV simulations from Eq. (2.74) for $N_2$(g) are plotted in Figure 24 for various smoothing lengths $h$ and particle masses $m_i$. For comparison, the values of $D_{AA}$ from Eq. (4.4) are evaluated and plotted in Figure 24 using $N_2$(g) molecular value. As it shown in Figure 24, the self-diffusivity of SDPD-DV is not scale free. The simulations show that $D$ decreases with increasing mass ratio. This is because that the smaller the fluid particle is, the larger is its stochastic agitation.

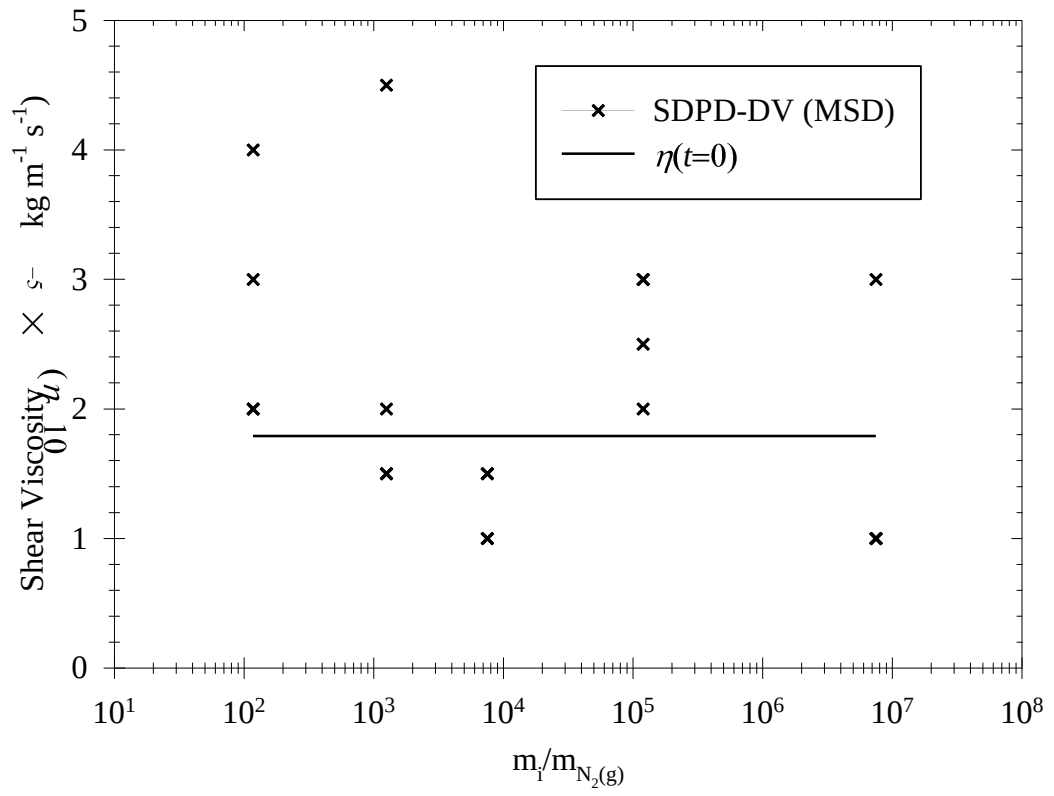

**Figure 25 : Shear viscosity from SDPD-DV simulations of $N_2$(g) for value of $m_i/m_{N2(g)}$ used in Cases 1-20. Initial input viscosity $\eta(t=0)$ is plotted for verification.**

For a given mass ratio, increasing the $h$, increases $D$, since that the stochastic agitation is stronger in the bigger support domain. The values for $\eta$ calculated from the SDPD-DV simulations using Eq. (2.77) are plotted in Figure 25. We plot for comparison the initial value $\eta(t=0)$ which is also $\eta_0$. It can be seen that shear viscosity is scale free and is not affected by the choice of particle mass or the smoothing length.

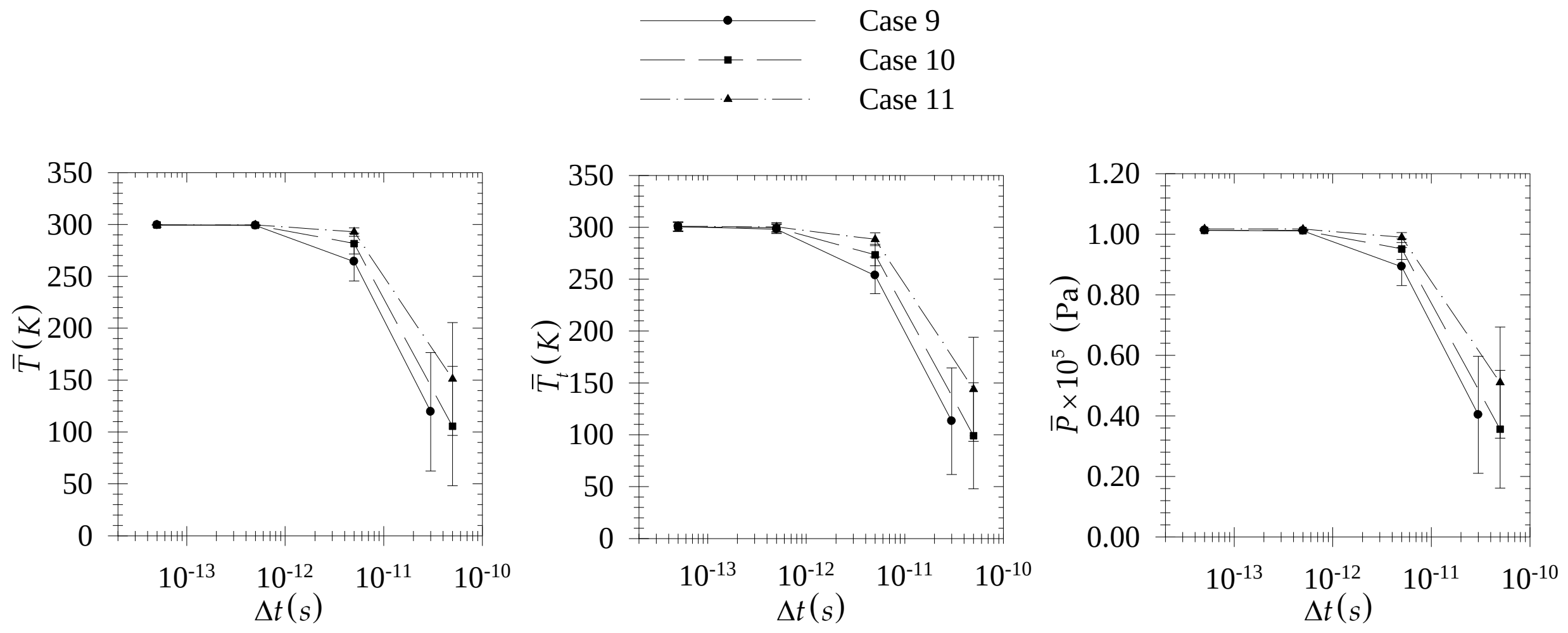

**Figure 26: Effects of time step on equilibrium temperature $\bar{T}$, translational temperature $\bar{T}_t$ and pressure $\bar{P}$ from SDPD-DV simulations of $N_2$(g). Results are for Case 9, 10, and 11.**

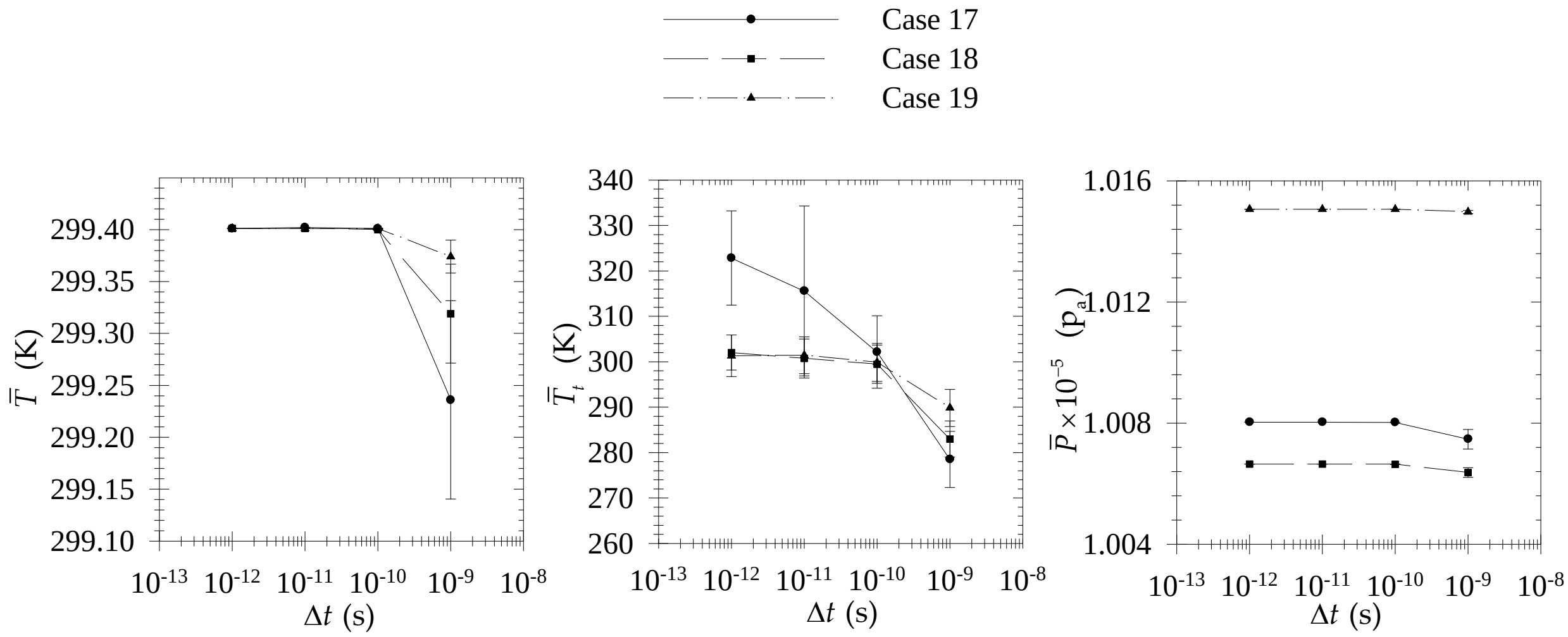

**Figure 27: Effects of time step on equilibrium temperature $\bar{T}$, translational temperature $\bar{T}_t$ and pressure $\bar{P}$ from SDPD-DV simulations of $N_2$(g). Results are for Case 17, 18, and 19.**

**Table 11: Fluctuations in temperature, density, pressure and velocity, from SDPD-DV and analytical expressions.**

| case | Analytical Variance | | | | Variance in SDPD-DV | | | |
|---|---|---|---|---|---|---|---|---|
| | $\left\langle (\delta T)^2 \right\rangle$ | $\left\langle \left(\overline{\delta\rho}\right)^2 \right\rangle$ | $\left\langle (\delta P)^2 \right\rangle$ | $\left\langle (\delta \mathbf{V})^2 \right\rangle$ | $\left\langle (\delta T)^2 \right\rangle$ | $\left\langle \left(\overline{\delta\rho}\right)^2 \right\rangle$ | $\left\langle (\delta P)^2 \right\rangle$ | $\left\langle \delta \mathbf{V}\ ^2 \right\rangle$ |
| 1 | 0.127 | $3.04\times10^{-6}$ | 22373 | 0.6714 | 0.1468 | $3.13\times10^{-7}$ | 19478 | 0.6733 |
| 2 | 0.127 | $3.01\times10^{-6}$ | 22208 | 0.6714 | 0.1467 | $1.42\times10^{-7}$ | 18134 | 0.6736 |
| 3 | 0.127 | $3.00\times10^{-6}$ | 22108 | 0.6714 | 0.1475 | $6.53\times10^{-8}$ | 17504 | 0.6733 |
| 4 | 0.127 | $2.99\times10^{-6}$ | 22086 | 0.6714 | 0.1456 | $2.94\times10^{-8}$ | 17042 | 0.6739 |
| 5 | $1.2\times10^{-2}$ | $2.81\times10^{-7}$ | 2074 | $6.3\times10^{-2}$ | $1.39\times10^{-2}$ | $3.53\times10^{-8}$ | 1866 | $6.28\times10^{-2}$ |
| 6 | $1.2\times10^{-2}$ | $2.83\times10^{-7}$ | 2090 | $6.3\times10^{-2}$ | $1.39\times10^{-2}$ | $1.88\times10^{-8}$ | 1749 | $6.35\times10^{-2}$ |
| 7 | $1.2\times10^{-2}$ | $2.81\times10^{-7}$ | 2079 | $6.3\times10^{-2}$ | $1.39\times10^{-2}$ | $9.84\times10^{-9}$ | 1682 | $6.3\times10^{-2}$ |
| 8 | $1.2\times10^{-2}$ | $2.81\times10^{-7}$ | 2075 | $6.3\times10^{-2}$ | $1.37\times10^{-2}$ | $3.85\times10^{-9}$ | 1615 | $6.26\times10^{-2}$ |
| 9 | $2\times10^{-3}$ | $4.63\times10^{-8}$ | 341 | $1.05\times10^{-2}$ | $2.34\times10^{-3}$ | $3.27\times10^{-8}$ | 392 | $1.05\times10^{-2}$ |
| 10 | $2\times10^{-3}$ | $4.63\times10^{-8}$ | 342 | $1.05\times10^{-2}$ | $2.31\times10^{-3}$ | $1.26\times10^{-8}$ | 329 | $1.04\times10^{-2}$ |
| 11 | $2\times10^{-3}$ | $3.11\times10^{-8}$ | 345 | $1.05\times10^{-2}$ | $2.33\times10^{-3}$ | $1.56\times10^{-8}$ | 359 | $1.05\times10^{-2}$ |
| 12 | $2\times10^{-3}$ | $4.67\times10^{-8}$ | 345 | $1.05\times10^{-2}$ | $2.33\times10^{-3}$ | $4.59\times10^{-9}$ | 345 | $1.06\times10^{-2}$ |
| 13 | $1.3\times10^{-4}$ | $2.96\times10^{-9}$ | 21.8 | $6.69\times10^{-4}$ | $1.49\times10^{-4}$ | $1.94\times10^{-8}$ | 159 | $6.92\times10^{-4}$ |
| 14 | $1.3\times10^{-4}$ | $2.95\times10^{-9}$ | 21.8 | $6.69\times10^{-4}$ | $1.49\times10^{-4}$ | $9.00\times10^{-9}$ | 82.2 | $6.84\times10^{-4}$ |
| 15 | $1.3\times10^{-4}$ | $2.95\times10^{-9}$ | 21.8 | $6.69\times10^{-4}$ | $1.49\times10^{-4}$ | $6.85\times10^{-8}$ | 524 | $6.73\times10^{-4}$ |
| 16 | $1.3\times10^{-4}$ | $2.93\times10^{-9}$ | 21.8 | $6.69\times10^{-4}$ | $1.49\times10^{-4}$ | $6.60\times10^{-8}$ | 500 | $6.69\times10^{-4}$ |
| 17 | $2.0\times10^{-6}$ | $4.59\times10^{-11}$ | 0.3387 | $1.05\times10^{-5}$ | $2.31\times10^{-6}$ | $1.67\times10^{-7}$ | 1231 | $1.13\times10^{-5}$ |
| 18 | $2.0\times10^{-6}$ | $4.57\times10^{-11}$ | 0.3377 | $1.05\times10^{-5}$ | $2.34\times10^{-6}$ | $1.42\times10^{-7}$ | 1049 | $1.06\times10^{-5}$ |
| 19 | $2.0\times10^{-6}$ | $4.65\times10^{-11}$ | 0.3434 | $1.05\times10^{-5}$ | $2.3\times10^{-6}$ | $1.17\times10^{-8}$ | 86.53 | $1.05\times10^{-5}$ |
| 20 | $2.0\times10^{-6}$ | $4.58\times10^{-11}$ | 0.3378 | $1.05\times10^{-5}$ | $2.3\times10^{-6}$ | $5.63\times10^{-8}$ | 416 | $1.05\times10^{-5}$ |

We examine also the effects of time step in the SDPD-DV simulations on the equilibrium characteristics. In Figure 26 and Figure 27, we plot the SDPD-DV temperature $\bar{T}$ averaged over the entire domain, the average translational temperature $\bar{T}_t$ from (3.5), and the average SDPD-DV pressure $\bar{P}$ for Case 9, 10, 11 and Case 17, 18, and 19. These results show that time step can lead to a significant error depending on the FP mass and smoothing length. The results show also that for larger FPs the stochastic agitation is smaller.

We compute also the variances in temperature, density, pressure and velocity for each case 1-20 according to Eq. (1.25), (1.22), (1.23) and (1.27). The results in

Table 11 show that SDPD-DV are in very good agreement with the analytical values.

### 4.2 Steady Planar Thermal Couette FLow

In this section, we continue the verification of SDPD-DV by performing simulations of steady planar non-isothermal Couette flow.

#### 4.2.1 Input Conditions and Computational Parameters

The test involves an incompressible flow with density $\rho_a = 1.184\ \text{kg/m}^3$ across two infinite parallel walls as depicted in Figure 28(a) with imposed constant wall temperatures and the top wall is moving with a constant velocity $V_{xw}$. The SDPD-DV simulation considers $N_2$(g) with density $\rho_a = 1.184\ \text{kg}\cdot\text{m}^{-3}$, initial viscosity $\eta(t=0) = 1.79\times10^{-5}\ \text{kg}\cdot\text{m}^{-1}\text{s}^{-1}$ and initial heat conductivity $\kappa(t=0) = 0.026\ \text{W}\cdot\text{m}^{-1}\text{K}^{-1}$. Closure for pressure and density is obtained by using the Eq.(2.47) and (2.46). Heat capacity $C_V$ is given by Eq.(2.48) with initial value $C_i(t=0) = 2.42\times10^{-14}\ \text{J/K}$.

The physical domain has $L_{X,Z} = 1\times10^{-4}\text{m}$ and $L_Y = 3\times10^{-5}\text{m}$ as shown in Figure 28. Periodic boundary are imposed along the $x$-axis and $y$-axis. The fluid is represented by 10,890 FPs and total mass $M_F = 3.552\times10^{-13}\text{kg}$. The upper wall is located at $Z = 1\times10^{-4}\text{m}$ and is assigned a velocity $V_{xw} = 30\ \text{ms}^{-1}$ and temperature $T_1 = 330\text{K}$. The temperature of the lower wall located at $Z = 0$ is set to $T_2 = 300\text{K}$. The solid walls are represented by 10,560 BPs. Input conditions used in the SDPD-DV simulation are shown in Table 12.

**Table 12: Input parameters used in SDPD-DV non-equilibrium simulations of Couette flow.**

| Input Parameters | Coutte |
|---|---|
| $L_X$ (m) | $1\times10^{-4}$ |
| $L_Y$ (m) | $3\times10^{-5}$ |
| $L_Z$ (m) | $1\times10^{-4}$ |
| $f_x$ (ms$^{-2}$) | N/A |
| $M_F$ kg | $3.552\times10^{-13}$ |
| Upper $T_1$ (K) | 330 |
| Lower $T_2$ (K) | 300 |
| $\rho_a$ $\text{kg}\cdot\text{m}^{-3}$ | 1.184 |
| $\eta$ $t = 0$ $\text{kg}\cdot\text{m}^{-1}\text{s}^{-1}$ | $1.79\times10^{-5}$ |
| $\zeta(t=0)$ $\text{kg}\cdot\text{m}^{-1}\text{s}^{-1}$ | 0 |
| $C_i(t=0)$ $\text{J}\cdot\text{K}^{-1}$ | $2.42\times10^{-14}$ |
| $\kappa$ $t = 0$ $\text{W}\cdot\text{m}^{-1}\text{K}^{-1}$ | 0.026 |
| $V_{xw}$ ms$^{-1}$ | 30 |
| $N_\text{F}$ | 10890 |
| $N_\text{B}$ | 10560 |
| $N_C$ | 300 |
| $h$ (m) | $9\times10^{-6}$ |
| $\Delta t$ (s) | $10^{-8}$ |

For verification we also performed a simulation using FLUENT in a domain shown in Figure 28 and input parameters shown in Table 12. The wall boundary conditions are assigned to both upper wall and lower wall with periodic boundary conditions to other boundaries.

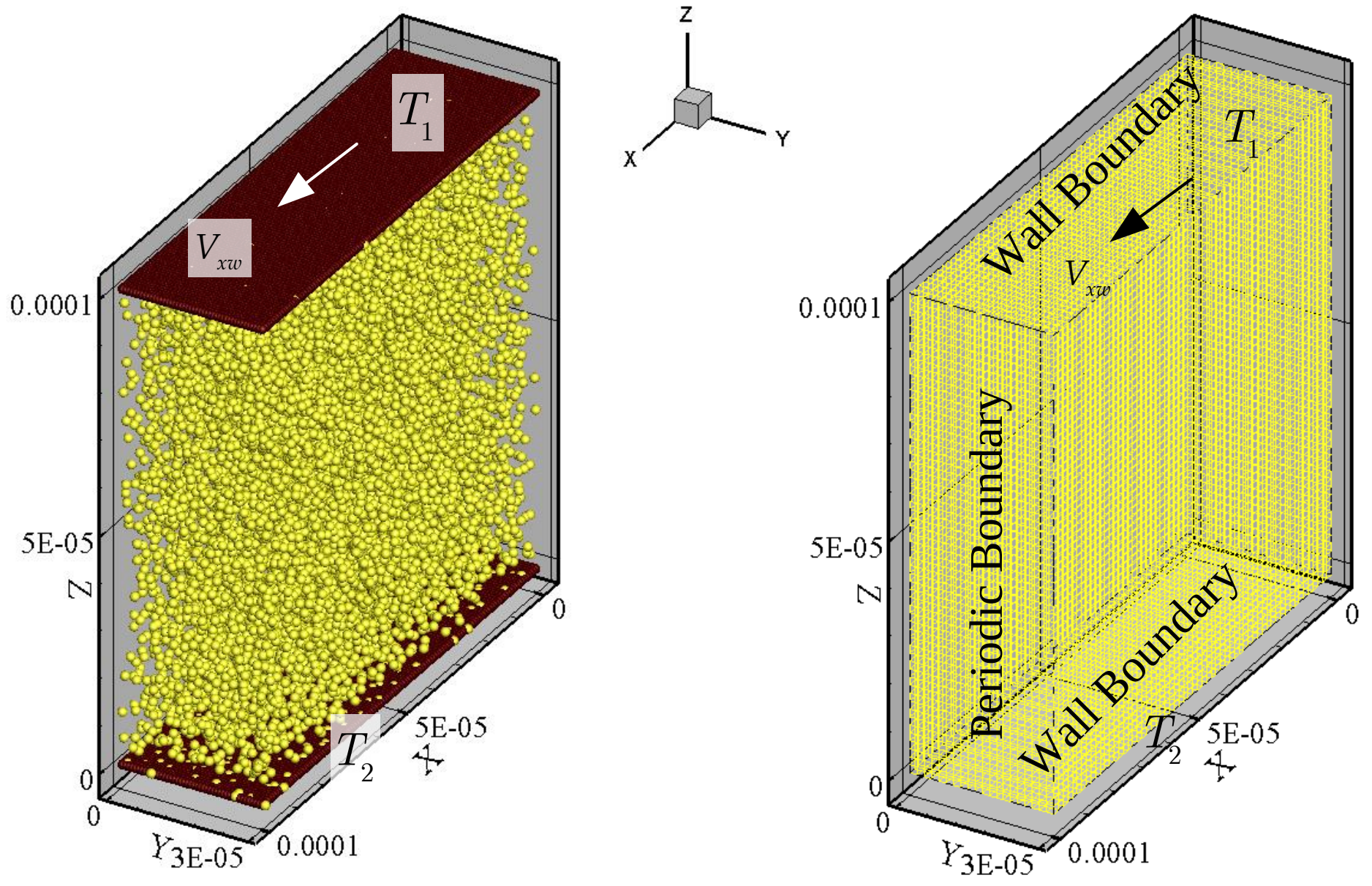

(a) SDPD-DV domain with $L_x$=$10^{-4}$m, $L_y$=$3\times10^{-5}$m, $L_z$=$10^{-4}$m showing the BPs on the top and bottom and FPs in the domain.

(b) FLUENT domain with $L_x$=$10^{-4}$m, $L_y$=$3\times10^{-5}$m, $L_z$=$10^{-4}$m showing imposed wall boundary conditions and periodic boundary conditions.

**Figure 28: Computational domains used in SDPD-DV and FLUENT simulations of steady $N_2$(g) Couette flow with $\rho_a$=1.184 kg/m$^3$, $\eta_0$=1.79×$10^{-5}$ kg·m$^{-1}$s$^{-1}$, $\kappa_0$=0.026 W·m$^{-1}$K$^{-1}$. The upper wall $T_1$= 330 K, lower wall $T_2$= 300 K and $V_{xw}$=30m/s.**

### 4.2.2 Results and Discussion

The overall flow field characteristics are shown in Figure 29. We plot the SDPD-DV sample-averaged steady temperature $T(\mathbf{r})$ and velocity $V_x(\mathbf{r})$ fields on planes $X=0\,\mathrm{m}$, $X=5\times10^{-5}\,\mathrm{m}$, and $X=1\times10^{-4}\,\mathrm{m}$ .

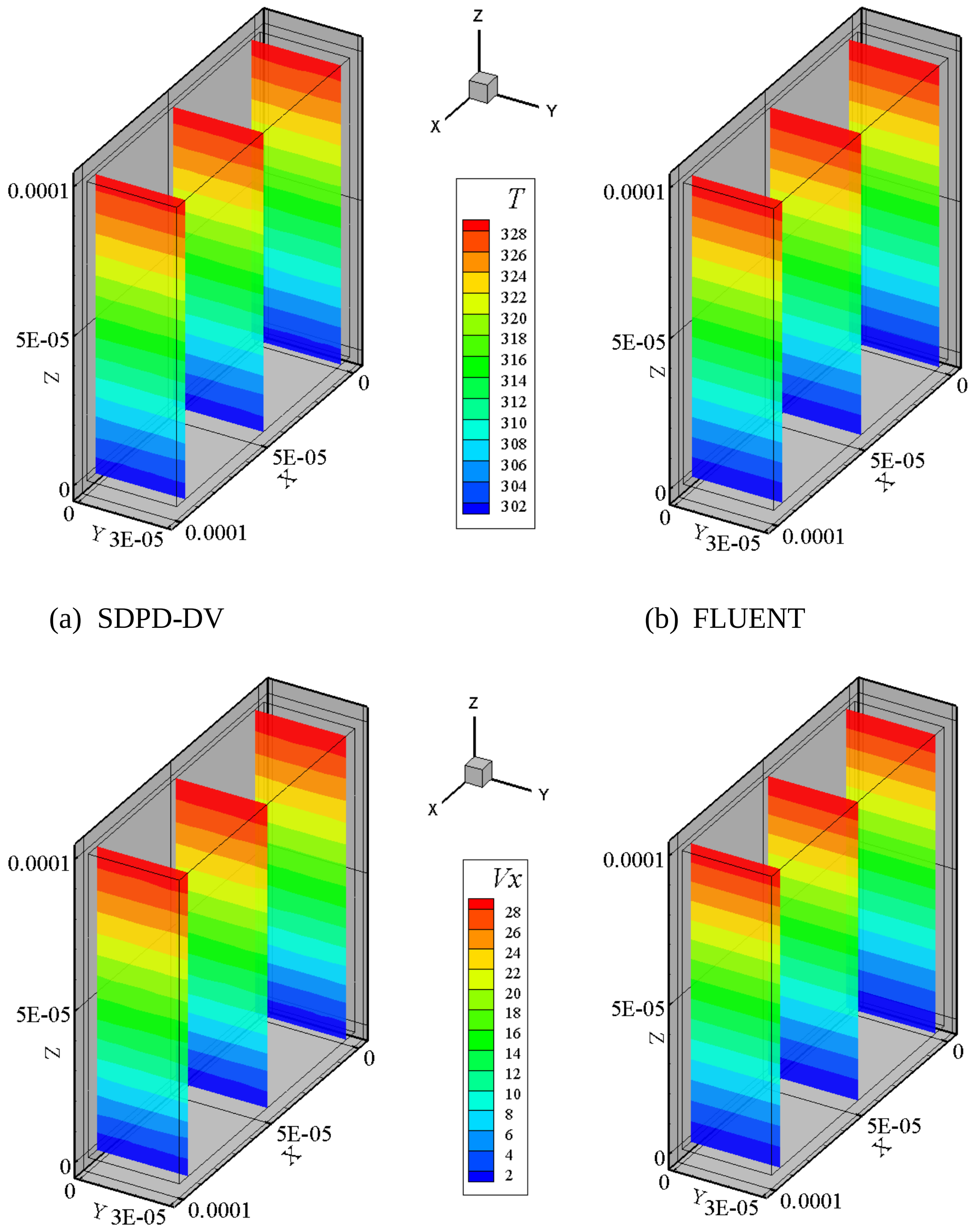

(a) SDPD-DV (b) FLUENT

(c) SDPD-DV (d) FLUENT

**Figure 29: Sample-averaged $V_x$(r) and $T$(r) on $y$ = 0 m, $y$ = 1.5×10$^{-5}$ m, $y$ = 3×10$^{-5}$ m planes from SDPD-DV and FLUENT simulations of steady state Couette flow. The domain has $L_X$ =10$^{-4}$m, $L_Y$ =3×10$^{-5}$m, $L_Z$ =10$^{-4}$m, upper wall $T_1$= 330 K, lower wall $T_2$= 300 K, and $V_{xw}$=30 m/s. The fluid is $N_2$(g) with $\rho_a$=1.184 kg/m$^3$, $\eta_0$=1.79×10$^{-5}$ kg·m$^{-1}$s$^{-1}$ and , $\kappa_0$=0.026 W·m$^{-1}$K$^{-1}$.**

We also plot for comparison the steady-state $T(\mathbf{r})$ and $V_x(\mathbf{r})$ from the FLUENT simulation. The velocity and temperature profiles are quantitatively similar to those obtained from FLUENT. For direct quantitative comparison, the sample-averaged field temperature $T(\mathbf{r}_d)$ and velocity $V_x(\mathbf{r}_d)$ profiles are plotted in Figure 30. The SDPD-DV properties are in excellent agreement with FLUENT solutions.

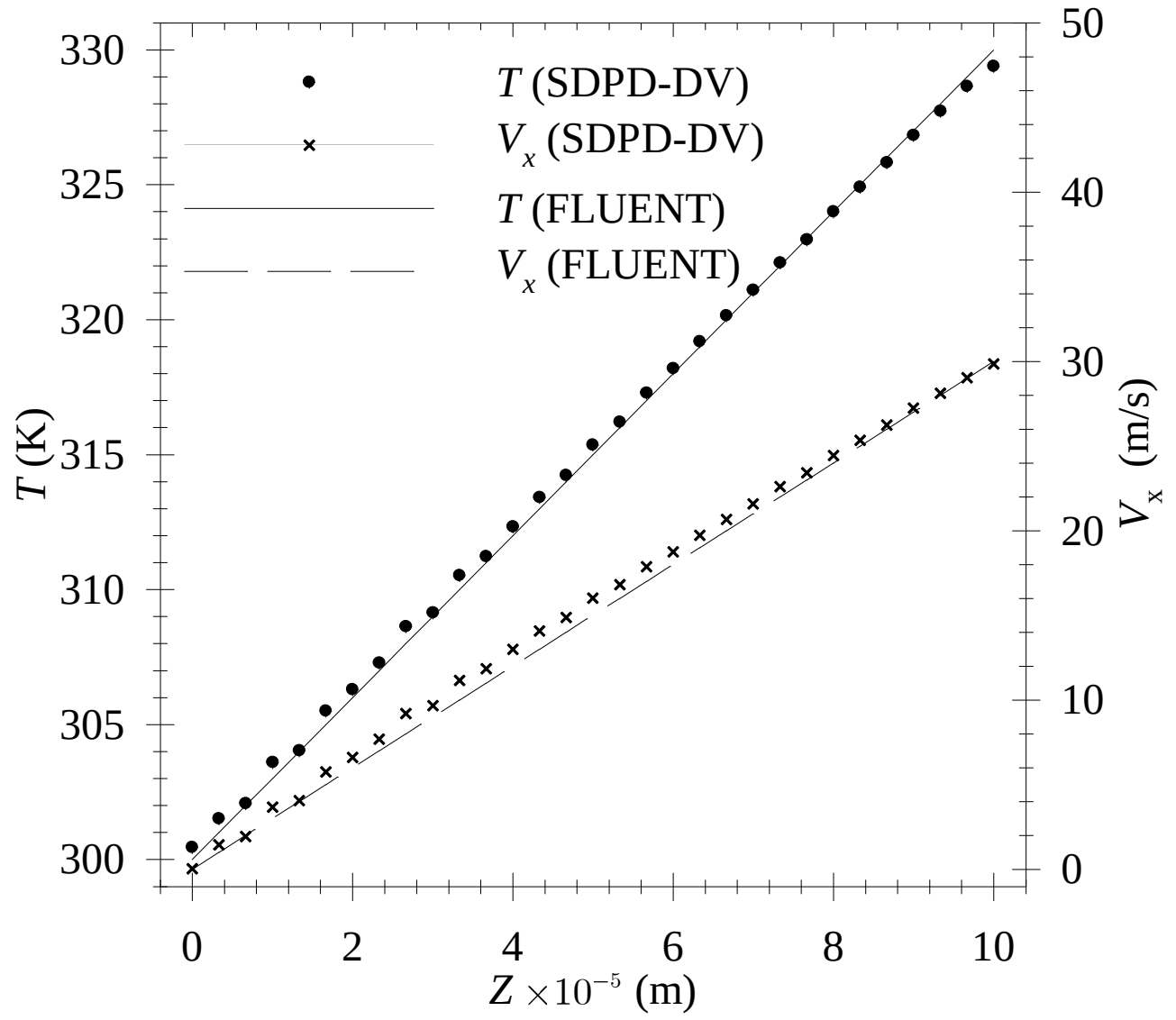

**Figure 30: Sample-averaged fluid temperature $T(\mathbf{r}_d)$ and velocity $V_x(\mathbf{r}_d)$ of steady-state Couette flow from SDPD-DV and FLUENT simulations. The domain has $L_X$ =$10^{-4}$m, $L_Y$ =$3\times10^{-5}$m, $L_Z$ =$10^{-4}$m, upper wall $T_1$= 330 K, lower wall $T_2$= 300 K and $V_{xw}$=30 m/s. The fluid is $N_2$(g) with $\rho_a$=1.184 kg/m$^3$, $\eta_0$=$1.79\times10^{-5}$ kg·m$^{-1}$s$^{-1}$ and , $\kappa_0$=0.026 W·m$^{-1}$K$^{-1}$.**

# 5. SUMMARY, CONCLUSIONS AND RECOMMENDATIONS

## 5.1 Summary and Conclusions

This work presented the mathematical formulation and computational implementation of a smoothed dissipative particle dynamics method for wall-bounded domains (SDPD-DV) under development at the Computational Gas&Plasma Dynamics Lab (CGPL) at WPI. The SDPD-DV allows simulation of liquid and gaseous flows at mesoscopic and hydrodynamic scales.

The SDPD-DV method and implementation utilizes fluid particles, boundary particles and dynamical allocated virtual particles. The boundary particles are placed on the physical boundary of the domain while fluid particles are used to discretize the fluid region of the domain. The algorithm for nearest neighbor particle search (NNPS) is based on a combination of the linked-cell and Verlet-list approaches. The NNPS utilizes large rectangular cells that organize the physical domain and increase computational efficiency. The dynamic virtual particle allocation (DVPA) method introduced in this work provides a formal way to minimize error in the density and force evaluation due to the truncated part of the support domain near a solid boundary. Models for the force components due to virtual particles are introduced near solid boundaries. A periodic boundary particle allocation method is used at periodic inlets and outlets. Integration of particle position and momentum equation is investigated with the implementation of the velocity-Verlet algorithm. The integration is supplemented by a bounce-forward algorithm to further prevent particle penetration. SDPD-DV outputs are sampled to obtain unsteady and steady particle properties. The self-diffusion coefficient is obtained by implementing the generalized Einstein (mean square displacement) and Green-Kubo (velocity auto-correlation) methods. Field properties are obtained by sampling particle properties on a post-processing Eulerian grid using the smoothing function evaluation.

This work featured several modifications and revisions of existing algorithms in the SDPD-DV code as well as implementation of new algorithms in order to achieve a fully functional SDPD-DV code for wall-bounded mesoscopic computations. Specifically this work:

1. Revised the periodic boundary conditions algorithm and the implementation of the periodic boundary cells searching list.
2. Modified the evaluation of smoothing function in order to include the contribution of the self-density for each fluid particle.
3. Modified the dynamic virtual particle allocation algorithm for the modeling of solid boundary in order to minimize the truncation error of density.
4. Developed and implemented the boundary normal vector algorithm used in the reflection of the dynamically allocated virtual particles.
5. Implemented the algorithm for the contribution to the boundary force from the virtual particles.
6. Implemented a constant-wall temperature boundary condition in the dynamic virtual particle allocation model.
7. Implemented a bounce-forward algorithm for a solid boundary with arbitrary shape and orientation.
8. Revised the portions of the Velocity Verlet integration method for the position and momentum equations.
9. Developed and implemented a Runge–Kutta integration algorithm for the entropy equation.

10. Implemented the artificial incompressibility method for liquid flows.
11. Implemented a temperature power law for the shear viscosity, bulk viscosity and heat conductivity that appear in the momentum and energy SDPD equations.
12. Developed and implemented algorithms for the evaluation of transport properties such as diffusion coefficient, shear viscosity and heat conductivity based on mean square displacement (MSD) and velocity autocorrelation function (VACF).
13. Developed analytical formulas of the self-diffusion coefficient based on the SDPD-fluid following Litvinov et al. (2009).

Several benchmark tests were performed for validation and verification of the SDPD-DV method. The first verification of SDPD-DV involves the simulation of transient, planar Poiseuille body-force driven liquid water flow. The parallel plates are separated by a $10^{-3}$ m and $H_2O(l)$ at $T = 300$ K is driven by a body force per unit mass $f_x = 10^{-4}$ ms$^{-2}$. Numerical results show that density error is less than 4% in the interior and less than 5% near the walls. The velocity profiles are in very close agreement with theoretical ones. The second benchmark test involves transient, planar Couette water liquid flow at $T = 300$ K. The parallel plates are separated by $10^{-3}$ m with the top plate moving at $V_{xw} = 1.25 \times 10^{-5}$ ms$^{-1}$. The numerical velocity profiles are in very good agreement with analytical solutions. These two benchmark tests verify major algorithmic parts of our SDPD-DV implementation and demonstrate the ability of SDPD-DV to enforce no-slip boundary conditions, avoid particle penetration of walls, and reduce the density error near boundaries.

Further verification SDPD-DV is accomplished with a more challenging 3D benchmark test involving the body-force driven flow of $H_2O(l)$ at $T = 300$ K over a cylinder of radius $R = 0.02$ m. The SDPD-DV simulations are performed in a domain with $L_X = 0.1$ m, $L_Y = 0.015$ m, $L_Z = 0.1$ m and a body-force per unit mass of $f_x = 1.5 \times 10^{-7}$ ms$^{-2}$. FLUENT simulations are also performed in a lattice of cylinders with the flow driven by a $g_x = 1.5 \times 10^{-7}$ ms$^{-2}$. The SDPD-DV results are compared directly with FLUENT results and are in excellent agreement. This test further verifies the ability of SDPD-DV to simulate curved solid boundaries and the implementation of the artificial compressibility method in SDPD-DV.

Additional SDPD-DV verification and validation features an extensive set of simulations used to obtain the equilibrium state of $H_2O$ (l) and $N_2$(g) at $T = 300$ K, and the self-diffusion coefficient of $H_2O$ (l). In order to examine the scale-effects in SDPD-DV, the mass of the fluid particles for the H2O(l) SDPD-DV simulations is varied between 1.24 and $3.3\times10^7$ real molecular masses and their corresponding size is between 1.08 and 323 physical length scales. For the $N_2$(g) SDPD-DV simulations the mass of the fluid particles is varied between $6.37\times10^3$ and $6.37\times10^6$ real molecular masses and their corresponding size is between $2.2\times10^2$ and $2.2\times10^3$ physical length scales. The particle speeds are shown to depend on mass (or size) of the fluid particle. The average translational temperature from the SDPD-DV results is close to 300K and is scale-free for the range of particle masses (or sizes) considered. The self-diffusion coefficient for $H_2O$(l) is obtained using the MSD and VAC methods. Additional estimates are obtained from analytical expressions for the SDPD-fluid and values derived from the SDPD-DV simulations. Comparisons are also given with experimental data. The results show that MSD and VAC results are in very good agreement with the analytical SDPD estimates. The results also show that the self-diffusion coefficient from SDPD-DV is not scale-free and reaches the experimental value of $H_2O$(l) when the fluid particle mass asymptotically approaches the actual molecular size. The Schmidt numbers obtained from SDPD-DV are within the range expected for liquid water.

The full-set of SDPD equations implemented in SDPD-DV model is verified and validated with simulations in bounded and periodic domains that cover the hydrodynamic and mesoscopic regimes. Validation and verification is achieved with an extensive set of SDPD-DV simulations of gaseous nitrogen in mesoscopic periodic domains in equilibrium. The simulations of $N_2$(g) were performed in rectangular domains with $L_X = L_Y = L_Z$ in the range $0.25\times10^{-6}$~$10\times10^{-6}$ m, with mass $M_F$ in the range $1.85\times10^{-20}$~$1.184\times10^{-15}$kg. The self-diffusion coefficient for $N_2$(g) at equilibrium states is obtained through the mean-square displacement for the range of fluid particle masses (or sizes) and smoothing length considered. The simulations show that both of the fluid particle mass and smoothing length affect the self–diffusion coefficient which is not scale free. This is because the smaller the fluid particle is, the larger is its stochastic agitation. For a given mass ratio, increasing the smoothing length, increases self-diffusivity, since the stochastic agitation is stronger in the bigger support domain. The shear viscosity for SDPD-DV simulation of $N_2$(g) is shown to be scale free and is not affected by the choice of particle mass or the smoothing length. The impact of fluid particle mass, smoothing length and timesteps are also presented. The results show also that for larger fluid particles the stochastic agitation is smaller.

Additional verification involves SDPD-DV simulations of steady Couette $N_2$(g) flow. The top plate is moving at $V_{xw}$=30m/s and separated by $10^{-4}$ m from the bottom stationary plate. The top plate has a constant $T_1 = 330\text{K}$ and the bottom plate $T_2 = 300\text{K}$. The SDPD-DV field velocity and temperature profiles are in excellent agreement with those obtained by FLUENT. This benchmark test further verifies the ability of our SDPD-DV to simulate non-isothermal flows and the implementation of the thermodynamics properties of virtual particles in SDPD-DV.

## 5.2 Recommendations for Future Work

The limitations of the SDPD-DV method and code developed in the course of this research indicate several areas for future work.

1. The dynamic virtual particle allocation method discussed in this work can lead to an imperfect representation of uneven boundary. In view of the need for wide applications with arbitrary geometries, there is need for modeling a solid boundary with dihedral or trihedral angle. With the implementation of the dynamic virtual particle allocation method, the reflection algorithm on flat wall is very mature. For curved boundary, higher resolution is required to avoid large errors from reflection distortion. However, for dihedral or trihedral angle, a new algorithm is needed to compensate for this distortion.
2. Implementation of variable smoothing length would improve the accuracy of estimation with enough particles in smoothing length. Several algorithms with VSL can be found in SPH applications. Nevertheless, on a curved boundary the varying smoothing length would be larger than the radius that lead to incorrect representation.
3. Development and implementation of a free surface boundary model with the DVPA method. A moveable boundary particle could be employed to form the free surface, as well as the virtual particle making up the truncation area. In turns, the model of boundary particle and fluid particle interaction will need to be added to simulate free surface.
4. In case of more complicated boundary conditions, it is considerable to combine the dynamic virtual particle method and frozen ghost/virtual particle method.

A wide application of SDPD-DV depends on the ability of handling complex flows. Considering the current SDPD-DV solver, future work could include:

1. Development and implementation of a multiphase flow model. Particle interaction model need to be added to basic SDPD model for multiphase flow.
2. Development and implementation of multispecies flow model. In the current solver, multispecies flow model capability can be developed by adding the multispecies particle interaction model. The force interpolation function of the fluid particle with another species fluid particle in support domain, the average coefficients of each force term are needed.
3. Implementation of other forms of the continuity equation instead of summation form.
4. Development and implementation of inflow and outflow boundary condition. In order to address pressure driven flows, the inlet and outlet boundary condition can be added by employing a buffer region or dynamic virtual particles. With a buffer region, the particle in the outflow can be re-injected into buffer region and will be assigned new properties.
5. Implementation of heat flux boundary model. The heat flux boundary can be modeled by assigning different temperature to virtual particles. For example, with the same temperature as fluid particle, the virtual particle will ensure the zero heat flux on boundary. Assigning a constant temperature to virtual particle can model the heat bath.

In addition to the algorithmic developments future work should include:

1. Further study of the scale dependence of transport coefficients including the thermal conductivity.
2. Implementation of a local diffusivity algorithm. This can be done either in sampling process or during the iterations.

# APPENDIX A. SDPD Analytic Self-diffusion Coefficient

We follow the derivation procedure introduced by Litvinov et al. (2009) to form the 3D SDPD self-diffusion coefficient with Lucy function as the interpolation function. Review the SDPD equations, the hydrodynamic momentum equation in a Lagrangian description is given by de Groot and Mazur (1962)

$$\rho \frac{d\mathbf{v}}{dt} = -\nabla P + \eta \nabla^2 \mathbf{v} + \left( \zeta + \frac{\eta}{3} \right) \nabla\nabla \cdot \mathbf{v} . \tag{A.1}$$

According to the derivation of self-diffusion coefficient for the SDPD fluid particles in Litvinov et al. (2009), one start from the following by neglecting the conservative forces

$$\frac{d\mathbf{v}_i}{dt} = \frac{1}{m_i} \sum_{i \neq j} \mathbf{F}_D + \frac{1}{m_i} \sum_{i \neq j} \mathbf{F}_R . \tag{A.2}$$

Since the dissipative part is linear in velocity difference, the Eq (A.2) can be written as

$$\frac{d\mathbf{v}_i}{dt} + \frac{\mathbf{v}_i}{\gamma} = \frac{\mathbf{F}_R}{m_i}, \tag{A.3}$$

where, the coefficient

$$\frac{1}{\gamma} = -\left( \frac{4\eta}{3} + \zeta \right) \sum_{i \neq j} \frac{1}{m_i d_i d_j} \frac{\partial W_{ij}}{r_{ij} \partial r_{ij}} . \tag{A.4}$$

Assume the density and temperature are uniformly distributed, one can write Eq. (A.4) as following

$$\frac{1}{\gamma} = \frac{1}{m_i d_i^2} \sum_{i \neq j} \frac{\partial W_{ij}}{r_{ij} \partial r_{ij}} , \tag{A.5}$$

where, $m_i d_i = \rho_i$. For three dimensional cases, the kernel takes the form

$$W_{ij} = W\left(\mathbf{r}_{ij}, h\right) = \frac{1}{h^3} f\left( \frac{\left|\mathbf{r}_{ij}\right|}{h} \right). \tag{A.6}$$

Replacing the summation in Eq. (A.5) with integration, the procedure is as following

$$
\begin{aligned}
\frac{1}{\gamma} &= -\frac{1}{m_i d_i^2}\left(\frac{4\eta}{3}+\zeta\right)\sum_{i\neq j}\frac{\partial W_{ij}}{r_{ij}\partial r_{ij}} \\
&= \frac{V_i}{\rho_i^2}\left(\frac{4\eta}{3}+\zeta\right)\sum_{i\neq j}F_{ij} = \frac{1}{\rho_i}\left(\frac{4\eta}{3}+\zeta\right)\sum_{i\neq j}F_{ij}V_i = \frac{1}{\rho_i}\left(\frac{4\eta}{3}+\zeta\right)\int_{\Omega}F(r)\,dV \\
&= \frac{1}{\rho_i}\left(\frac{4\eta}{3}+\zeta\right)\int_0^\infty 4\pi r^2 F(r)\,dr = -\frac{4\pi}{\rho_i}\left(\frac{4\eta}{3}+\zeta\right)\int_0^\infty r\left[-rF(r)\right]dr \\
&= -\frac{4\pi}{\rho_i}\left(\frac{4\eta}{3}+\zeta\right)\left[rW(r)\Big|_0^\infty - \int_0^\infty W(r)\,dr\right] = \frac{4\pi}{\rho_i}\left(\frac{4\eta}{3}+\zeta\right)\int_0^\infty W(r)\,dr
\end{aligned}
$$

Then the coefficient would be

$$\frac{1}{\gamma} = \frac{4\pi}{\rho_i}\left(\frac{4\eta}{3}+\zeta\right)\int_0^\infty W(r)\,dr \tag{A.7}$$

For Lucy function, we have

$$\int_0^\infty W(r)\,dr = \int_0^\infty \frac{105}{16\pi h^3}\left(1+3\frac{r}{h}\right)\left(1-\frac{r}{h}\right)^3 dr = \frac{105}{16\pi h^2}\int_0^\infty \left(1+3\frac{r}{h}\right)\left(1-\frac{r}{h}\right)^3 d\,\frac{r}{h} = \frac{21}{8\pi h^2}$$

$$\int_0^\infty W(r)\,dr = \frac{21}{8\pi h^2} \tag{A.8}$$

yield

$$\frac{1}{\gamma} = \frac{4\pi}{\rho_i}\left(\frac{4\eta}{3}+\zeta\right)\frac{21}{8\pi h^2} \tag{A.9}$$

The solution of Eq. (A.3) gives the following expression for the diffusion coefficient

$$D = \frac{\gamma k_B T}{3m} \tag{A.10}$$

Substitute Eq. (A.9) to Eq.(A.10), we obtain

$$D = \frac{2h^2 k_B T \rho_i}{63 m_i\left(\frac{4\eta}{3}+\zeta\right)} \tag{A.11}$$

For incompressible flow, we set

$$\zeta = \frac{2}{3}\eta \qquad \text{(A.12)}$$

Then, Eq.(A.11) can be written as

$$D = \frac{h^2 k_B T \rho_i}{63 m_i} \qquad \text{(A.13)}$$

## APPENDIX B. Boundary Method in 3D with Lucy's Smoothing Function

In order to figure out a better way to correct truncation error, we used Vazquez-Quesada boundary condition method (2009a) and derived the formula for 3D using the Lucy function. The procedure is shown bellow.

For 3D SDPD, the corrected density $\bar{d}_i$ of the particle $i$ is given by:

$$\bar{d}_i = \sum_{j\in fluid} W\left(\mathbf{r}_{ij}\right) + \sum_{k\in wall} W\left(\mathbf{r}_i\right) = \sum_{j\in fluid} W\left(\mathbf{r}_{ij}\right) + \int_{\Omega} d\mathbf{r} n_r W\left(\left|\mathbf{r}-\mathbf{r}_i\right|\right) \tag{A.14}$$

where $\Omega$ denotes the missing volume of the particle $i$ or the region belongs to the wall in the smoothing region of particle $i$. And $n_{\mathbf{r}}$ is the distribution of wall particles.

$$n_{\mathbf{r}} = \sum_{k\in wall} \delta\left(\mathbf{r}_k - \mathbf{r}\right) \tag{A.15}$$

If $n_{\mathbf{r}}$ is equal to the corrected density $\bar{d}_i$, then

$$\begin{aligned} \bar{d}_i &= \sum_{j\in fluid} W\left(\mathbf{r}_{ij}\right) + \int_{\Omega} d\mathbf{r} n_r W\left(\left|\mathbf{r}-\mathbf{r}_i\right|\right) \\ &= \sum_{j\in fluid} W\left(\mathbf{r}_{ij}\right) + \bar{d}_i \int_{\Omega} d\mathbf{r} W\left(\left|\mathbf{r}-\mathbf{r}_i\right|\right) \end{aligned} \tag{A.16}$$

$$\bar{d}_i = \frac{\sum_{j\in fluid} W\left(\mathbf{r}_{ij}\right)}{1-\int_{\Omega} d\mathbf{r} W\left(\left|\mathbf{r}-\mathbf{r}_i\right|\right)} = \frac{d_i}{1-\Delta\left(h_i\right)} = \frac{1}{\bar{\mathcal{V}}_i} \tag{A.17}$$

where

$$\Delta\left(h_i\right) = \int_{\Omega} d\mathbf{r} W\left(\left|\mathbf{r}-\mathbf{r}_i\right|\right) \tag{A.18}$$

**Graph function $\Delta\left(h_i\right)$**

Integral equation (A.18)in the missing region $\Omega$

$$\Delta(h_i)=\iiint_{\Omega} d\boldsymbol{V}W\left(|\boldsymbol{r}-\boldsymbol{r}_i|\right)=\int_0^{2\pi}d\theta\int_0^{\arccos\bar{h}}\sin\varphi d\varphi\int_{r_0}^{r}drr^2W\left(|r-r_i|\right) \quad \text{(A.19)}$$

Set the origin at the particle $i$ and $R=\dfrac{|r-r_i|}{H}$, where $H$ is the smoothing length. Then

$$\begin{aligned}
\Delta(h_i)&=\int_0^{2\pi}d\varphi\int_0^{\arccos\bar{h}}\sin\theta d\theta\int_{\bar{h}/\cos\theta}^{1}dR\, H\left(RH\right)^2\ W\left(R\right)\\
&=\int_0^{2\pi}d\varphi\int_0^{\arccos\bar{h}}\sin\theta d\theta\int_{\bar{h}/\cos\theta}^{1}dR\, H\left(RH\right)^2\frac{105}{16\pi H^3}\left(1+3R\right)\left(1-R\right)^3\\
&=2\pi\frac{105}{16\pi H^3}H^3\int_0^{\arccos\bar{h}}\sin\theta d\theta\int_{\bar{h}/\cos\theta}^{1}dRR^2\left(1+3R\right)\left(1-R\right)^3\\
&=\frac{105}{8}\int_0^{\arccos\bar{h}}\sin\theta d\theta\int_{\bar{h}/\cos\theta}^{1}dRR^2\left(1-6R^2+8R^3-3R^4\right)\\
&=\frac{105}{8}\int_0^{\arccos\bar{h}}\sin\theta d\theta\int_{\bar{h}/\cos\theta}^{1}dR\left(R^2-6R^4+8R^5-3R^6\right)\\
&=\frac{105}{8}\int_0^{\arccos\bar{h}}d\theta\ \sin\theta\left(\frac{R^3}{3}-\frac{6R^5}{5}+\frac{4R^6}{3}-\frac{3R^7}{7}\right)\Bigg|_{\bar{h}/\cos\theta}^{1}\\
&=\frac{105}{8}\int_0^{\arccos\bar{h}}d\theta\ \sin\theta\left(\frac{4}{105}-\frac{\bar{h}^3}{3\cos^3\theta}+\frac{6\bar{h}^5}{5\cos^5\theta}-\frac{4\bar{h}^6}{3\cos^6\theta}+\frac{3\bar{h}^7}{7\cos^7\theta}\right)\\
&=\frac{105}{8}\int_0^{\arccos\bar{h}}-d\cos\theta\ \left(\frac{4}{105}-\frac{\bar{h}^3}{3\cos^3\theta}+\frac{6\bar{h}^5}{5\cos^5\theta}-\frac{4\bar{h}^6}{3\cos^6\theta}+\frac{3\bar{h}^7}{7\cos^7\theta}\right)\\
&=\frac{105}{8}\int_{\bar{h}}^{1}dr\ \left(\frac{4}{105}-\frac{\bar{h}^3}{3r^3}+\frac{6\bar{h}^5}{5r^5}-\frac{4\bar{h}^6}{3r^6}+\frac{3\bar{h}^7}{7r^7}\right)\\
&=\frac{105}{8}\left(\frac{4r}{105}+\frac{\bar{h}^3}{6r^2}-\frac{3\bar{h}^5}{10r^4}+\frac{4\bar{h}^6}{15r^5}-\frac{\bar{h}^7}{14r^6}\right)\Bigg|_{\bar{h}}^{1}\\
&=\frac{105}{8}\left(\frac{4}{105}-\frac{\bar{h}}{10}+\frac{\bar{h}^3}{6}-\frac{3\bar{h}^5}{10}+\frac{4\bar{h}^6}{15}-\frac{\bar{h}^7}{14}\right)
\end{aligned}$$

$$\Delta(h_i)=\begin{cases}\dfrac{1}{2}-\dfrac{21\bar{h}}{16}+\dfrac{35\bar{h}^3}{16}-\dfrac{63\bar{h}^5}{16}+\dfrac{7\bar{h}^6}{2}-\dfrac{15\bar{h}^7}{16} & \text{for } 0<r<H\\ 0 & \text{for } r>H\end{cases} \quad \text{(A.20)}$$

where $\bar{h}=h_i/H$

For 2D case, we rewrite and correct the derivation

$$
\begin{aligned}
\Delta(h_i) &= 2\int_0^{\arccos\bar{h}} d\theta \int_{\bar{h}/\cos\varphi}^{1} dR\ R\ W(R) \\
&= 2\frac{5}{\pi H^2} H^2 \int_0^{\arccos\bar{h}} d\theta \int_{\bar{h}/\cos\varphi}^{1} dR\ R(1+3R)(1-R)^3 \\
&= \frac{10}{\pi}\int_0^{\arccos\bar{h}} d\theta \left( \frac{1}{10} - \frac{\bar{h}^2}{2\cos^2\theta} + \frac{3\bar{h}^4}{2\cos^4\theta} - \frac{8\bar{h}^5}{5\cos^5\theta} + \frac{\bar{h}^6}{2\cos^6\theta} \right) \\
&= \frac{10}{\pi}\left\{ \begin{array}{l} \frac{\theta}{10} - \frac{\bar{h}^2\sin\theta}{2\cos\theta} + \left(\frac{\sin\theta}{2\cos^3\theta} + \frac{\sin\theta}{\cos\theta}\right)\bar{h}^4 - \left(\frac{2\sin\theta}{5\cos^4\theta} + \frac{3\sin\theta}{5\cos^2\theta} + \frac{3}{5}\ln\frac{1+\sin\theta}{\cos\theta}\right)\bar{h}^5 \\ +\left(\frac{\sin\theta}{10\cos^5\theta} + \frac{2\sin\theta}{15\cos^3\theta} + \frac{4\sin\theta}{15\cos\theta}\right)\bar{h}^6 \end{array} \right\}\Bigg|_0^{\arccos\bar{h}} \\
&= \frac{10}{\pi}\left( \begin{array}{l} \frac{\arccos\bar{h}}{10} - \frac{\bar{h}\sqrt{1-\bar{h}^2}}{2} + \frac{\bar{h}\sqrt{1-\bar{h}^2}}{2} + \bar{h}^3\sqrt{1-\bar{h}^2} - \frac{2\bar{h}\sqrt{1-\bar{h}^2}}{5} - \frac{3\bar{h}^3\sqrt{1-\bar{h}^2}}{5} \\ -\frac{3}{5}\bar{h}^5\ln\frac{1+\sqrt{1-\bar{h}^2}}{\bar{h}} + \frac{\bar{h}\sqrt{1-\bar{h}^2}}{10} + \frac{2\bar{h}^3\sqrt{1-\bar{h}^2}}{15} + \frac{4\bar{h}^5\sqrt{1-\bar{h}^2}}{15} \end{array} \right) \\
&= \frac{10}{\pi}\left( \frac{\arccos\bar{h}}{10} - \frac{3\bar{h}\sqrt{1-\bar{h}^2}}{10} + \frac{8\bar{h}^3\sqrt{1-\bar{h}^2}}{15} + \frac{4\bar{h}^5\sqrt{1-\bar{h}^2}}{15} - \frac{3}{5}\bar{h}^5\ln\frac{1+\sqrt{1-\bar{h}^2}}{\bar{h}} \right) \\
&= \frac{1}{6\pi}\left( 6\arccos\bar{h} - 18\bar{h}\sqrt{1-\bar{h}^2} + 32\bar{h}^3\sqrt{1-\bar{h}^2} + 16\bar{h}^5\sqrt{1-\bar{h}^2} - 36\bar{h}^5\ln\frac{1+\sqrt{1-\bar{h}^2}}{\bar{h}} \right)
\end{aligned}
$$

This can also be written as

$$
\Delta(h_i) = \begin{cases} \frac{1}{6\pi}\left[ \begin{array}{l} \left(-18\bar{h} + 32\bar{h}^3 + 16\bar{h}^5\right)\sqrt{1-\bar{h}^2} + 3\pi - 6\arcsin\bar{h} \\ +36\bar{h}^5\left(\ln\bar{h} - \ln\left(1+\sqrt{1-\bar{h}^2}\right)\right) \end{array} \right] & \text{for } 0 < r < H \\ 0 & \text{for } r > H \end{cases} \tag{A.21}
$$

which is exactly the same as Vazquez-Quesada $\Delta(h_i)$ function. We plot the comparison the value of $\Delta(h_i)$ in 2D and 3D.

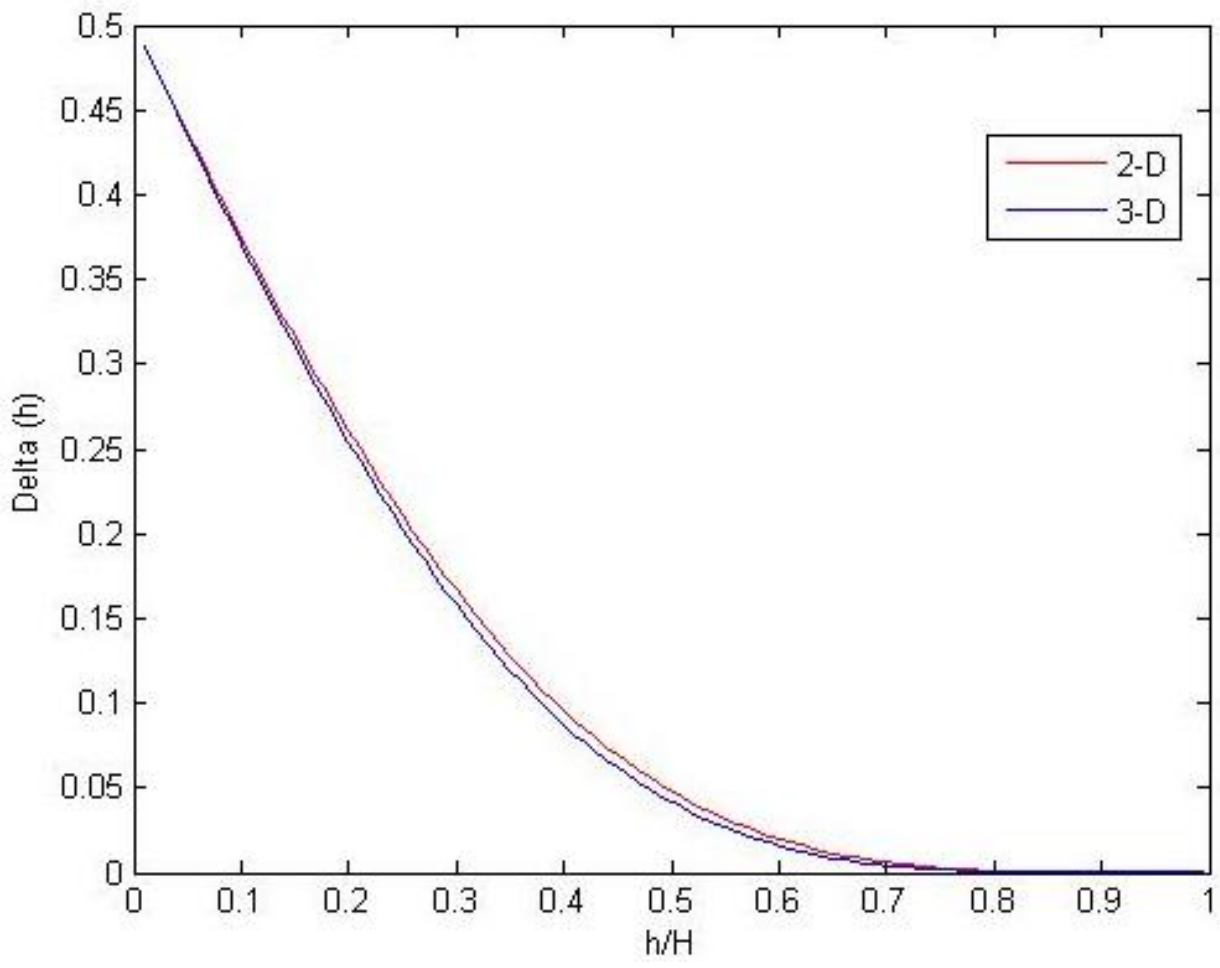

**Figure 31: Function $\Delta(h_i)$ for Lucy kernel in 2-D and 3-D.**

**Function $\psi(h)$**

Given that

$$\psi(h) = -\frac{\partial \Delta(h)}{\partial h} \tag{A.22}$$

In three dimension

$$\frac{\partial \Delta(h)}{\partial h} = \frac{\partial \Delta(h)}{\partial \bar{h}} \frac{\partial \bar{h}}{\partial h} = \frac{1}{H} \frac{\partial \Delta(h)}{\partial \bar{h}}$$

According to (A.20)

$$\frac{\partial \Delta(h)}{\partial \bar{h}} = \begin{cases} -\frac{21}{16} + \frac{105\bar{h}^2}{16} - \frac{315\bar{h}^4}{16} + 21\bar{h}^5 - \frac{105\bar{h}^6}{16} & \text{for } 0 < r < H \\ 0 & \text{for } r > H \end{cases}$$

Then the function $\psi(h)$ turns into

$$\psi(h) = \begin{cases} \frac{1}{H}\left( \frac{21}{16} - \frac{105\bar{h}^2}{16} + \frac{315\bar{h}^4}{16} - 21\bar{h}^5 + \frac{105\bar{h}^6}{16} \right) & \text{for } 0 < r < H \\ 0 & \text{for } r > H \end{cases} \tag{A.23}$$

The derivation is shown below:

$$
\begin{aligned}
\psi(h) &= \int_{\Omega} d\boldsymbol{r}\ F\left(\left|\boldsymbol{r}-\boldsymbol{r}_i\right|\right)\left(\boldsymbol{r}-\boldsymbol{r}_i\right)\cdot\boldsymbol{n} \\
&= \int_0^{2\pi} d\varphi \int_0^{\arccos\bar{h}} \sin\theta d\theta \int_{\bar{h}/\cos\theta}^{1} dR\ F(R)H(RH)^2(RH)\cos\theta \\
&= 2\pi \int_0^{\arccos\bar{h}} \sin\theta d\theta \int_{\bar{h}/\cos\theta}^{1} dR\ \frac{315}{4\pi H^5}(1-R)^2 H(RH)^2(RH)\cos\theta \\
&= \frac{315}{2H}\int_0^{\arccos\bar{h}} \sin\theta\cos\theta d\theta \int_{\bar{h}/\cos\theta}^{1} dR\ (1-R)^2 R^3 \\
&= \frac{315}{2H}\int_0^{\arccos\bar{h}} d\theta\ \sin\theta\cos\theta\left(\frac{1}{60}-\frac{\bar{h}^4}{4\cos^4\theta}+\frac{2\bar{h}^5}{5\cos^5\theta}-\frac{\bar{h}^6}{6\cos^6\theta}\right) \\
&= \frac{315}{2H}\int_{\bar{h}}^{1} dr\ r\left(\frac{1}{60}-\frac{\bar{h}^4}{4r^4}+\frac{2\bar{h}^5}{5r^5}-\frac{\bar{h}^6}{6r^6}\right) \\
&= \frac{315}{2H}\left(\frac{r^2}{120}+\frac{\bar{h}^4}{8r^2}-\frac{2\bar{h}^5}{15r^3}+\frac{\bar{h}^6}{24r^4}\right)\Bigg|_{\bar{h}}^{1} \\
&= \frac{315}{2H}\left(\frac{1}{120}-\frac{\bar{h}^2}{24}+\frac{\bar{h}^4}{8}-\frac{2\bar{h}^5}{15}+\frac{\bar{h}^6}{24}\right) \\
&= \frac{1}{H}\left(\frac{21}{16}-\frac{105\bar{h}^2}{16}+\frac{315\bar{h}^4}{16}-21\bar{h}^5+\frac{105\bar{h}^6}{16}\right)
\end{aligned}
$$

which is exactly equation (A.23)

**Function $\boldsymbol{\Psi}(h_i)$**

After substituting the velocity field of the wall particles, obtaining the irreversible part as

$$\dot{P}_i\Big|_{wall} \approx -a\psi(h_i)\frac{(v_i-V_{wall})}{\bar{d}_i h_i} - b\boldsymbol{\Psi}(h_i)\cdot\frac{(v_i-V_{wall})}{\bar{d}_i h_i} \tag{A.24}$$

$\psi(h)$ is the equation (A.23), and $\boldsymbol{\Psi}(h_i)$ is a tensor given by

$$\boldsymbol{\Psi}(h_i) = \int_{\Omega} d\boldsymbol{r}\ F\left(\left|\boldsymbol{r}_i-\boldsymbol{r}\right|\right)\ \frac{\boldsymbol{r}_i-\boldsymbol{r}}{\left|\boldsymbol{r}_i-\boldsymbol{r}\right|}\frac{\boldsymbol{r}_i-\boldsymbol{r}}{\left|\boldsymbol{r}_i-\boldsymbol{r}\right|}\ \left(\boldsymbol{r}_i-\boldsymbol{r}\right)\cdot\boldsymbol{n} \tag{A.25}$$

The tensor $\frac{\boldsymbol{r}_i-\boldsymbol{r}}{\left|\boldsymbol{r}_i-\boldsymbol{r}\right|}\frac{\boldsymbol{r}_i-\boldsymbol{r}}{\left|\boldsymbol{r}_i-\boldsymbol{r}\right|}$ in equation (A.25)can be expressed as $\frac{\boldsymbol{r}}{|\boldsymbol{r}|}\frac{\boldsymbol{r}}{|\boldsymbol{r}|}$, and

$$\frac{\boldsymbol{r}}{|\boldsymbol{r}|}\frac{\boldsymbol{r}}{|\boldsymbol{r}|}=\sum_{i=1}^{3}\sum_{j=1}^{3}\boldsymbol{\delta}_i\boldsymbol{\delta}_j r_i r_j$$

$$\begin{aligned}&=\boldsymbol{\delta}_1\boldsymbol{\delta}_1\sin^2\theta\cos^2\phi+\boldsymbol{\delta}_1\boldsymbol{\delta}_2\sin^2\theta\sin\phi\cos\phi+\boldsymbol{\delta}_1\boldsymbol{\delta}_3\sin\theta\cos\theta\cos\phi \\ &\quad+\boldsymbol{\delta}_2\boldsymbol{\delta}_1\sin^2\theta\sin\phi\cos\phi+\boldsymbol{\delta}_2\boldsymbol{\delta}_2\sin^2\theta\sin^2\phi+\boldsymbol{\delta}_2\boldsymbol{\delta}_3\sin\theta\cos\theta\sin\phi \\ &\quad+\boldsymbol{\delta}_3\boldsymbol{\delta}_1\sin\theta\cos\theta\cos\phi+\boldsymbol{\delta}_3\boldsymbol{\delta}_2\sin\theta\cos\theta\sin\phi+\boldsymbol{\delta}_3\boldsymbol{\delta}_3\cos^2\theta\end{aligned} \tag{A.26}$$

where $\boldsymbol{\delta}_i\boldsymbol{\delta}_j$ is the unit dyad. We express $\boldsymbol{\Psi}(h_i)$ as

$$\begin{aligned}\boldsymbol{\Psi}(h_i)&=\boldsymbol{\Psi}_{\boldsymbol{\delta}_1\boldsymbol{\delta}_1}(h_i)+\boldsymbol{\Psi}_{\boldsymbol{\delta}_2\boldsymbol{\delta}_2}(h_i)+\boldsymbol{\Psi}_{\boldsymbol{\delta}_3\boldsymbol{\delta}_3}(h_i) \\ &\quad+\boldsymbol{\Psi}_{\boldsymbol{\delta}_1\boldsymbol{\delta}_2}(h_i)+\boldsymbol{\Psi}_{\boldsymbol{\delta}_2\boldsymbol{\delta}_1}(h_i)+\boldsymbol{\Psi}_{\boldsymbol{\delta}_1\boldsymbol{\delta}_3}(h_i)+\boldsymbol{\Psi}_{\boldsymbol{\delta}_3\boldsymbol{\delta}_1}(h_i) \\ &\quad+\boldsymbol{\Psi}_{\boldsymbol{\delta}_2\boldsymbol{\delta}_3}(h_i)+\boldsymbol{\Psi}_{\boldsymbol{\delta}_3\boldsymbol{\delta}_2}(h_i)\end{aligned} \tag{A.27}$$

The integral form is

$$\begin{aligned}\boldsymbol{\Psi}(h_i)&=\int_{\Omega}d\boldsymbol{r}\;F(|\boldsymbol{r}_i-\boldsymbol{r}|)\;\frac{\boldsymbol{r}_i-\boldsymbol{r}}{|\boldsymbol{r}_i-\boldsymbol{r}|}\frac{\boldsymbol{r}_i-\boldsymbol{r}}{|\boldsymbol{r}_i-\boldsymbol{r}|}\;(\boldsymbol{r}_i-\boldsymbol{r})\cdot\boldsymbol{n} \\ &=\int_0^{2\pi}d\phi\int_0^{\arccos\bar{h}}\sin\theta d\theta\int_{\bar{h}/\cos\theta}^{1}dR\;H(RH)^2(RH)\cos\theta\;F(R)\frac{\boldsymbol{r}}{|\boldsymbol{r}|}\frac{\boldsymbol{r}}{|\boldsymbol{r}|} \\ &=\frac{315}{4\pi H}\int_0^{2\pi}d\phi\int_0^{\arccos\bar{h}}\sin\theta\cos\theta d\theta\int_{\bar{h}/\cos\theta}^{1}dR\;R^3\;(1-R)^2\frac{\boldsymbol{r}}{|\boldsymbol{r}|}\frac{\boldsymbol{r}}{|\boldsymbol{r}|}\end{aligned}$$

$\frac{\boldsymbol{r}}{|\boldsymbol{r}|}\frac{\boldsymbol{r}}{|\boldsymbol{r}|}$ is only the function of $\theta$ and $\phi$, so

$$\begin{aligned}\boldsymbol{\Psi}(h_i)&=\frac{315}{4\pi H}\int_0^{2\pi}d\phi\int_0^{\arccos\bar{h}}\frac{\boldsymbol{r}}{|\boldsymbol{r}|}\frac{\boldsymbol{r}}{|\boldsymbol{r}|}\sin\theta\cos\theta d\theta\int_{\bar{h}/\cos\theta}^{1}dR\;R^3\;(1-R)^2 \\ &=\frac{315}{4\pi H}\int_0^{2\pi}d\phi\int_0^{\arccos\bar{h}}d\theta\;\frac{\boldsymbol{r}}{|\boldsymbol{r}|}\frac{\boldsymbol{r}}{|\boldsymbol{r}|}\sin\theta\cos\theta\;\left(\frac{1}{60}-\frac{\bar{h}^4}{4\cos^4\theta}+\frac{2\bar{h}^5}{5\cos^5\theta}-\frac{\bar{h}^6}{6\cos^6\theta}\right)\end{aligned} \tag{A.28}$$

Integrate equation (A.28) by each dyad. For $\boldsymbol{\delta}_1\boldsymbol{\delta}_1$

$$\boldsymbol{\Psi}_{\delta_1\delta_1}(h_i)=\frac{315}{4\pi H}\int_0^{2\pi}d\phi\int_0^{\arccos\bar{h}}d\theta\ \sin^2\theta\cos^2\phi\sin\theta\cos\theta\ \left(\frac{1}{60}-\frac{\bar{h}^4}{4\cos^4\theta}+\frac{2\bar{h}^5}{5\cos^5\theta}-\frac{\bar{h}^6}{6\cos^6\theta}\right)$$

$$=\frac{315}{4\pi H}\int_0^{2\pi}\cos^2\phi d\phi\int_0^{\arccos\bar{h}}d\theta\ \sin^3\theta\cos\theta\ \left(\frac{1}{60}-\frac{\bar{h}^4}{4\cos^4\theta}+\frac{2\bar{h}^5}{5\cos^5\theta}-\frac{\bar{h}^6}{6\cos^6\theta}\right)$$

$$=\frac{315}{4\pi H}\left[\frac{1}{2}(\phi+\sin\phi\cos\phi)\right]_0^{2\pi}\int_0^{\arccos\bar{h}}-d\cos\theta\ \left(1-\cos^2\theta\right)\cos\theta\ \left(\frac{1}{60}-\frac{\bar{h}^4}{4\cos^4\theta}+\frac{2\bar{h}^5}{5\cos^5\theta}-\frac{\bar{h}^6}{6\cos^6\theta}\right)$$

Where $r=\cos\theta$

$$\boldsymbol{\Psi}_{\delta_1\delta_1}(h_i)=\frac{315}{4\pi H}\pi\int_{\bar{h}}^1 dr\ \left(1-r^2\right)r\ \left(\frac{1}{60}-\frac{\bar{h}^4}{4r^4}+\frac{2\bar{h}^5}{5r^5}-\frac{\bar{h}^6}{6r^6}\right)$$

$$=\frac{315}{4H}\int_{\bar{h}}^1 dr\ \left(1-r^2\right)\left(\frac{r}{60}-\frac{\bar{h}^4}{4r^3}+\frac{2\bar{h}^5}{5r^4}-\frac{\bar{h}^6}{6r^5}\right)$$

$$=\frac{315}{4H}\int_{\bar{h}}^1 dr\ \left(\frac{r}{60}-\frac{\bar{h}^4}{4r^3}+\frac{2\bar{h}^5}{5r^4}-\frac{\bar{h}^6}{6r^5}-\frac{r^3}{60}+\frac{\bar{h}^4}{4r}-\frac{2\bar{h}^5}{5r^2}+\frac{\bar{h}^6}{6r^3}\right)$$

$$=\frac{315}{4H}\int_{\bar{h}}^1 dr\ \left(\frac{r}{60}-\frac{r^3}{60}+\frac{\bar{h}^4}{4r}-\frac{2\bar{h}^5}{5r^2}+\left(\frac{\bar{h}^6}{6}-\frac{\bar{h}^4}{4}\right)\frac{1}{r^3}+\frac{2\bar{h}^5}{5r^4}-\frac{\bar{h}^6}{6r^5}\right)$$

$$=\frac{315}{4H}\left(\frac{r^2}{120}-\frac{r^4}{240}+\frac{\bar{h}^4}{4}\ln r+\frac{2\bar{h}^5}{5r}-\left(\frac{\bar{h}^6}{6}-\frac{\bar{h}^4}{4}\right)\frac{1}{2r^2}-\frac{2\bar{h}^5}{15r^3}+\frac{\bar{h}^6}{24r^4}\right)_{\bar{h}}^1$$

$$=\frac{315}{4H}\left(\begin{array}{l}\frac{1}{120}-\frac{1}{240}+\frac{2\bar{h}^5}{5}-\left(\frac{\bar{h}^6}{12}-\frac{\bar{h}^4}{8}\right)-\frac{2\bar{h}^5}{15}+\frac{\bar{h}^6}{24}\\ -\frac{\bar{h}^2}{120}+\frac{\bar{h}^4}{240}-\frac{\bar{h}^4}{4}\ln\bar{h}-\frac{2\bar{h}^4}{5}+\frac{\bar{h}^4}{12}-\frac{\bar{h}^2}{8}+\frac{2\bar{h}^2}{15}-\frac{\bar{h}^2}{24}\end{array}\right)$$

Yield, one can get

$$\boldsymbol{\Psi}_{\delta_1\delta_1}(h_i)=\frac{315}{4H}\left(\frac{1}{240}-\frac{\bar{h}^2}{24}-\frac{3\bar{h}^4}{16}+\frac{4\bar{h}^5}{15}-\frac{\bar{h}^6}{24}-\frac{\bar{h}^4}{4}\ln\bar{h}\right)\boldsymbol{\delta}_1\boldsymbol{\delta}_1 \qquad \text{(A.29)}$$

For $\boldsymbol{\delta}_2\boldsymbol{\delta}_2$

$$\boldsymbol{\Psi}_{\delta_2\delta_2}(h_i) = \frac{315}{4\pi H}\int_0^{2\pi} d\phi \int_0^{\arccos\bar{h}} d\theta\ \sin^2\theta \sin^2\phi \sin\theta\cos\theta \left(\frac{1}{60} - \frac{\bar{h}^4}{4\cos^4\theta} + \frac{2\bar{h}^5}{5\cos^5\theta} - \frac{\bar{h}^6}{6\cos^6\theta}\right)$$

$$= \frac{315}{4\pi H}\int_0^{2\pi} \sin^2\phi d\phi \int_0^{\arccos\bar{h}} d\theta\ \sin^3\theta\cos\theta \left(\frac{1}{60} - \frac{\bar{h}^4}{4\cos^4\theta} + \frac{2\bar{h}^5}{5\cos^5\theta} - \frac{\bar{h}^6}{6\cos^6\theta}\right)$$

$$= \frac{315}{4\pi H}\left[\frac{1}{2}(\phi - \sin\phi\cos\phi)\right]_0^{2\pi} \int_0^{\arccos\bar{h}} -d\cos\theta\ \left(1-\cos^2\theta\right)\cos\theta \begin{pmatrix} \frac{1}{60} - \frac{\bar{h}^4}{4\cos^4\theta} \\ + \frac{2\bar{h}^5}{5\cos^5\theta} - \frac{\bar{h}^6}{6\cos^6\theta} \end{pmatrix}$$

$$= \frac{315}{4\pi H}\pi\int_{\bar{h}}^{1} dr\ \left(1-r^2\right) r \left(\frac{1}{60} - \frac{\bar{h}^4}{4r^4} + \frac{2\bar{h}^5}{5r^5} - \frac{\bar{h}^6}{6r^6}\right) \qquad r = \cos\theta$$

$$= \frac{315}{4H}\int_{\bar{h}}^{1} dr\ \left(1-r^2\right) \left(\frac{r}{60} - \frac{\bar{h}^4}{4r^3} + \frac{2\bar{h}^5}{5r^4} - \frac{\bar{h}^6}{6r^5}\right)$$

$$= \frac{315}{4H}\left(\frac{1}{240} - \frac{\bar{h}^2}{24} - \frac{3\bar{h}^4}{16} + \frac{4\bar{h}^5}{15} - \frac{\bar{h}^6}{24} - \frac{\bar{h}^4}{4}\ln\bar{h}\right)$$

The integral result is

$$\boldsymbol{\Psi}_{\delta_2\delta_2}(h_i) = \frac{315}{4H}\left(\frac{1}{240} - \frac{\bar{h}^2}{24} - \frac{3\bar{h}^4}{16} + \frac{4\bar{h}^5}{15} - \frac{\bar{h}^6}{24} - \frac{\bar{h}^4}{4}\ln\bar{h}\right)\boldsymbol{\delta}_2\boldsymbol{\delta}_2 \qquad \text{(A.30)}$$

which gives $\boldsymbol{\Psi}_{\delta_2\delta_2}(h_i) = \boldsymbol{\Psi}_{\delta_1\delta_1}(h_i)$

For $\boldsymbol{\delta}_3\boldsymbol{\delta}_3$

$$\boldsymbol{\Psi}_{\delta_3\delta_3}(h_i) = \frac{315}{4\pi H}\int_0^{2\pi} d\phi \int_0^{\arccos\bar{h}} d\theta\ \cos^2\theta \sin\theta\cos\theta \left(\frac{1}{60} - \frac{\bar{h}^4}{4\cos^4\theta} + \frac{2\bar{h}^5}{5\cos^5\theta} - \frac{\bar{h}^6}{6\cos^6\theta}\right)$$

$$= \frac{315}{4\pi H}\int_0^{2\pi} d\phi \int_0^{\arccos\bar{h}} d\theta\ \sin\theta\cos^3\theta \left(\frac{1}{60} - \frac{\bar{h}^4}{4\cos^4\theta} + \frac{2\bar{h}^5}{5\cos^5\theta} - \frac{\bar{h}^6}{6\cos^6\theta}\right)$$

$$= \frac{315}{4\pi H}2\pi\int_0^{\arccos\bar{h}} -d\cos\theta\ \cos^3\theta \left(\frac{1}{60} - \frac{\bar{h}^4}{4\cos^4\theta} + \frac{2\bar{h}^5}{5\cos^5\theta} - \frac{\bar{h}^6}{6\cos^6\theta}\right)$$

Rearranging

$$\boldsymbol{\Psi}_{\delta_3\delta_3}(h_i) = \frac{315}{2H}\left(\frac{1}{240} + \frac{5\bar{h}^4}{16} - \frac{2\bar{h}^5}{5} + \frac{\bar{h}^6}{12} + \frac{\bar{h}^4}{4}\ln\bar{h}\right)\boldsymbol{\delta}_3\boldsymbol{\delta}_3 \qquad \text{(A.31)}$$

For $\boldsymbol{\delta_1\delta_2}$

$$\boldsymbol{\Psi}_{\delta_1\delta_2}\left(h_i\right)=\frac{315}{4\pi H}\int_0^{2\pi}d\phi\int_0^{\arccos\bar{h}}d\theta\ \sin^2\theta\sin\phi\cos\phi\sin\theta\cos\theta\ \left(\frac{1}{60}-\frac{\bar{h}^4}{4\cos^4\theta}+\frac{2\bar{h}^5}{5\cos^5\theta}-\frac{\bar{h}^6}{6\cos^6\theta}\right)$$
$$=\frac{315}{4\pi H}\int_0^{2\pi}\sin\phi\cos\phi d\phi\int_0^{\arccos\bar{h}}d\theta\ \sin^3\theta\cos\theta\ \left(\frac{1}{60}-\frac{\bar{h}^4}{4\cos^4\theta}+\frac{2\bar{h}^5}{5\cos^5\theta}-\frac{\bar{h}^6}{6\cos^6\theta}\right)$$
$$=\frac{315}{4\pi H}0\int_0^{\arccos\bar{h}}d\theta\ \sin^3\theta\cos\theta\ \left(\frac{1}{60}-\frac{\bar{h}^4}{4\cos^4\theta}+\frac{2\bar{h}^5}{5\cos^5\theta}-\frac{\bar{h}^6}{6\cos^6\theta}\right)$$
$$=0$$

for $r=\cos\theta$

$$\boldsymbol{\Psi}_{\delta_1\delta_2}\left(h_i\right)=\frac{315}{2H}\int_{\bar{h}}^{1}dr\ r^3\left(\frac{1}{60}-\frac{\bar{h}^4}{4r^4}+\frac{2\bar{h}^5}{5r^5}-\frac{\bar{h}^6}{6r^6}\right)$$
$$=\frac{315}{2H}\int_{\bar{h}}^{1}dr\ \left(\frac{r^3}{60}-\frac{\bar{h}^4}{4r}+\frac{2\bar{h}^5}{5r^2}-\frac{\bar{h}^6}{6r^3}\right)$$
$$=\frac{315}{2H}\left(\frac{r^4}{240}-\frac{\bar{h}^4}{4}\ln r-\frac{2\bar{h}^5}{5r}+\frac{\bar{h}^6}{12r^2}\right)_{\bar{h}}^{1}$$
$$=\frac{315}{2H}\left(\frac{1}{240}-\frac{2\bar{h}^5}{5}+\frac{\bar{h}^6}{12}-\frac{\bar{h}^4}{240}+\frac{\bar{h}^4}{4}\ln\bar{h}+\frac{2\bar{h}^4}{5}-\frac{\bar{h}^4}{12}\right)$$

For $\boldsymbol{\delta_2\delta_1}$

$$\boldsymbol{\Psi}_{\delta_2\delta_1}\left(h_i\right)=\boldsymbol{\Psi}_{\delta_1\delta_2}\left(h_i\right)=0 \qquad \text{(A.32)}$$

For $\boldsymbol{\delta_1\delta_3}$

$$\boldsymbol{\Psi}_{\delta_1\delta_3}\left(h_i\right)=\frac{315}{4\pi H}\int_0^{2\pi}d\phi\int_0^{\arccos\bar{h}}d\theta\ \sin\theta\cos\theta\cos\phi\sin\theta\cos\theta\ \left(\frac{1}{60}-\frac{\bar{h}^4}{4\cos^4\theta}+\frac{2\bar{h}^5}{5\cos^5\theta}-\frac{\bar{h}^6}{6\cos^6\theta}\right)$$
$$=\frac{315}{4\pi H}\int_0^{2\pi}\cos\phi d\phi\int_0^{\arccos\bar{h}}d\theta\ \sin^2\theta\cos^2\theta\ \left(\frac{1}{60}-\frac{\bar{h}^4}{4\cos^4\theta}+\frac{2\bar{h}^5}{5\cos^5\theta}-\frac{\bar{h}^6}{6\cos^6\theta}\right)$$
$$=\frac{315}{4\pi H}0\int_0^{\arccos\bar{h}}d\theta\ \sin^2\theta\cos^2\theta\ \left(\frac{1}{60}-\frac{\bar{h}^4}{4\cos^4\theta}+\frac{2\bar{h}^5}{5\cos^5\theta}-\frac{\bar{h}^6}{6\cos^6\theta}\right)$$
$$=0$$

For $\boldsymbol{\delta}_3\boldsymbol{\delta}_1$

$$\boldsymbol{\Psi}_{\boldsymbol{\delta}_3\boldsymbol{\delta}_1}\left(h_i\right)=\boldsymbol{\Psi}_{\boldsymbol{\delta}_1\boldsymbol{\delta}_3}\left(h_i\right)=0 \tag{A.33}$$

For $\boldsymbol{\delta}_2\boldsymbol{\delta}_3$

$$\begin{aligned}
\boldsymbol{\Psi}_{\boldsymbol{\delta}_1\boldsymbol{\delta}_3}\left(h_i\right)&=\frac{315}{4\pi H}\int_0^{2\pi}d\phi\int_0^{\arccos\bar{h}}d\theta\ \sin\theta\cos\theta\sin\phi\sin\theta\cos\theta\ \left(\frac{1}{60}-\frac{\bar{h}^4}{4\cos^4\theta}+\frac{2\bar{h}^5}{5\cos^5\theta}-\frac{\bar{h}^6}{6\cos^6\theta}\right)\\
&=\frac{315}{4\pi H}\int_0^{2\pi}\sin\phi d\phi\int_0^{\arccos\bar{h}}d\theta\ \sin^2\theta\cos^2\theta\ \left(\frac{1}{60}-\frac{\bar{h}^4}{4\cos^4\theta}+\frac{2\bar{h}^5}{5\cos^5\theta}-\frac{\bar{h}^6}{6\cos^6\theta}\right)\\
&=\frac{315}{4\pi H}0\int_0^{\arccos\bar{h}}d\theta\ \sin^2\theta\cos^2\theta\ \left(\frac{1}{60}-\frac{\bar{h}^4}{4\cos^4\theta}+\frac{2\bar{h}^5}{5\cos^5\theta}-\frac{\bar{h}^6}{6\cos^6\theta}\right)\\
&=0
\end{aligned}$$

For dyad $\boldsymbol{\delta}_3\boldsymbol{\delta}_2$

$$\boldsymbol{\Psi}_{\boldsymbol{\delta}_3\boldsymbol{\delta}_2}\left(h_i\right)=\boldsymbol{\Psi}_{\boldsymbol{\delta}_2\boldsymbol{\delta}_3}\left(h_i\right)=0 \tag{A.34}$$

Substitute all components of $\boldsymbol{\Psi}\left(h_i\right)$ into equation(A.27)

$$\begin{aligned}
\boldsymbol{\Psi}\left(h_i\right)&=\boldsymbol{\Psi}_{\boldsymbol{\delta}_1\boldsymbol{\delta}_1}\left(h_i\right)+\boldsymbol{\Psi}_{\boldsymbol{\delta}_2\boldsymbol{\delta}_2}\left(h_i\right)+\boldsymbol{\Psi}_{\boldsymbol{\delta}_3\boldsymbol{\delta}_3}\left(h_i\right)+\boldsymbol{\Psi}_{\boldsymbol{\delta}_1\boldsymbol{\delta}_2}\left(h_i\right)+\boldsymbol{\Psi}_{\boldsymbol{\delta}_2\boldsymbol{\delta}_1}\left(h_i\right)\\
&\quad+\boldsymbol{\Psi}_{\boldsymbol{\delta}_1\boldsymbol{\delta}_3}\left(h_i\right)+\boldsymbol{\Psi}_{\boldsymbol{\delta}_3\boldsymbol{\delta}_1}\left(h_i\right)+\boldsymbol{\Psi}_{\boldsymbol{\delta}_2\boldsymbol{\delta}_3}\left(h_i\right)+\boldsymbol{\Psi}_{\boldsymbol{\delta}_3\boldsymbol{\delta}_2}\left(h_i\right)\\
&=\frac{315}{4H}\left(\frac{1}{240}-\frac{\bar{h}^2}{24}-\frac{3\bar{h}^4}{16}+\frac{4\bar{h}^5}{15}-\frac{\bar{h}^6}{24}-\frac{\bar{h}^4}{4}\ln\bar{h}\right)\boldsymbol{\delta}_1\boldsymbol{\delta}_1\\
&\quad+\frac{315}{4H}\left(\frac{1}{240}-\frac{\bar{h}^2}{24}-\frac{3\bar{h}^4}{16}+\frac{4\bar{h}^5}{15}-\frac{\bar{h}^6}{24}-\frac{\bar{h}^4}{4}\ln\bar{h}\right)\boldsymbol{\delta}_2\boldsymbol{\delta}_2\\
&\quad+\frac{315}{2H}\left(\frac{1}{240}+\frac{5\bar{h}^4}{16}-\frac{2\bar{h}^5}{5}+\frac{\bar{h}^6}{12}+\frac{\bar{h}^4}{4}\ln\bar{h}\right)\boldsymbol{\delta}_3\boldsymbol{\delta}_3
\end{aligned}$$

Here we set $\boldsymbol{nn}=\boldsymbol{\delta}_3\boldsymbol{\delta}_3$, and $\boldsymbol{\Psi}\left(h_i\right)$ can be written as

$$\boldsymbol{\Psi}\left(h_i\right)=\psi_1\left(h_i\right)\boldsymbol{1}+\psi_2\left(h_i\right)\boldsymbol{nn} \tag{A.35}$$

where

$$\psi_1(h_i)=\begin{cases}\frac{315}{4H}\left(\frac{1}{240}-\frac{\bar{h}^2}{24}-\frac{3\bar{h}^4}{16}+\frac{4\bar{h}^5}{15}-\frac{\bar{h}^6}{24}-\frac{\bar{h}^4}{4}\ln\bar{h}\right) & 0<r<H\\ 0 & r>H\end{cases} \qquad \text{(A.36)}$$

and

$$\psi_2(h_i)=\begin{cases}\frac{315}{2H}\left(\frac{1}{480}+\frac{\bar{h}^2}{48}+\frac{13\bar{h}^4}{32}-\frac{8\bar{h}^5}{15}+\frac{5\bar{h}^6}{48}+\frac{3\bar{h}^4}{8}\ln\bar{h}\right) & 0<r<H\\ 0 & r>H\end{cases} \qquad \text{(A.37)}$$

The functions $\psi(h_i)$, $\psi_1(h_i)$, and $\psi_2(h_i)$ are plotted in Figure 32-Figure 34.

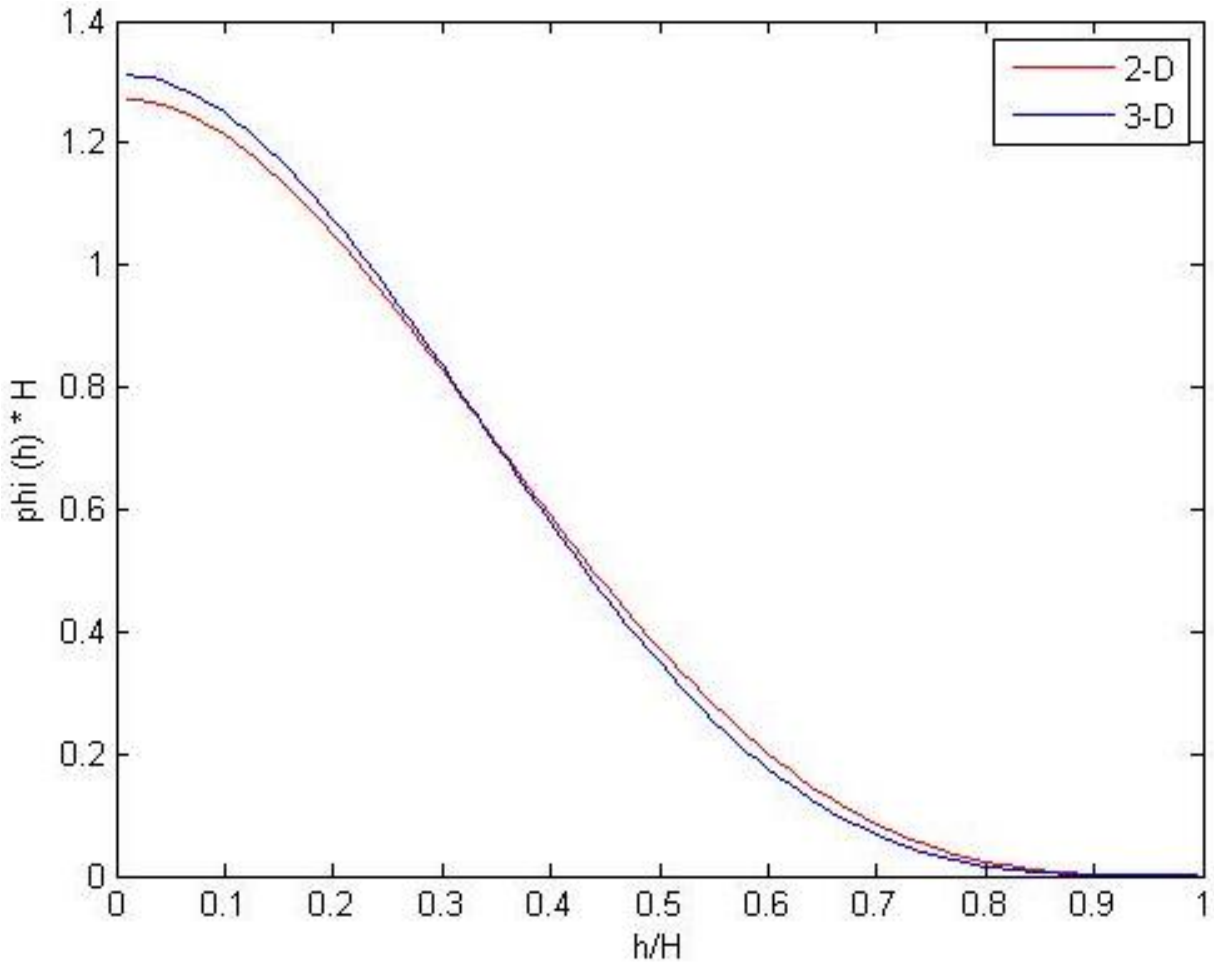

**Figure 32: Function $\psi(h_i)$ for Lucy kernel in 2-D and 3-D.**

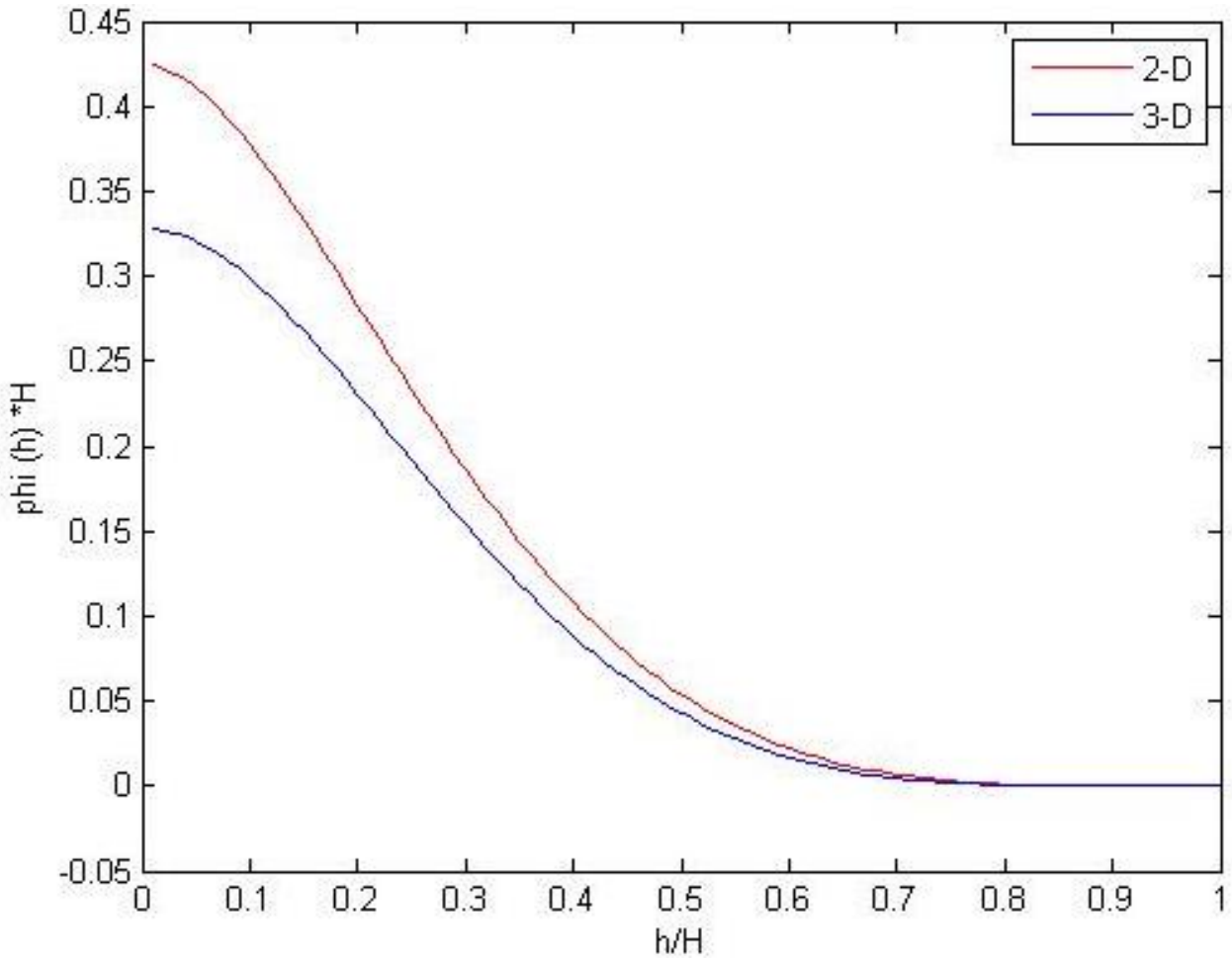

**Figure 33: Function $\psi_1(h_i)$ for Lucy kernel in 2-D and 3-D.**

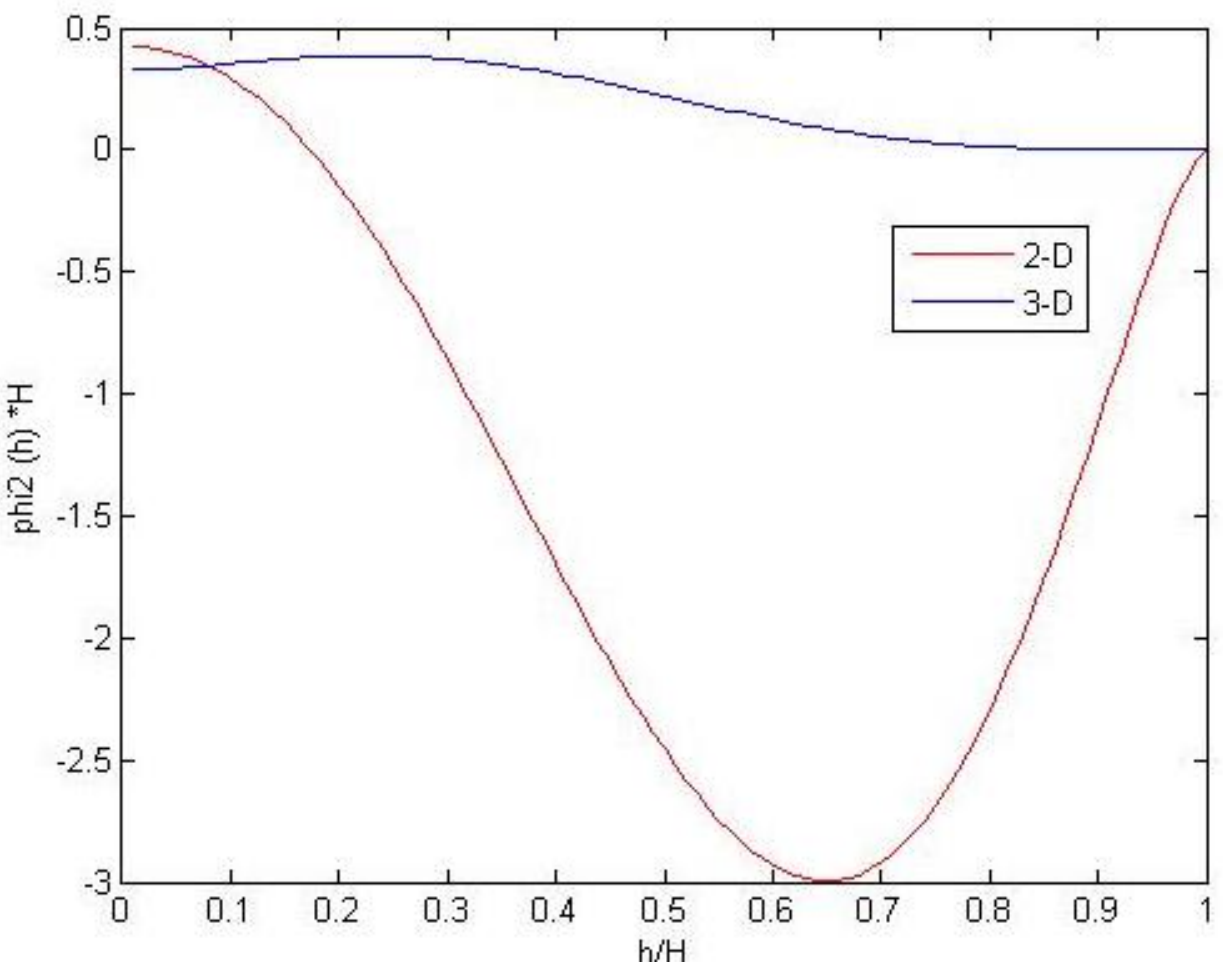

**Figure 34: Function $\psi_2\left(h_i\right)$ for Lucy kernel in 2-D (Vazquez-Quesada formulation) and in 3-D.**

For the 2-D case

$$
\begin{aligned}
rF\left(r\right) &= -\frac{\partial W\left(r\right)}{\partial r} = -\frac{\partial\left(\frac{5}{\pi h^2}\left(1+3\frac{r}{h}\right)\left(1-\frac{r}{h}\right)^3\right)}{\partial r} \\
&= \frac{5}{\pi h^2}\frac{12r}{h^2}\left(1-\frac{r}{h}\right)^2 \\
&= \frac{60r}{\pi h^4}\left(1-\frac{r}{h}\right)^2
\end{aligned}
$$

$$
\begin{aligned}
\boldsymbol{\Psi}_{\delta_1\delta_1}\left(h_i\right) &= 2\int_0^{\arccos\bar{h}} d\theta \int_{\bar{h}/\cos\theta}^{1} d\left(RH\right)\ \left(RH\right)\ \left(RH\right)\cos\theta F\left(R\right)\sin^2\theta \\
&= 2\int_0^{\arccos\bar{h}} \sin^2\theta\cos\theta d\theta \int_{\bar{h}/\cos\theta}^{1} dR\ R^2 H^3\ F\left(R\right) \\
&= 2\frac{60H^3}{\pi H^4}\int_0^{\arccos\bar{h}} \sin^2\theta\cos\theta d\theta \int_{\bar{h}/\cos\theta}^{1} dR\ R^2\ \left(1-R\right)^2 \\
&= \frac{120}{\pi H}\int_0^{\arccos\bar{h}} \sin^2\theta\cos\theta d\theta \int_{\bar{h}/\cos\theta}^{1} dR\ R^2\ \left(1-R\right)^2 \\
&= \frac{120}{\pi H}\int_0^{\arccos\bar{h}} d\theta\left(1-\cos^2\theta\right)\cos\theta\left(\frac{1}{30}-\frac{\bar{h}^3}{3\cos^3\theta}+\frac{\bar{h}^4}{2\cos^4\theta}-\frac{\bar{h}^5}{5\cos^5\theta}\right)
\end{aligned}
$$

$$\mathbf{\Psi}_{\boldsymbol{\delta}_1\boldsymbol{\delta}_1}\left(h_i\right) = \frac{120}{\pi H}\int_0^{\arccos\bar{h}} d\theta \left( \begin{array}{l} \dfrac{\bar{h}^3}{3} + \dfrac{\cos\theta}{30} - \dfrac{\cos^3\theta}{30} - \dfrac{\bar{h}^4}{2\cos\theta} + \left(\dfrac{\bar{h}^5}{5} - \dfrac{\bar{h}^3}{3}\right)\dfrac{1}{\cos^2\theta} \\ + \dfrac{\bar{h}^4}{2\cos^3\theta} - \dfrac{\bar{h}^5}{5\cos^4\theta} \end{array} \right)$$

$$= \frac{120}{\pi H}\left( \begin{array}{l} \dfrac{\theta\bar{h}^3}{3} + \dfrac{\sin\theta}{30} - \dfrac{\sin\theta}{30} + \dfrac{\sin^3\theta}{90} - \dfrac{\bar{h}^4}{2}\ln\left(\dfrac{1+\sin\theta}{\cos\theta}\right) + \dfrac{\bar{h}^5\sin\theta}{5\cos\theta} - \dfrac{\bar{h}^3\sin\theta}{3\cos\theta} \\ + \dfrac{\bar{h}^4\sin\theta}{4\cos^2\theta} + \dfrac{\bar{h}^4}{4}\ln\left|\dfrac{1+\sin\theta}{\cos\theta}\right| - \dfrac{\bar{h}^5\sin\theta}{15\cos^3\theta} - \dfrac{2\bar{h}^5\sin\theta}{15\cos\theta} \end{array} \right)_0^{\arccos\bar{h}}$$

$$= \frac{120}{\pi H}\left( \begin{array}{l} \dfrac{\bar{h}^3\arccos\bar{h}}{3} + \dfrac{\left(\sqrt{1-\bar{h}^2}\right)^3}{90} - \dfrac{\bar{h}^4}{4}\ln\left(\dfrac{1+\sqrt{1-\bar{h}^2}}{\bar{h}}\right) + \dfrac{\bar{h}^4\sqrt{1-\bar{h}^2}}{5} - \dfrac{\bar{h}^2\sqrt{1-\bar{h}^2}}{3} \\ + \dfrac{\bar{h}^2\sqrt{1-\bar{h}^2}}{4} - \dfrac{\bar{h}^2\sqrt{1-\bar{h}^2}}{15} - \dfrac{2\bar{h}^4\sqrt{1-\bar{h}^2}}{15} \end{array} \right)$$

$$= \frac{120}{\pi H}\left( \dfrac{\bar{h}^3\arccos\bar{h}}{3} + \dfrac{\left(\sqrt{1-\bar{h}^2}\right)^3}{90} - \dfrac{\bar{h}^4}{4}\ln\left(\dfrac{1+\sqrt{1-\bar{h}^2}}{\bar{h}}\right) + \dfrac{\bar{h}^4\sqrt{1-\bar{h}^2}}{15} - \dfrac{3\bar{h}^2\sqrt{1-\bar{h}^2}}{20} \right)$$

$$= \frac{120}{\pi H}\left( \dfrac{\bar{h}^3\arccos\bar{h}}{3} + \dfrac{\sqrt{1-\bar{h}^2}}{90} - \dfrac{29\bar{h}^2\sqrt{1-\bar{h}^2}}{180} + \dfrac{\bar{h}^4\sqrt{1-\bar{h}^2}}{15} - \dfrac{\bar{h}^4}{4}\ln\left(\dfrac{1+\sqrt{1-\bar{h}^2}}{\bar{h}}\right) \right)$$

$$\begin{aligned} \mathbf{\Psi}_{\boldsymbol{\delta}_2\boldsymbol{\delta}_2}\left(h_i\right) &= 2\int_0^{\arccos\bar{h}} d\theta \int_{\bar{h}/\cos\theta}^{1} d\left(RH\right)\ \left(RH\right)\ \left(RH\right)\cos\theta F\left(R\right)\cos^2\theta \\ &= 2\int_0^{\arccos\bar{h}} \cos^3\theta d\theta \int_{\bar{h}/\cos\theta}^{1} dR\ R^2 H^3\ F\left(R\right) \\ &= 2\frac{60H^3}{\pi H^4}\int_0^{\arccos\bar{h}} \cos^3\theta d\theta \int_{\bar{h}/\cos\theta}^{1} dR\ R^2\ \left(1-R\right)^2 \\ &= \frac{120}{\pi H}\int_0^{\arccos\bar{h}} \cos^3\theta d\theta \int_{\bar{h}/\cos\theta}^{1} dR\ R^2\ \left(1-R\right)^2 \end{aligned}$$

$$\mathbf{\Psi}_{\boldsymbol{\delta}_2\boldsymbol{\delta}_2}(h_i)=\frac{120}{\pi H}\int_0^{\arccos\bar{h}} d\theta\cos^3\theta\left(\frac{1}{30}-\frac{\bar{h}^3}{3\cos^3\theta}+\frac{\bar{h}^4}{2\cos^4\theta}-\frac{\bar{h}^5}{5\cos^5\theta}\right)$$

$$=\frac{120}{\pi H}\int_0^{\arccos\bar{h}} d\theta\left(\frac{\cos^3\theta}{30}-\frac{\bar{h}^3}{3}+\frac{\bar{h}^4}{2\cos\theta}-\frac{\bar{h}^5}{5\cos^2\theta}\right)$$

$$=\frac{120}{\pi H}\left(\frac{\sin\theta}{30}-\frac{\sin^3\theta}{90}-\frac{\theta\bar{h}^3}{3}+\frac{\bar{h}^4}{2}\ln\left(\frac{1+\sin\theta}{\cos\theta}\right)-\frac{\bar{h}^5\sin\theta}{5\cos\theta}\right)_0^{\arccos\bar{h}}$$

$$=\frac{120}{\pi H}\left(-\frac{\bar{h}^3\arccos\bar{h}}{3}+\frac{\sqrt{1-\bar{h}^2}}{30}-\frac{\left(\sqrt{1-\bar{h}^2}\right)^3}{90}+\frac{\bar{h}^4}{2}\ln\left(\frac{1+\sqrt{1-\bar{h}^2}}{\bar{h}}\right)-\frac{\bar{h}^4\sqrt{1-\bar{h}^2}}{5}\right)$$

$$=\frac{120}{\pi H}\left(-\frac{\bar{h}^3\arccos\bar{h}}{3}+\frac{\sqrt{1-\bar{h}^2}}{30}-\frac{\sqrt{1-\bar{h}^2}}{90}+\frac{\bar{h}^2\sqrt{1-\bar{h}^2}}{90}+\frac{\bar{h}^4}{2}\ln\left(\frac{1+\sqrt{1-\bar{h}^2}}{\bar{h}}\right)-\frac{\bar{h}^4\sqrt{1-\bar{h}^2}}{5}\right)$$

$$=\frac{120}{\pi H}\left(-\frac{\bar{h}^3\arccos\bar{h}}{3}+\frac{\sqrt{1-\bar{h}^2}}{45}+\frac{\bar{h}^2\sqrt{1-\bar{h}^2}}{90}-\frac{\bar{h}^4\sqrt{1-\bar{h}^2}}{5}+\frac{\bar{h}^4}{4}\ln\left(\frac{1+\sqrt{1-\bar{h}^2}}{\bar{h}}\right)\right)$$

$$\psi_1(h_i)=\begin{cases}\dfrac{120}{\pi H}\left(\begin{array}{l}\dfrac{\bar{h}^3\arccos\bar{h}}{3}+\dfrac{\sqrt{1-\bar{h}^2}}{90}-\dfrac{29\bar{h}^2\sqrt{1-\bar{h}^2}}{180}\\+\dfrac{\bar{h}^4\sqrt{1-\bar{h}^2}}{15}-\dfrac{\bar{h}^4}{4}\ln\left(\dfrac{1+\sqrt{1-\bar{h}^2}}{\bar{h}}\right)\end{array}\right) & 0<r<H\\ 0 & r>H\end{cases}$$

$$\psi_2(h_i)=\begin{cases}\dfrac{120}{\pi H}\left(\begin{array}{l}-\dfrac{2\bar{h}^3\arccos\bar{h}}{3}+\dfrac{\sqrt{1-\bar{h}^2}}{90}+\dfrac{31\bar{h}^2\sqrt{1-\bar{h}^2}}{180}\\-\dfrac{4\bar{h}^4\sqrt{1-\bar{h}^2}}{15}+\dfrac{\bar{h}^4}{2}\ln\left(\dfrac{1+\sqrt{1-\bar{h}^2}}{\bar{h}}\right)\end{array}\right) & 0<r<H\\ 0 & r>H\end{cases}$$

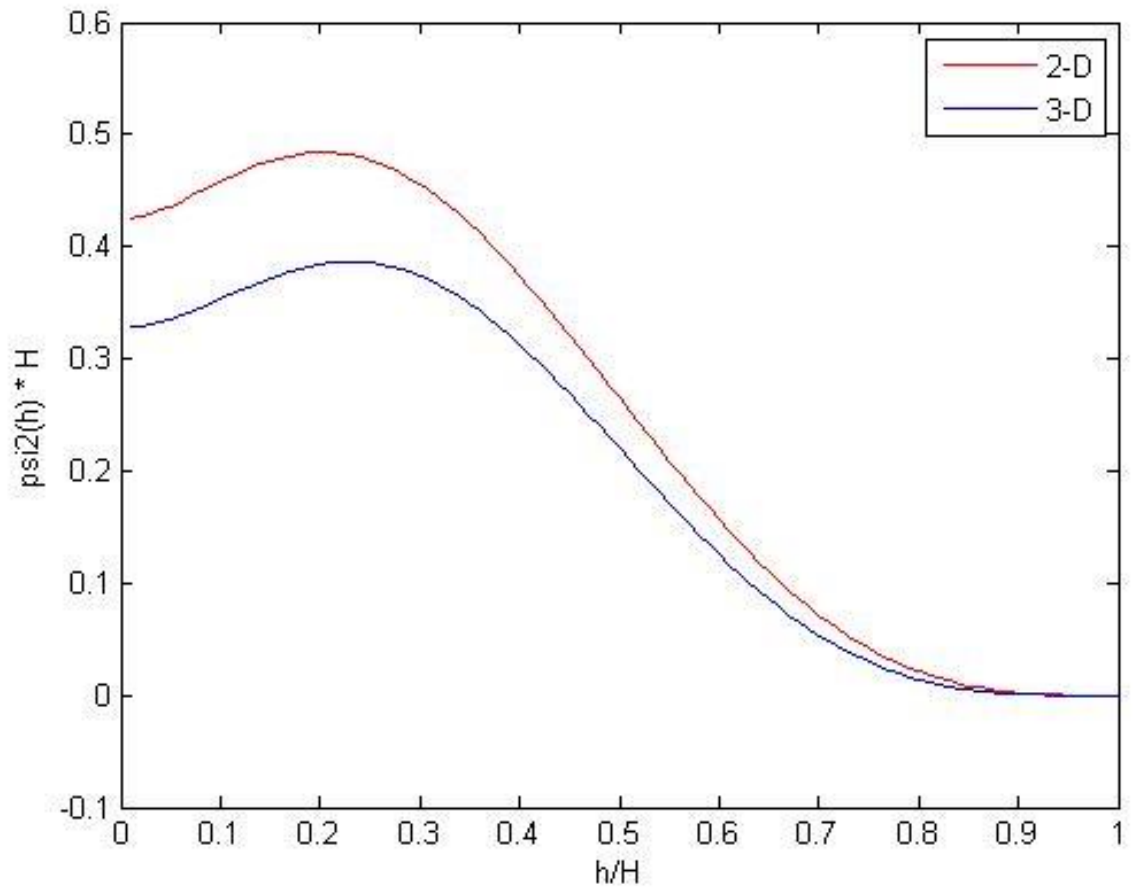

**Figure 35: Function $\psi_2\left(h_i\right)$ for Lucy kernel in 2-D (corrected) and 3-D.**

1. Table used in the integral

$$\tan\left(\frac{\alpha \pm \beta}{2}\right) = \frac{\sin\alpha \pm \sin\beta}{\cos\alpha + \cos\beta}$$

$$\int \frac{dx}{\cos ax} = \frac{1}{a}\ln\left|\tan\left(\frac{ax}{2} + \frac{\pi}{4}\right)\right| + C$$

$$\int \frac{dx}{\cos^n ax} = \frac{\sin ax}{a(n-1)\cos^{n-1} ax} + \frac{n-2}{n-1}\int \frac{dx}{\cos^{n-2} ax} \quad (\mathit{for}\ \text{n>1})$$

$$\int \frac{dx}{\cos x} = \ln\left|\tan\left(\frac{x}{2} + \frac{\pi}{4}\right)\right| + C = \ln\left|\frac{1+\sin x}{\cos x}\right| + C$$

$$\int \frac{dx}{\cos^2 x} = \frac{\sin x}{\cos x} + C$$

$$\int \frac{dx}{\cos^3 x} = \frac{\sin x}{2\cos^2 x} + \frac{1}{2}\ln\left|\frac{1+\sin x}{\cos x}\right| + C$$

$$\int \frac{dx}{\cos^4 x} = \frac{\sin x}{3\cos^3 x} + \frac{2\sin x}{3\cos x} + C$$

$$\int \frac{dx}{\cos^5 x} = \frac{\sin x}{4\cos^4 x} + \frac{3\sin x}{8\cos^2 x} + \frac{3}{8}\ln\left|\frac{1+\sin x}{\cos x}\right| + C$$

$$\int \frac{dx}{\cos^6 x} = \frac{\sin x}{5\cos^5 x} + \frac{4\sin x}{15\cos^3 x} + \frac{8\sin x}{15\cos x} + C$$

$$\int \cos^3 ax\, dx = \frac{2\sin x}{3} + \frac{\sin x \cos^2 x}{3} + C$$

2. Tensor $\frac{\mathbf{r}}{|\mathbf{r}|}\frac{\mathbf{r}}{|\mathbf{r}|}$

For 3-D

The unit vector $\frac{\mathbf{r}}{|\mathbf{r}|}$ is given as

$$\frac{\mathbf{r}}{|\mathbf{r}|} = \begin{bmatrix} \sin\theta \cos\phi \\ \sin\theta \sin\phi \\ \cos\theta \end{bmatrix}$$

Then the tensor $\frac{\mathbf{r}}{|\mathbf{r}|}\frac{\mathbf{r}}{|\mathbf{r}|}$ is in the following expression

$$\begin{aligned} \frac{\mathbf{r}}{|\mathbf{r}|}\frac{\mathbf{r}}{|\mathbf{r}|} &= \begin{bmatrix} \sin\theta \cos\phi \\ \sin\theta \sin\phi \\ \cos\theta \end{bmatrix} \begin{bmatrix} \sin\theta\cos\phi & \sin\theta\sin\phi & \cos\theta \end{bmatrix} \\ &= \begin{bmatrix} \sin^2\theta\cos^2\phi & \sin^2\theta\sin\phi\cos\phi & \sin\theta\cos\theta\cos\phi \\ \sin^2\theta\sin\phi\cos\phi & \sin^2\theta\sin^2\phi & \sin\theta\cos\theta\sin\phi \\ \sin\theta\cos\theta\cos\phi & \sin\theta\cos\theta\sin\phi & \cos^2\theta \end{bmatrix} \end{aligned}$$

This can be expressed in the sum of each $\boldsymbol{\delta}_i\boldsymbol{\delta}_j$

$$\begin{aligned} \frac{\mathbf{r}}{|\mathbf{r}|}\frac{\mathbf{r}}{|\mathbf{r}|} &= \sum_{i=1}^{3}\sum_{j=1}^{3} \boldsymbol{\delta}_i\boldsymbol{\delta}_j r_i r_j \\ &= \boldsymbol{\delta}_1\boldsymbol{\delta}_1 \sin^2\theta\cos^2\phi + \boldsymbol{\delta}_1\boldsymbol{\delta}_2 \sin^2\theta\sin\phi\cos\phi + \boldsymbol{\delta}_1\boldsymbol{\delta}_3 \sin\theta\cos\theta\cos\phi \\ &\quad + \boldsymbol{\delta}_2\boldsymbol{\delta}_1 \sin^2\theta\sin\phi\cos\phi + \boldsymbol{\delta}_2\boldsymbol{\delta}_2 \sin^2\theta\sin^2\phi + \boldsymbol{\delta}_2\boldsymbol{\delta}_3 \sin\theta\cos\theta\sin\phi \\ &\quad + \boldsymbol{\delta}_3\boldsymbol{\delta}_1 \sin\theta\cos\theta\cos\phi + \boldsymbol{\delta}_3\boldsymbol{\delta}_2 \sin\theta\cos\theta\sin\phi + \boldsymbol{\delta}_3\boldsymbol{\delta}_3 \cos^2\theta \end{aligned}$$

where $\theta \in [0,\pi]$ and $\phi \in [0,2\pi]$

For the 2-D case

The unit vector $\frac{\mathbf{r}}{|\mathbf{r}|}$ is given as

$$\frac{\mathbf{r}}{|\mathbf{r}|} = \begin{bmatrix} \sin\theta \\ \cos\theta \end{bmatrix}$$

Then the tensor $\frac{\mathbf{r}}{|\mathbf{r}|}\frac{\mathbf{r}}{|\mathbf{r}|}$ becomes

$$\begin{aligned} \frac{\mathbf{r}}{|\mathbf{r}|}\frac{\mathbf{r}}{|\mathbf{r}|} &= \begin{bmatrix} \sin\theta \\ \cos\theta \end{bmatrix} \begin{bmatrix} \sin\theta & \cos\theta \end{bmatrix} \\ &= \begin{bmatrix} \sin^2\theta & \sin\theta\cos\theta \\ \sin\theta\cos\theta & \cos^2\theta \end{bmatrix} \end{aligned}$$

This can be expressed as the sum of each dyad

$$\begin{aligned} \frac{\mathbf{r}}{|\mathbf{r}|}\frac{\mathbf{r}}{|\mathbf{r}|} &= \sum_{i=1}^{2}\sum_{j=1}^{2} \boldsymbol{\delta}_i \boldsymbol{\delta}_j r_i r_j \\ &= \boldsymbol{\delta}_1\boldsymbol{\delta}_1 \sin^2\theta + \boldsymbol{\delta}_1\boldsymbol{\delta}_2 \sin\theta\cos\theta + \boldsymbol{\delta}_2\boldsymbol{\delta}_1 \sin\theta\cos\theta + \boldsymbol{\delta}_2\boldsymbol{\delta}_2 \cos^2\theta \end{aligned}$$

Where, $\theta \in [0, \pi]$ and $\phi \in [0, 2\pi]$.